\documentclass[%
 reprint,
superscriptaddress,
nofootinbib,
 amsmath,amssymb,
 aps,
prb,
floatfix,
longbibliography,
]{revtex4-2}

\usepackage{float}
\makeatletter
\let\newfloat\newfloat@ltx
\makeatother

\usepackage{graphicx}
\usepackage{dcolumn}
\usepackage{bm}
\usepackage{makecell}
\usepackage{xcolor}
\usepackage{amsmath}
\usepackage{float}
\usepackage{caption}
\usepackage{subcaption}
\usepackage{multirow}
\usepackage{amsfonts}
\usepackage{textcomp, gensymb}
\usepackage{algpseudocode}
\usepackage{algorithm}
\usepackage{hyperref}
\hypersetup{
    colorlinks=true,
    linkcolor=blue,
    filecolor=magenta,
    urlcolor=cyan,
    pdfpagemode=FullScreen,
    }
\usepackage{xr}
\usepackage[bottom]{footmisc}

\newcommand{\mM}{\mathbb{M}}

\newcommand{\bG}{\mathbf{G}}
\newcommand{\bD}{\mathbf{D}}
\newcommand{\bd}{\mathbf{d}}
\newcommand{\br}{\mathbf{r}}
\newcommand{\bth}[1]{\mathbf{\Theta}_{#1}}

\newcommand{\bI}{\mathbf{I}}
\newcommand{\bQ}{\mathbf{Q}}
\newcommand{\angstrom}{\mbox{\normalfont\AA}}

\newcommand{\ihatnospace}{\bm\hat{\textbf{\i}}}
\newcommand{\jhatnospace}{\bm\hat{\textbf{\j}}}
\newcommand{\khatnospace}{\bm\hat{\textbf{k}}}

\newcommand{\ihat}{\,\ihatnospace}
\newcommand{\jhat}{\,\jhatnospace}
\newcommand{\khat}{\,\khatnospace}

\newcommand{\Nlay}{N_{\rm layer}}
\newcommand{\Ntbg}{N_{\rm TBG}}

\newcommand{\Eqn}[1]{Eqn.~(\ref{#1})}
\newcommand{\eqn}[1]{eqn.~(\ref{#1})}
\newcommand{\eqns}[1]{eqns.~(\ref{#1})}
\newcommand{\Eqns}[1]{Eqns.~(\ref{#1})}
\newcommand{\eqnx}[1]{(\ref{#1})}
\newcommand{\sect}[1]{Section~\ref{#1}}
\newcommand{\sects}[1]{Sections~\ref{#1}}
\newcommand{\sectx}[1]{\ref{#1}}
\newcommand{\fig}[1]{Fig.~\ref{#1}}
\newcommand{\figs}[1]{Figs.~\ref{#1}}
\newcommand{\figx}[1]{\ref{#1}}
\newcommand{\tab}[1]{Table~\ref{#1}}
\newcommand{\tabs}[1]{Tables~\ref{#1}}
\newcommand{\tabx}[1]{\ref{#1}}
\newcommand{\app}[1]{Appendix~\ref{#1}}

\renewcommand{\arraystretch}{1.1}

\graphicspath{{figures/}{./}}

\begin{document}

\title{Elastic plate basis for the deformation and electron diffraction of twisted bilayer graphene on a substrate}

\author{Moon-ki Choi}
\affiliation{Department of Aerospace Engineering and Mechanics, University of Minnesota, Minneapolis, Minnesota 55455, USA}

\author{Suk Hyun Sung}
\affiliation{Department of Materials Science and Engineering, University of Michigan, Ann Arbor, Michigan 48109, USA}

\author{Robert Hovden}
\affiliation{Department of Materials Science and Engineering, University of Michigan, Ann Arbor, Michigan 48109, USA}

\author{Ellad B. Tadmor}
\email[Corresponding author: ]{tadmor@umn.edu}
\affiliation{Department of Aerospace Engineering and Mechanics, University of Minnesota, Minneapolis, Minnesota 55455, USA}


\begin{abstract}
A basis is derived from elastic plate theory that quantifies equilibrium and dynamic deformation and electron diffraction patterns of twisted bilayer graphene (TBG). The basis is derived by solving in-plane and out-of-plane normal modes of an unforced parallelogram elastic plate. We show that a combination of only a few basis terms successfully captures the relaxed TBG structure with and without an underlying substrate computed using atomistic simulations. The results are validated by comparison with electron diffraction experiments.
A code for extracting the elastic plate basis coefficients from an experimental electron diffraction image accompanies this paper.
TBG dynamics are also studied by computing the phonon band structure from atomistic simulations. Low-energy phonons at the $\Gamma$ point are examined in terms of the mode shape and frequency. These modes are captured by simple elastic plate models with uniformly distributed springs for inter-layer and substrate interactions.
\end{abstract}

\maketitle

\section{Introduction}
Van der Waals (vdW) multilayers (or ``heterostructures'') are stacked two-dimensional (2D) materials with a controlled order and orientation between layers. These materials have novel properties compared to their monolayer and bulk counterparts that depend sensitively on the type and ordering of the layers and on their mechanical deformation 
\cite{nimbalkar2020,Peng2020,yu2021,lemme2022}. Given the large number of potential 2D materials \cite{mounet:gibertini:2018}, 2D heterostructures have great potential for tunable nanotechnology applications, such as ultra-sensitive sensors, flexible/low-power electronics, and semiconductor devices \cite{cui2015,glavin2020}.

The prototypical example of a 2D heterostructure is twisted graphene bilayer (TBG), which consists of two layers of graphene that are stacked and rotated relative to each other by a specified angle. The twist between layers in a TBG generates a moir\'e pattern in which circular high energy domains of AA stacking are surrounded by low-energy AB and BA domains separated by SP solitons \cite{zhang2018,yoo2019}. The size of the pattern depends on the magnitude of the twist angle, with the separation between AA domains increasing with decreasing twist angle. Energy relaxation leads to an additional localized twisting centered on the AA domains that reduces their diameter while increasing the size of the low-energy AB and BA domains creating a triangular domain pattern that is observed in electron microscopy experiments \cite{alden2013,wijk:schuring:2015,zhang2018,guinea2019,sung2022}. The restructuring associated with the twist changes the electronic structure \cite{nam2017,pathak2022}, and at a special ``magic angle'' of $1.1\degree$ even leads to an unconventional form of superconductivity \cite{cao2018,lu2019,yankowitz2019}.

The dynamics of TBG also exhibit unique characteristics due to the periodic moir\'e structure and the presence of the different stackings. Alden et al.\ \cite{alden2013} observed dynamic motion and reconstruction of SP solitons in dark-field transmission electron microscopy (TEM) at a high temperature. de Jong et al.\ \cite{de2022} studied the moir\'e pattern in a TBG using low-energy electron microscopy and found that the local disorder in moir\'e supercells decreases with thermal annealing. Gadelha et al.\ \cite{gadelha2021} measured the phonon band structure of graphene unit cells within the moir\'e supercell in a TBG using Raman spectroscopy and found that the phonon band exhibits divergent frequencies depending on the location of the unit cell.

Given the fundamental importance of TBG, this heterostructure has been studied extensively using a variety of theoretical approaches including continuum mechanics \cite{guinea2019}, ab initio calculations \cite{cantele2020}, multiscale modeling \cite{zhang2018,leconte2022,lu2022}, dislocation theory \cite{annevelink2020}, and the periodic lattice distortion (PLD) model \cite{sung2022}. In this paper, we study the static and dynamic deformation of TBG using atomistic simulations and propose a new approach for characterizing TBG deformation and its effect on electron diffraction patterns in terms of a modal basis obtained from a continuum analysis using the theory of elastic plates. This approach provides a highly efficient framework for characterizing TBG behavior, which can be readily generalized to other 2D heterostructures.

Atomistic simulations are performed for a range of twist angles for both a free-standing TBG and a TBG suspended over two common substrates: hBN and Si$_3$N$_4$. Two different interatomic potentials are used: a physics-based AIREBO potential with DRIP interlayer interactions \cite{stuart2000,wen2018}, and a machine learning based hybrid Neural Network (hNN) potential \cite{wen2019}. The interactions with the substrate are modeled using a using Lennard--Jones (LJ) potential. The in-plane and out-of-plane deformations of the TBG are quantified using a normal mode basis obtained from a continuum elastic plate theory for monolayer graphene for the same parallelogram periodic cell used in the atomistic simulations. It is found that TBG deformation is well represented by a linear combination of a small number of modes from this basis. To directly compare the atomistic simulation results with experimental electron diffraction images, optimized basis coefficients for a given experimental electron diffraction image are computed. This is done by simulating the electron diffraction of the TBG deformation, where degrees of freedom are reduced by employing the basis. A Matlab code for computing these optimized coefficients is provided.

The effects of the substrate type and twist angle on the dynamics of TBG are examined by computing a phonon band structure via atomistic simulations. To reduce the cost of computing the dynamical matrix at low twist angles where the TBG supercell can be very large, we propose an algorithm that allows us to compute the dynamical matrix at any value of $\mathbf{k}$ from the dynamical matrix at $\mathbf{k}=0$ by using a cutoff distance that is small compared with the simulation box size. In the obtained phonon band structure, we focus on low-energy phonons at the $\Gamma$ point that appears in every TBG supercell regardless of boundary conditions. We analyze these phonons in terms of the frequency and mode shape. Additionally, we predict the frequencies of these $\Gamma$-point phonons using a simplified double elastic plate model that has uniformly distributed springs to account for the weak interlayer and substrate interactions.

\section{TBG atomistic simulation}
\subsection{TBG model}\label{sec:tbg_model}
The atomistic structure of TBG is constructed by using the Atomic Simulation Environment package \cite{larsen2017}. Two graphene layers are initially arranged in AA stacking, and then the top and bottom layers are twisted by $\theta/2$ and $-\theta/2$, respectively. This process generates a periodic nonorthogonal moir\'e supercell defined by the vectors
\begin{equation}\label{eqn:cellvecs}
\mathbf{a}_1 = L\left(\frac{\sqrt{3}}{2}\ihat-\frac{1}{2}\jhat\right), \quad
\mathbf{a}_2 = L\jhat,
\end{equation}
where $\ihatnospace$ and $\jhatnospace$ are unit vectors along the Cartesian directions, $L$ is the distance between AA domain centers \cite{shallcross2010},
\begin{equation}\label{eqn:LAA}
    L = \frac{\sqrt{3}a}{2 \sin{\left(\theta/2\right)}},
\end{equation}
and $a$ is a graphene bond length. The atomistic simulation box is taken to be coincident with the moir\'{e} supercell with the origin located at the center of the AA domain. The initial interlayer distance between the top and bottom layers is set to 3.44 $\angstrom$. The interactions between the atoms are modeled using two interatomic potentials (IPs) denoted IP$_1$ and IP$_2$ (see \tab{tab:ip_tbg}). IP$_1$ is the physics-based Adaptive Intermolecular Reactive Empirical Bond Order (AIREBO) potential \cite{stuart2000} with interlayer interactions modeled using the Dihedral-angle-corrected Registry-dependent Interlayer Potential (DRIP) \cite{wen2018}. IP$_2$ is a machine learning based hybrid Neural Network (hNN) potential that models both the intra- and inter-layer interactions \cite{wen2019}. 
Both IP$_1$ and IP$_2$ successfully capture registry effects between the graphene layers, which is crucial for TBG relaxation \cite{wen2018,wen2019}. A detailed comparison of the predictions of IP$_1$ and IP$_2$ with other IPs and DFT and experimental results is presented in Table~2 of \cite{wen2019}.
In each case, the TBG is generated using the value of $a$ predicted by the IP as shown in \tab{tab:ip_tbg}. For comparison, the experimental value is $a=1.42$~\AA\;\cite{cooper2012}.

\begin{table}\centering
\begin{tabular}{c|c|c|c}
\hline
& Intra-layer interaction & Inter-layer interaction & $a$ \\
\hline
IP$_1$ & AIREBO & DRIP & 1.397 $\angstrom$\\
\hline
IP$_2$ & \multicolumn{2}{c|}{hNN} &  1.425 $\angstrom$ \\
\hline
\end{tabular}
\captionsetup{justification=raggedright, singlelinecheck=false}
\caption{IPs used to model TBG interactions and their prediction for $a$, the monolayer graphene lattice constant.}
\label{tab:ip_tbg}
\end{table}

Atomistic simulations of the TBG are performed via the Large-scale Atomic/Molecular Massively Parallel Simulator (LAMMPS) \cite{plimpton1995}.
The initial relaxed TBG structure is obtained by allowing the simulation box to change until the in-plane pressure becomes zero and the forces on all atoms are zero. This is done in two steps.
First, the Fast Inertial Relaxation Engine (FIRE) method \cite{guenole2020} is used to minimize the energy with respect to the positions of the atom while keeping the simulation box fixed, then the conjugate-gradient (CG) method is used to minimize the energy with respect to both atom positions and box size.
FIRE and CG are used alternatively until the force norm is less than 10$^{-8}$ eV/$\angstrom$.
This approach is necessary because FIRE is not implemented to minimize the box shape, and minimization with CG alone does not achieve sufficient accuracy. The results are presented in \sect{sec:relaxation} after discussing the elastic plate basis in \sect{sec:normal_mode_basis}.

\subsection{TBG-substrate interaction}\label{sec:tbg_sub_interaction}
In experiments, TBGs can be free standing or supported on a substrate. There are several common substrates used including SiO$_2$/Si, hBN, Si$_3$N$_4$, and Cu depending on the synthesis technique \cite{cai2021,cao2021,yoo2019}.\footnote{In practice, TBGs can be sandwiched between a substrate and an overlayer, e.g., between Si$_3$N$_4$ and hBN \cite{sung2022}. Here we only consider TBGs suspended over a single substrate.} To simulate the influence of the substrate on the TBG, we use a LJ potential to model the vdW interactions between carbon atoms in the graphene layers and atoms in the substrate. The LJ potential between two atoms at distance $r$ is given by
\begin{equation}\label{eqn:LJ}
    \phi(r)= 4\epsilon
    \left[\left(\frac{\sigma}{r}\right)^{12} - \left(\frac{\sigma}{r}\right)^6\right],
\end{equation}
where $\epsilon$ and $\sigma$ are energy and length parameters that depend on the species of the interacting atoms. In order not to explicitly model all atoms in the substrate, the substrate is approximated as a continuum of atoms at a specified density\footnote{Note that in the limiting cases of near alignment of graphene with the hBN substrate, a registry-dependent potential (such as that of Leven et al.\cite{leven:maaravi:2016}) should be used for accurate results, precluding this continuum approximation. Such calculations were not pursued here due to their expense, and since our focus is on developing a method for representing TBG deformation rather than performing the most accurate possible atomistic calculations. Repeating our calculations with a more accurate modeling of the substrate interactions is an interesting direction for future work.}. The interaction of a single carbon atom in a graphene layer with the substrate is then given by an effective potential $U_{\infty}$ obtained by integrating \eqn{eqn:LJ} over all atoms in the substrate represented by a three-dimensional half plane:
\begin{multline}\label{eqn:LJ_infinite_int}
    U_{\infty}(\sigma,\epsilon,z_0)= \\ \int_{-\infty}^{\infty}\int_{\infty}^{\infty}\int_{-\infty}^{0}
    \rho_{\rm sub}\phi\left(r\left(x,y,z,z_0\right)\right)dxdydz,
\end{multline}
where $\rho_{\rm sub}$ is the number density of a substrate atom per unit volume, $r(x,y,z,z_0)=\sqrt{x^2+y^2+(z_0-z)^2}$, and $z_0$ is the height of a carbon atom above the top surface of the substrate. Switching to cylindrical coordinates, \eqn{eqn:LJ_infinite_int} becomes
\begin{multline}\label{eqn:LJ_infinite_int_cyl}
    U_{\infty}(\sigma,\epsilon,z_0)= \\ \int_{0}^{\infty}\int_{0}^{2\pi}\int_{-\infty}^{0}
    \rho_{\rm sub}\phi\left(r(R,\theta,z,z_0)\right)RdRd\theta dz,
\end{multline}
where $r(R,\theta,z,z_0) = \sqrt{R^2+(z-z_0)^2}$. \Eqn{eqn:LJ_infinite_int_cyl} has a closed-fofrm solution as
\begin{multline}\label{eqn:LJ_infinite}
    U_{\infty}(\rho_{\rm sub},\sigma,\epsilon,z_0)= \\ \frac{4}{3}\pi\rho_{\rm{sub}}\epsilon\sigma^3
    \left[\frac{1}{15}\left(\frac{\sigma}{z_0}\right)^9 - \frac{1}{2}\left(\frac{\sigma}{z_0}\right)^3\right].
\end{multline}
The force on a carbon atom in the $z$-direction is the negative derivative of $U_\infty$ with respect to $z_0$, given by
\begin{multline}\label{eqn:sub_LJ}
    f_{\infty}(\rho_{\rm sub},\sigma,\epsilon,z_0)= \\ 
    -\frac{4}{3}\pi\rho_{\rm{sub}}\epsilon\sigma^2
    \left[\frac{3}{5}\left(\frac{\sigma}{z_0}\right)^{10} - \frac{3}{2}\left(\frac{\sigma}{z_0}\right)^{4}\right].
\end{multline}
In this paper, we focus on two substrates commonly used in experiments: hBN and Si$_3$N$_4$ \cite{sung2022,yoo2019}.\footnote{Since graphene and hBN have a similar lattice constant, registry effects between them can influence TBG relaxation. We do not consider this effect in the present paper.} Since these substrates contain two different kinds of atoms, the total energy of a carbon atom above the substrate is given by
\begin{multline}\label{eqn:ext_energy}
U_{\rm ext}(z_0) = \\ U_{\infty}(\rho_a,\sigma_{{\rm C}-a},\epsilon_{{\rm C}-a},z_0)+U_{\infty}(\rho_b,\sigma_{{\rm C}-b},\epsilon_{{\rm C}-b},z_0),
\end{multline}
where $a$ and $b$ are the substrate atom type, and ``ext'' indicates that this energy is external to the carbon--carbon interactions in the TBG. The corresponding force in the $z$ direction is
\begin{multline}
\label{eqn:ext_force}
f_{\rm ext}(z_0) = \\ f_{\infty}(\rho_a,\sigma_{{\rm C}-a},\epsilon_{{\rm C}-a},z_0)+f_{\infty}(\rho_b,\sigma_{{\rm C}-b},\epsilon_{{\rm C}-b},z_0).
\end{multline}

\begin{table}\centering
\begin{tabular}{c|c|c}
\hline
 & $\epsilon$ (meV) & $\sigma$ (nm)\\
\hline
C-B & 3.24 & 0.34\\
C-N$^1$ & 4.07 & 0.34\\
C-Si & 10.46 & 0.39\\
C-N$^2$ & 21.36 & 0.37\\
\hline
\end{tabular}
\captionsetup{justification=raggedright, singlelinecheck=false}
\caption{LJ parameters for the interaction of carbon and substrate atoms (B, N$^1$, Si, and N$^2$)\cite{baowan2007,mayo1990,heinz2013}.}
\label{tab:LJ_parameters}
\end{table}

The LJ parameters ($\epsilon$ and $\sigma$) used in the atomistic simulations with the hBN and Si$_3$N$_4$ substrates are given in \tab{tab:LJ_parameters} \cite{baowan2007,mayo1990}. Although nitrogen (N) appears in both substrates, different coefficients are used in each case. To differentiate between the two cases, we denote the nitrogen atoms as N$^1$ and N$^2$ for the hBN and Si$_3$N$_4$ substrates, respectively. The number densities of B, N$^1$, Si, and  N$^2$ are 0.077, 0.077, 0.042, and 0.056 atoms/$\angstrom^3$, respectively 
\section{In-plane and out-of-plane elastic plate basis }\label{sec:normal_mode_basis}
To quantify TBG deformation, an elastic plate basis is derived in this section. In \sects{sec:in_plane_basis} and \sectx{sec:out_of_plane_basis}, in-plane and out-of-plane bases are derived from the normal modes of a periodic continuum elastic plate with a parallelogram supercell. Using the derived bases, the deformation of each layer of a relaxed TBG with and without underlying substrates is quantified in \sect{sec:relaxation}.

\subsection{In-plane elastic plate basis} \label{sec:in_plane_basis}
The in-plane wave equations of an elastic plate are \cite{arreola2015}:
\begin{align}
    \frac{\partial N_{xx}}{\partial x}+\frac{\partial N_{xy}}{\partial y}=\rho h\frac{\partial^2u}{\partial t^2}, \label{eqn:in_plane_gov1}\\
    \frac{\partial N_{xy}}{\partial x}+\frac{\partial N_{yy}}{\partial y}=\rho h\frac{\partial^2v}{\partial t^2},
    \label{eqn:in_plane_gov2}
\end{align}
where $N_{ij}$ are the stress resultants with units of force per length, $h$ is the plate thickness, $\rho$ is the mass density, $u$ and $v$ are the displacements along the $x$ and $y$ Cartesian directions, and $t$ is the time. The relation between the in-plane stress residual components and the in-plane strain components ($\epsilon_{xx}$, $\epsilon_{yy}$, $\epsilon_{xy}$) follows from Hooke's law,
\begin{equation} \label{eqn:stress_strain}
\begin{bmatrix}
N_{xx} \\
N_{yy} \\
N_{xy}
\end{bmatrix}
=
C
\begin{bmatrix}
1 & \nu & 0 \\
\nu & 1 & 0 \\
0 & 0 & 1-\nu
\end{bmatrix}
\begin{bmatrix}
\epsilon_{xx} \\
\epsilon_{yy} \\
\epsilon_{xy}
\end{bmatrix},
\end{equation}
where
\begin{equation}
C=\frac{Eh}{1-\nu^2},
\end{equation}
and $E$ and $\nu$ are Young's modulus and Poisson's ratio, respectively. The strain--displacement relations are
\begin{equation} \label{eqn:strain_disp}
    \epsilon_{xx}=\frac{\partial u}{\partial x}, \quad
    \epsilon_{yy}=\frac{\partial v}{\partial y}, \quad
    \epsilon_{xy}=\frac{1}{2}\left(\frac{\partial u}{\partial y}+\frac{\partial v}{\partial x}\right).
\end{equation}
A series solution to \eqns{eqn:in_plane_gov1} and \eqnx{eqn:in_plane_gov2} for the normal modes of an elastic plate subject to different boundary conditions can be obtained using the plane wave expansion method (PWEM) \cite{griffin1974,manzanares2010,arreola2015}. For the periodic supercell in \eqn{eqn:cellvecs}, the in-plane displacements can be written as a Fourier expansion,
\begin{align}
        u(\br,t) &= 
\sum_{\bG}\exp(-i \omega^{\rm in} (\bG) t)\zeta_{\bG}\exp(i(\bG\cdot\br+\phi_{\bG})), 
\label{eqn:disp_expansions_1}
\\
    v(\br,t) &= 
\sum_{\bG}\exp(-i \omega^{\rm in} (\bG) t)\xi_{\bG}\exp(i(\bG\cdot\br+\phi_{\bG})),
\label{eqn:disp_expansions_2}
\end{align}
where $\omega^{\rm in}(\bG)$ is the in-plane angular frequency (related to the ordinary frequency $f^{\rm in}$ through $\omega^{\rm in}=2\pi f^{\rm in}$), $\mathbf{r}=x\ihat+y\jhat$ is a position in 2D space, $\phi_{\bG}$ is a phase shift, $\zeta_{\bG}$ and $\xi_{\bG}$ are coefficients associated with the reciprocal lattice vectors $\bG$ defined on the supercell in terms of integers $m$ and $n$,
\begin{multline} \label{eqn:Gmn}
\bG(m,n)= \\ G_x \ihat + G_y \jhat = \frac{4\pi}{\sqrt{3}L}\Big(\big(m+\frac{1}{2}n\big)\ihat + \frac{\sqrt{3}}{2}n\jhat\Big),
\end{multline}
where $L$ is defined in \eqn{eqn:LAA}.

Substituting \eqns{eqn:disp_expansions_1} and \eqnx{eqn:disp_expansions_2} into the governing equations in \eqns{eqn:in_plane_gov1} and \eqnx{eqn:in_plane_gov2} after using the strain--displacement relations and Hooke's law, give for each $\bG$:
\begin{equation} \label{eqn:eigen_matrix}
\begin{bmatrix}
\mM_{xx}(\bG) && \mM_{xy}(\bG) \\
\mM_{yx}(\bG) && \mM_{yy}(\bG)
\end{bmatrix}
\begin{bmatrix}
\zeta_{\bG} \\
\xi_{\bG}
\end{bmatrix}
= \rho h (\omega^{\rm in}(\bG))^2
\begin{bmatrix}
\zeta_{\bG} \\
\xi_{\bG}
\end{bmatrix},
\end{equation}
where
\begin{multline}\label{eqn:Mmatrix}
\begin{bmatrix}
 \mM_{xx}(\bG) &  \mM_{xy}(\bG) \\
 \mM_{yx}(\bG) &  \mM_{yy}(\bG)
\end{bmatrix}
= \\
\frac{1}{2}C
\begin{bmatrix}
2G_x^2 + (1-\nu)G_y^2 & (1+\nu) G_x G_y \\
(1+\nu) G_x G_y & (1-\nu) G_x^2 + 2 G_y^2
\end{bmatrix}.
\end{multline}

The eigenproblem in \eqn{eqn:eigen_matrix} gives two eigenvalues and eigenvectors for each vector $\bG(m,n)$ defined in \eqn{eqn:Gmn}. The eigenvalues (which define the in-plane frequencies) are:
\begin{multline}\label{eqn:eigenvalue_G}
\lambda(m,n,\ell)=\rho h (\omega^{\rm in}(\bG(m,n),\ell))^2
= \\ 
\begin{cases}
C(G_x(m,n)^2+G_y(m,n)^2) & \text{for}\ \ell=1, \\[6pt]
\frac{1}{2}C(1-\nu)\big[G_x(m,n)^2+G_y(m,n)^2\big] & \text{for}\ \ell=2,
\end{cases}
\end{multline}
and the eigenvectors are for $n\ne0$:
\begin{multline}
\bd(m,n,\ell) = \zeta_{\bG(m,n)}(\ell) \ihat + \xi_{\bG(m,n)}(\ell) \jhat
= \\ 
\begin{cases}
\frac{G_x(m,n)\ihat + G_y(m,n)\jhat}{\sqrt{G_x(m,n)^2+G_y(m,n)^2}}
 & \text{for}\ \ell=1, \\[12pt]
\frac{-G_y(m,n)\ihat + G_x(m,n)\jhat}{\sqrt{G_x(m,n)^2+G_y(m,n)^2}}
 & \text{for}\ \ell=2,
\end{cases}
\end{multline}
for $n=0$:
\begin{multline}
\bd(m,n,\ell) =  \\ \zeta_{\bG(m,n)}(\ell) \ihat + \xi_{\bG(m,n)}(\ell) \jhat
=
\begin{cases}
-\ihat & \text{for}\ \ell=1, \\[6pt]
-\jhat & \text{for}\ \ell=2, \\
\end{cases}
\end{multline}
for $2m+n=0$:
\begin{multline}
\bd(m,n,\ell) = \\ \zeta_{\bG(m,n)}(\ell) \ihat + \xi_{\bG(m,n)}(\ell) \jhat
=
\begin{cases}
-\jhat & \text{for}\ \ell=1, \\[6pt]
-\ihat & \text{for}\ \ell=2. \\
\end{cases}
\end{multline}
We note that the eigenvectors are independent of the elastic constants.

We construct 120 $\bG$ vectors for combinations of the integers $m$ and $n$ in the range [-5,5] except for $m=n=0$. Retaining only unique vectors (up to inversion) leaves 60 $\bG$ vectors that are used to compute the corresponding eigenvectors and eigenvalues. We order the results be increasing values of the eigenvalues (and hence increasing frequency) using the simplified notation $\lambda_k$ (and $\omega^{\rm in}_k$) with associated eigenvectors $\bd_k$, where $k\in[1,120]$. To make it easier to compare IP results, we use the same eigenvalue ordering for AIREBO, hNN, and experiment using the elastic properties of AIREBO. A specific $\bG(m,n)$ associated with $k$ is denoted as $\bar{\bG}_k$. Note that there are two $\bar{\bG}_k$'s for each $\bG(m,n)$ since there are two eigenvalues for each $\bG$ vector.

\begin{table}\centering
\begin{tabular}{c|c|c|c|c}
\hline
 Source & $E$ (GHz) & $h\;(\angstrom)$ & $\nu$ & D (eV) \\
\hline
AIREBO & 827.99 & 3.44 & 0.36 & 1.22 \\
hNN & 961.11 & 3.44 & 0.24 & 1.234\\
Experiments & 809.9-928.2  & 3.44 & 0.3$^*$ & 1.2-1.7 \\
\hline
\end{tabular}
\captionsetup{justification=raggedright, singlelinecheck=false}
\caption{Elastic properties of monolayer graphene for the IPs used in this work \cite{choi2023} and experiments \cite{lee2008,zhang2012,han2020,blees2015}. $^*$Note that Poisson's ratio is not measured in the reported experiments. We set $\nu = 0.3$.}
\label{tab:e_ip_test}
\end{table}

The eigenvalues are computed using the elastic properties ($E$ and $\nu$) and the effective thickness $h$ of monolayer graphene given in \tab{tab:e_ip_test}. The table also includes the bending stiffness $D$ used in \sect{sec:out_of_plane_basis}.  The experimental values are from \cite{lee2008,zhang2012,han2020,blees2015}. The properties for the IPs AIREBO and hNN are computed via atomistic simulations of uniaxial tension and bending of monolayer MoS$_2$ \cite{choi2023}. The density of the monolayer graphene ($\rho$) is 2273.04 kg/m$^3$.

The first 12 elements of $f_k^{\rm in}=\omega_k^{\rm in}/2\pi$, $\bd_k$ and $\bar{\bG}_k$ are presented in \tab{tab:in_plane_mode} for a TBG twist angle of $2.28 \degree$ for which $L=61.74\;\angstrom$. The results are arranged in sets of three that have the same frequency $f_k^{\rm in}$ (i.e., $f_{1,2,3}^{\rm in}$, $f_{4,5,6}^{\rm in}$, $f_{7,8,9}^{\rm in}$, and $f_{10,11,12}^{\rm in}$). The three $\bar{\bG}_{k}$ vectors sharing the same frequency exhibit a 3-fold symmetry with each $\bar{\bG}_k$ rotating by $120\degree$ as a result of the periodic supercell. 

The in-plane modes in \tab{tab:in_plane_mode} can be used as a basis for characterizing the in-plane deformation of graphene layers in a TBG.  Referring to the displacements in \eqns{eqn:disp_expansions_1} and \eqnx{eqn:disp_expansions_2}, mode $k$ in vector form is
\begin{equation}\label{eqn:in_plane_mode}
\bth{k}^{\rm in}(\br)=\bd_k\exp{(i(\bar{\bG}_k\cdot\mathbf{r}+\phi_{\bar{\bG}_k}))},
\end{equation}
where $\phi_{\bar{\bG}_k}$ is a phase shift. Basis vectors with the same frequency, $k=1,2,3$, $k=4,5,6$, etc.\ (see \tab{tab:in_plane_mode}) will have the same coefficients and are therefore combined into single summed terms, e.g., $\sum_{k=1}^3 \bth{k}^{\rm in}(\br)$, $\sum_{k=4}^6 \bth{k}^{\rm in}(\br)$, etc. Imposing the 6-fold symmetry of the TBG structure on the combined basis terms leads to the result that the phase angle for all in-plane basis terms is $\phi_{\bar{\bG}_k}=\phi^{\rm in}=-\pi/2$ (see \app{app:phasederiv} for the proof).

The real parts of in-plane modes with different $k$ are orthogonal to each other in terms of an inner product defined as an integral over the supercell domain (see Supplementary material (SM) \cite{SM}). The vectors $\bd_k$ and $\bar{\bG}_k$ are given in \tab{tab:in_plane_mode}.

\begin{table}\centering
\renewcommand{\arraystretch}{1.3}
\begin{tabular}{c|c|c|c}
\hline
\multicolumn{2}{c|}{$f^{\rm in}_{k}$ (GHz)} & $\bd_{k}$ & $\bar{\bG}_{k}$ \\
\hline
\multirow{3}{*}{$f^{\rm in}_{1,2,3}$} & \multirow{3}{10em}{IP: 2157.16$^{\rm{a}}$, 2436.77$^{\rm{b}}$ Exp: 2187.91--2342.26$^{\rm{c}}$}  & $[-\frac{\sqrt{3}}{2},\frac{1}{2}]^\mathrm{T}$ & $\bG(0,1)$ \\
& & $[0,-1]^\mathrm{T}$ & $\bG(-1,0)$
\\
& & $[\frac{\sqrt{3}}{2},\frac{1}{2}]^\mathrm{T}$ & $\bG(1,-1)$
\\
\hline
\multirow{3}{*}{$f^{\rm in}_{4,5,6}$} & \multirow{3}{10em}{IP: 3736.31$^{\rm{a}}$, 3962.62$^{\rm{b}}$ Exp: 3698.25--3959.14$^{\rm{c}}$}   & $[\frac{1}{2},\frac{\sqrt{3}}{2}]^\mathrm{T}$ & $\bG(0,1)$  \\
& & $[-1,0]^\mathrm{T}$ & $\bG(-1,0)$
\\
& & $[\frac{1}{2},-\frac{\sqrt{3}}{2}]^\mathrm{T}$ & $\bG(1,-1)$
\\
\hline
\multirow{3}{*}{$f^{\rm in}_{7,8,9}$} &
\multirow{3}{10em}{IP: 3834.99$^{\rm{a}}$, 4220.61$^{\rm{b}}$ Exp: 3789.58--4056.92$^{\rm{c}}$}    & $[\frac{1}{2},\frac{\sqrt{3}}{2}]^\mathrm{T}$ & $\bG(2,-1)$ \\
& & $[-1,0]^\mathrm{T}$ & $\bG(-1,2)$
\\
& & $[\frac{1}{2},-\frac{\sqrt{3}}{2}]^\mathrm{T}$ & $\bG(-1,-1)$
\\
\hline
\multirow{3}{*}{$f^{\rm in}_{10,11,12}$} & \multirow{3}{10em}{IP: 4314.32$^{\rm{a}}$, 4873.54$^{\rm{b}}$ Exp: 4375.83--4684.53$^{\rm{c}}$}    & $[-\frac{\sqrt{3}}{2},\frac{1}{2}]^\mathrm{T}$ & $\bG(0,2)$ \\
& & $[0,-1]^\mathrm{T}$ & $\bG(-2,0)$
\\
& & $[\frac{\sqrt{3}}{2},\frac{1}{2}]^\mathrm{T}$ & $\bG(2,-2)$
\\
\hline
\end{tabular}
\captionsetup{justification=raggedright, singlelinecheck=false}
\caption{First 12 elements of $f_k^{\rm in}$, $\bd_k$, and $\bar{\bG}_k$ for a twist angle of $2.28 \degree$. The frequncies $f_k^{\rm in}$ are computed using the elastic properties of monolayer graphene in \tab{tab:e_ip_test} for a) AIREBO, b) hNN, and c) experiments
\cite{lee2008,zhang2012,lee2012}. The components of $\mathbf{d}_k$ are given relative to a Cartesian basis ($\ihat$ and $\jhat$).}
\label{tab:in_plane_mode}
\end{table}
\begin{figure}\centering
     \begin{subfigure}[b]{0.22\textwidth}
         \centering
         \includegraphics[trim = 0mm 0mm 0mm 0mm, clip=true,width=\textwidth]{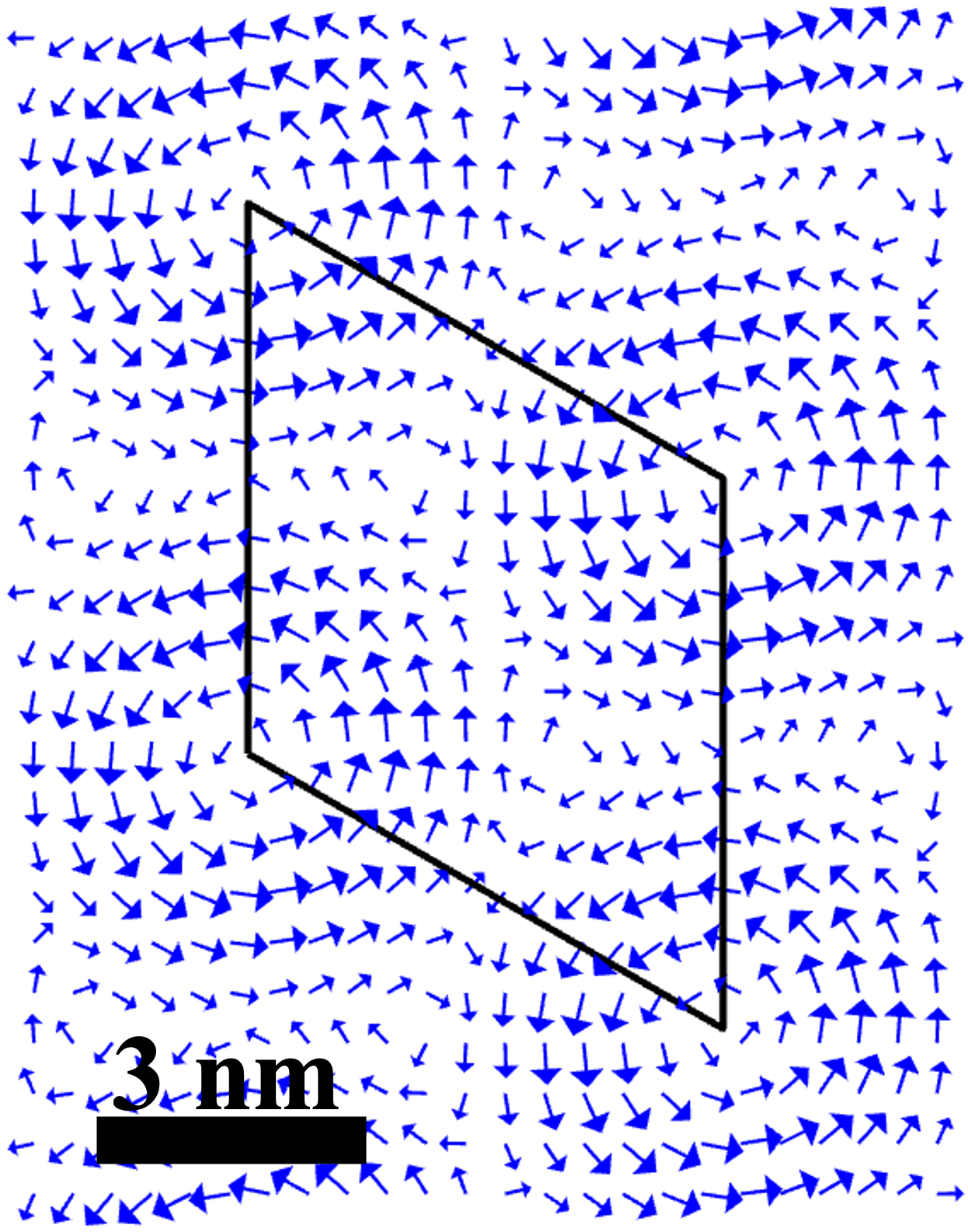}
         \caption{}
         \label{fig:in_plane_mode:1}
     \end{subfigure}
    \begin{subfigure}[b]{0.22\textwidth}
         \centering
         \includegraphics[trim = 0mm 0mm 0mm 0mm, clip=true,width=\textwidth]{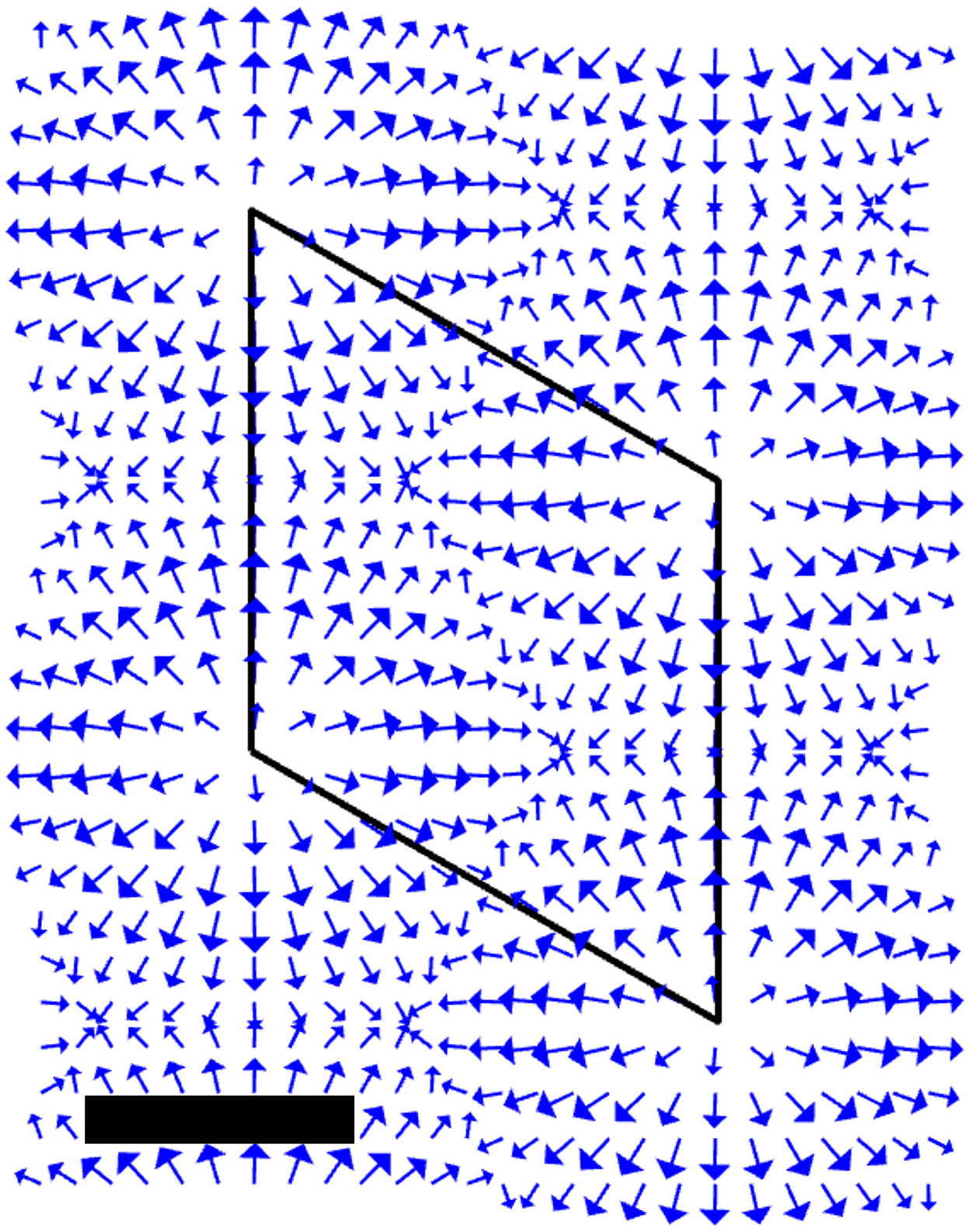}
         \caption{}
         \label{fig:in_plane_mode:2}
     \end{subfigure}
         \begin{subfigure}[b]{0.22\textwidth}
         \centering
         \includegraphics[trim = 0mm 0mm 0mm 0mm, clip=true,width=\textwidth]{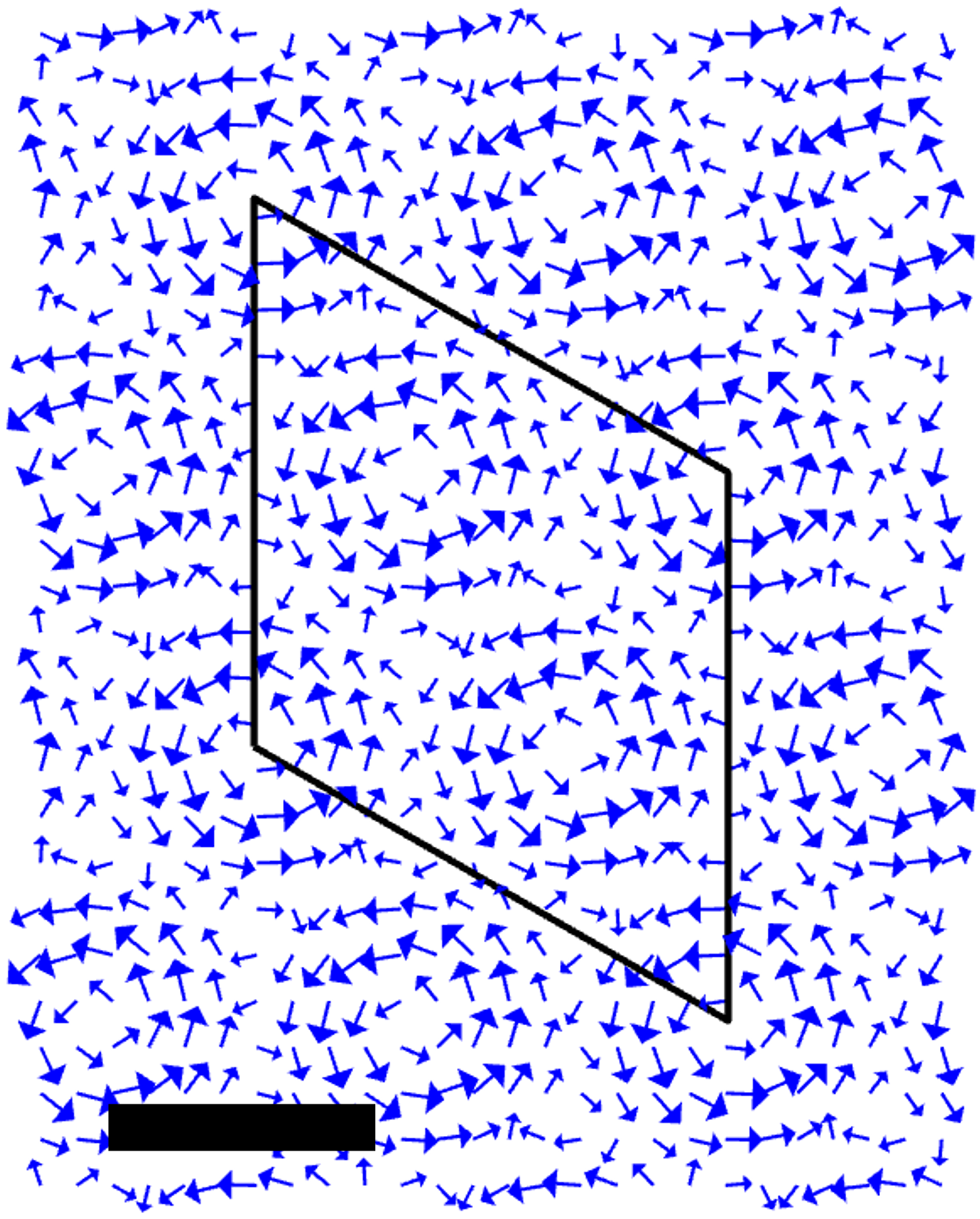}
         \caption{}
         \label{fig:in_plane_mode:3}
     \end{subfigure}
         \begin{subfigure}[b]{0.22\textwidth}
         \centering
         \includegraphics[trim = 0mm 0mm 0mm 0mm, clip=true,width=\textwidth]{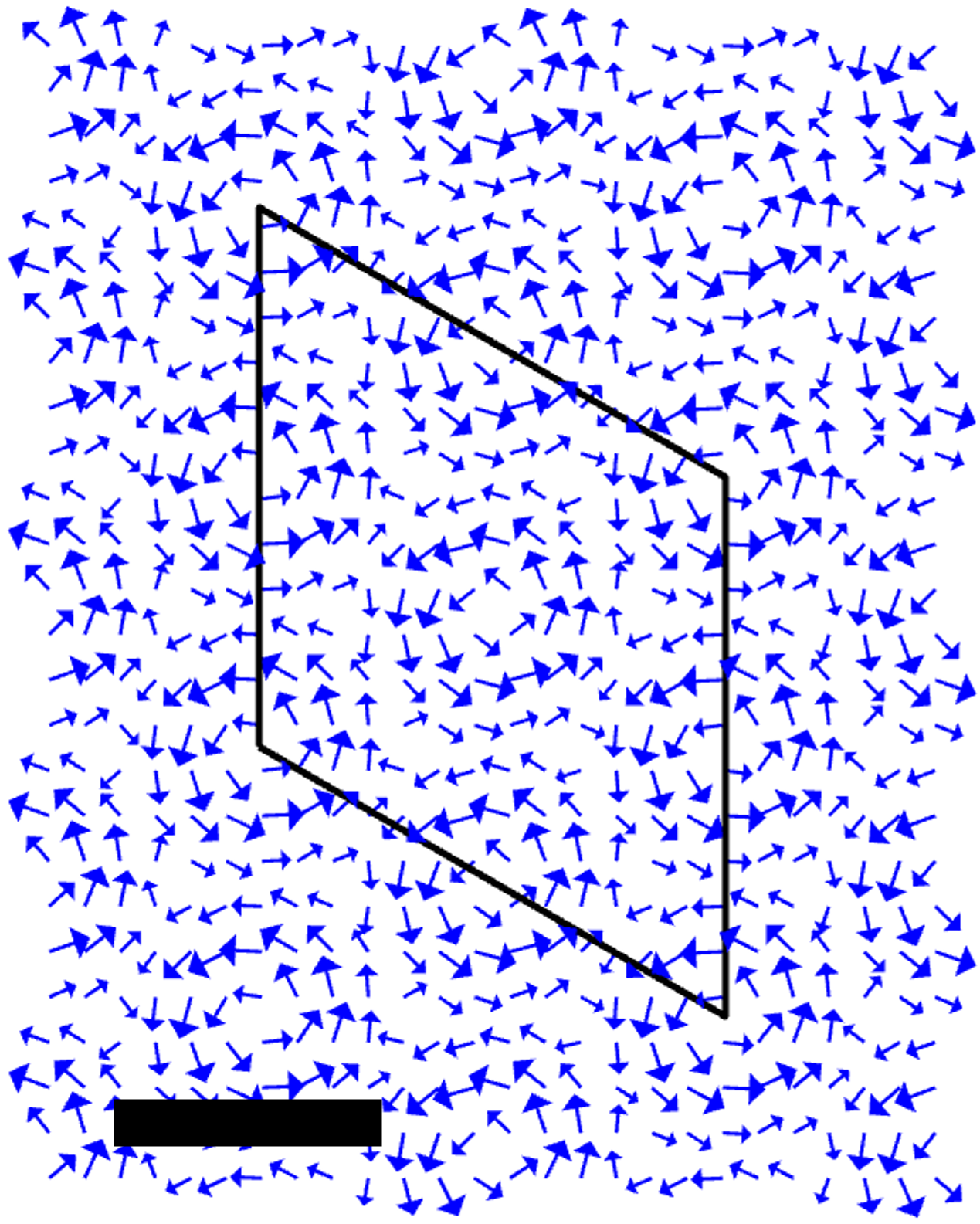}
         \caption{}
         \label{fig:in_plane_mode:4}
     \end{subfigure}
\captionsetup{justification=raggedright, singlelinecheck=false}
\caption{The combined in-plane displacement fields for sets of modes with the same frequency are represented as blue arrows. 
The modes are normalized (see \sect{sec:relaxation}) so an absolute length is not specified. The solid black line is the boundary of the TBG supercell ($\theta=2.28 \degree$), coinciding with $\mathbf{a}_1$ and $\mathbf{a}_2$. (a) $\bth{1}^{\rm in}(\br)$ $+$ $\bth{2}^{\rm in}(\br)$ $+$ $\bth{3}^{\rm in}(\br)$, (b) $\bth{4}^{\rm in}(\br)$ $+$ $\bth{5}^{\rm in}(\br)+\bth{6}^{\rm in}(\br)$, (c) $\bth{7}^{\rm in}(\br)+\bth{8}^{\rm in}(\br)$ $+$ $\bth{9}^{\rm in}(\br)$, and (d) $\bth{10}^{\rm in}(\br)$ $+$ $\bth{11}^{\rm in}(\br)$ $+$ $\bth{12}^{\rm in}(\br)$.}
\label{fig:in_plane_mode}
\end{figure}

The displacement fields associated with the in-plane basis vectors are shown in \fig{fig:in_plane_mode} summed over the sets of modes with the same frequency (see \tab{tab:in_plane_mode}). 
The orthogonality of $\bd_k$ and $\bar{\bG}_k$ (i.e., $\bd_k\cdot\bar{\bG}_k=0$) in each $\bth{1,2,3}^{\rm in}(\br)$ leads to a transverse wave in which the displacement field of $\bth{1}^{\rm in}(\br)+\bth{2}^{\rm in}(\br)+\bth{3}^{\rm in}(\br)$ exhibits localized twisting at the origin (the center of AA domain) of the TBG supercell. Thus the (1,2,3) mode set captures the localized twisting at AA domains observed in TBG relaxation \cite{zhang2018,sung2022}. Indeed, $\bth{1}^{\rm in}(\br)+\bth{2}^{\rm in}(\br)+\bth{3}^{\rm in}(\br)$ was also used as the distortion direction in the PLD model (Eqn. (1) in \cite{sung2022} with $n=1$), showing that the (1,2,3) mode set captures the majority of the in-plane deformation in the TBG relaxation \cite{sung2022}.
The other mode sets also create distinct deformation signatures. Mode set (4,5,6) shown in \fig{fig:in_plane_mode:2} forms a longitudinal wave since each $\bth{4,5,6}^{\rm in}$ has the same directions for $\bd_k$ and $\bar{\bG}_k$ (see \tab{tab:in_plane_mode}). The (4,5,6) mode set captures radial displacements emanating from the AA domain centers. Mode set (7,8,9) in \fig{fig:in_plane_mode:3} corresponds to three smaller localized twistings centered on the AA, AB, and BA domains.
Finally for mode set (10,11,12), we note from \tab{tab:in_plane_mode} that $\bd_{10}$, $\bd_{11}$, and $\bd_{12}$ are the same as $\bd_{1}$, $\bd_{2}$, and $\bd_{3}$, whereas $\bar{\bG}_{10}$, $\bar{\bG}_{11}$, and $\bar{\bG}_{12}$ are twice $\bar{\bG}_{1}$, $\bar{\bG}_{2}$, and $\bar{\bG}_{3}$. Thus, $\bth{10}^{\rm in}(\br)+\bth{11}^{\rm in}(\br)+\bth{12}^{\rm in}(\br)$ is a transverse wave with a more localized displacement field having four small localized twistings, as shown in \fig{fig:in_plane_mode:4}.
Higher order modes (i.e., $k>12$) are associated with more localized displacements due to larger $\bar{\bG}_{k}$ magnitudes. We show in \sect{sec:relaxation} that retaining the first 12 in-plane modes is sufficient to represent TBG relaxation in atomistic simulations with good accuracy. 

\subsection{Out-of-plane elastic plate basis}\label{sec:out_of_plane_basis}
The out-of-plane wave equation of an elastic plate is
\begin{equation}\label{eqn:out_of_plane_wave}
    D\nabla^4w = -\rho h \frac{\partial^2w}{\partial t^2},
\end{equation}
where $D$ is the bending stiffness of the plate (see \tab{tab:e_ip_test} for $D$ for monolayer graphene), and $w$ is the out-of-plane displacement along the $z$ Cartesian direction. Using the PWEM, the solution for the out-of-plane displacement $w$ has the form
\begin{equation}\label{eqn:out_of_plane_displacement}
    w = \exp(-i \omega^{\rm out} t)\sum_{\bG}\exp{(i(\bG\cdot\br + \phi_{\bG}))},
\end{equation}
where $\omega^{\rm out}$ is the out-of-plane angular frequency (related to the ordinary frequency $f^{\rm out}$ through $\omega^{\rm out}=2\pi f^{\rm out}$). Inserting \eqn{eqn:out_of_plane_displacement} into \eqn{eqn:out_of_plane_wave} gives, for each $\bG$,
\begin{equation}\label{eqn:out_of_plane_eigenvalue}
    \rho h \left(\omega^{\rm out}\right)^2 = D\left(G_x^2+G_y^2\right)^2.
\end{equation}
Solving this relation gives the frequency:
\begin{equation}
\omega^{\rm out}(\bG)=\sqrt{\frac{D}{\rho h}}\left(G_x^2+G_y^2\right).
\end{equation}
Since the frequency depends on $\bG$, the time-dependent term must be included in the sum in \eqn{eqn:out_of_plane_displacement}, and so we have
\begin{equation}\label{eqn:out_of_plane_displacement_omega}
    w = \exp{(i(\bG\cdot\br + \phi_{\bG}-\omega^{\rm out}(\bG) t))}.
\end{equation}

\begin{table}[!htb]\centering
\begin{tabular}{c|c|c}
\hline
\multicolumn{2}{c|}{$f^{\rm out}_{o}$ (GHz)} & $\hat{\bG}_l$\\
\hline
\multirow{3}{*}{$f^{\rm out}_{1,2,3}$} & \multirow{3}{12em}{IP: 111.59$^{\rm{a}}$,112$^{\rm{b}}$ \\ Exp: 110.44--131.45$^{\rm{c}}$}&  $\bG(0,1)$  \\
\cline{3-3}
& &  $\bG(-1,0)$
\\
\cline{3-3}
& & $\bG(1,-1)$
\\
\hline
\multirow{3}{*}{$f^{\rm out}_{4,5,6}$} &\multirow{3}{12em}{IP: 334.78$^{\rm{a}}$, 336.01$^{\rm{b}}$ \\ Exp: 331.34--394.37$^{\rm{c}}$} & $\bG(-2,1)$  \\
\cline{3-3}
& & $\bG(1,-2)$
\\
\cline{3-3}
& & $\bG(1,1)$
\\
\hline
\multirow{3}{*}{$f^{\rm out}_{7,8,9}$} & \multirow{3}{12em}{IP: 446.37$^{\rm{a}}$, 448.01$^{\rm{b}}$ \\ Exp: 441.79--525.83$^{\rm{c}}$} & $\bG(0,2)$ \\
\cline{3-3}
& & $\bG(-2,0)$ \\
\cline{3-3}
& & $\bG(2,-2)$ \\
\hline
\multirow{3}{*}{$f^{\rm out}_{10,11,12}$} & \multirow{3}{12em}{IP: 777.5$^{\rm{a}}$, 780.36$^{\rm{b}}$ \\ Exp: 769.52--915.91$^{\rm{c}}$} & $\bG(1,2)$ \\
\cline{3-3}
& & $\bG(2,-3)$ \\
\cline{3-3}
& & $\bG(3,-1)$ \\
\hline
\multirow{3}{*}{$f^{\rm out}_{13,14,15}$} & \multirow{3}{12em}{IP: 777.5$^{\rm{a}}$, 780.36$^{\rm{b}}$ \\ Exp: 769.52--915.91$^{\rm{c}}$} & $\bG(1,-3)$ \\
\cline{3-3}
& & $\bG(2,1)$ \\
\cline{3-3}
& & $\bG(3,-2)$ \\
\hline
\end{tabular}
\captionsetup{justification=raggedright, singlelinecheck=false}
\caption{First 15 elements of $f_{l}^{\rm out}$ and $\hat{\bG}_l$ for a twist angle of $2.28\degree$ (L = 61.74 \AA) in sets of three that have the same frequency. The frequencies $f_{l}^{\rm out}$ are computed using the elastic properties of monolayer graphene in \tab{tab:e_ip_test} for a) AIREBO, b) hNN, and c) experiments \cite{han2020,blees2015}.}
\label{tab:out_of_plane_mode}
\end{table}

The out-of-plane frequencies $\omega^{\rm out}_l$ are computed for the $\bG_l$ vectors ($l\in[1,66]$) used for in-plane modes, ordered by increasing value. Note that each $\bG$ has a single $\omega^{\rm out}$, unlike the in-plane normal mode where each $\bG$ has two $\omega^{\rm in}$. Similar to the in-plane normal modes, we use the same eigenvalue ordering for both AIREBO, hNN, and experiment using the bending stiffness $D$ of AIREBO in \tab{tab:e_ip_test}. The first 15 elements of $f_l^{\rm out}=\omega_{l}^{\rm out}/2\pi$ and associated $\hat{\bG}_l$ are listed in \tab{tab:out_of_plane_mode} for a TBG twist angle of 2.28$\degree$ for which  $L = 61.74$~\AA. Similar to the in-plane modes (see \tab{tab:in_plane_mode}), the results are arranged in sets of three that have the same frequency $f_{l}^{\rm out}$ (i.e., $f_{1,2,3}^{\rm out}, f_{4,5,6}^{\rm out}, f_{7,8,9}^{\rm out}$, $f_{10,11,12}^{\rm out}$, and $f_{13,14,15}^{\rm out}$). We omit $\bd_o$ for the out-of-plane modes since it is simply $\khat$. We note that frequency range of the out-of-plane modes (111.59--777.5 GHz for AIREBO in \tab{tab:out_of_plane_mode}) is much lower than the in-plane frequencies (2157.16--4314.32 GHz for AIREBO in \tab{tab:in_plane_mode}); this is because monolayer graphene is sigificantly stiffer in stretching than in bending.

The out-of-plane modes in \tab{tab:in_plane_mode} can be used as a basis for characterizing the out-of-plane deformation of graphene layers in a TBG. Mode vector $l$ is
\begin{equation}\label{eqn:out_of_plane_mode}
\bth{l}^{\rm out}(\br)=\exp{(i(\hat{\bG}_l\cdot\br+\phi_{\hat{\bG}_l}))}\khat,
\end{equation}
where $\phi_{\hat{\bG}_l}$ is the phase shift for $l$th mode. Similar to the in-plane basis, triplets of out-of-plane bases that share the same frequency are summed. Imposing 6-fold symmetry leads to the result that the phase angle for all out-of-plane basis terms is $\phi_{\hat{\bG}_l}=\phi^{\rm out}=0$ (see \app{app:phasederiv}).

\begin{figure}[!htbp]\centering
     \begin{subfigure}[b]{0.15\textwidth}
         \centering
         \includegraphics[trim = 0mm 0mm 0mm 0mm, clip=true,width=\textwidth]{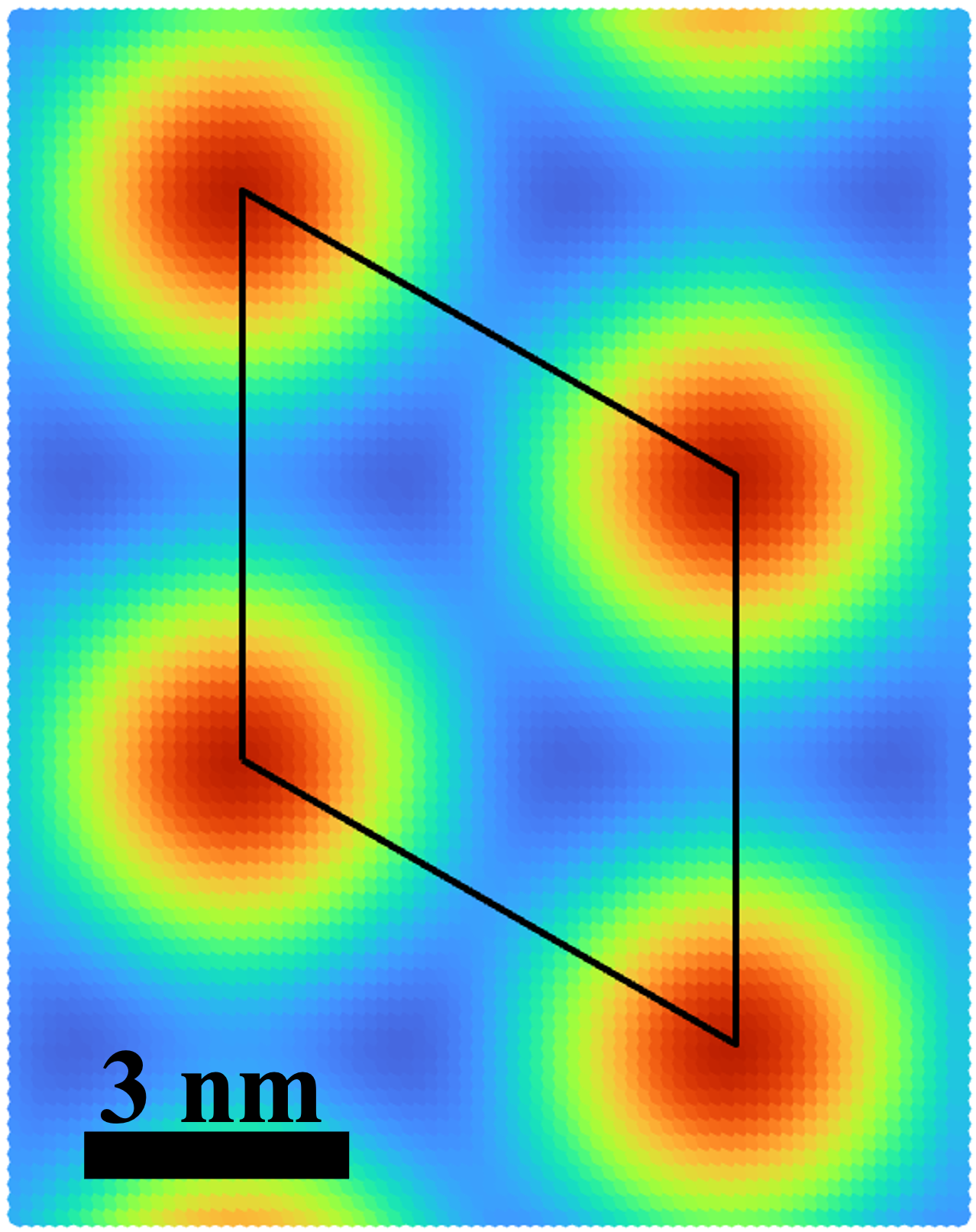}
         \caption{}
         \label{fig:out_of_plane_mode:1}
     \end{subfigure}
    \begin{subfigure}[b]{0.15\textwidth}
         \centering
         \includegraphics[trim = 0mm 0mm 0mm 0mm, clip=true,width=\textwidth]{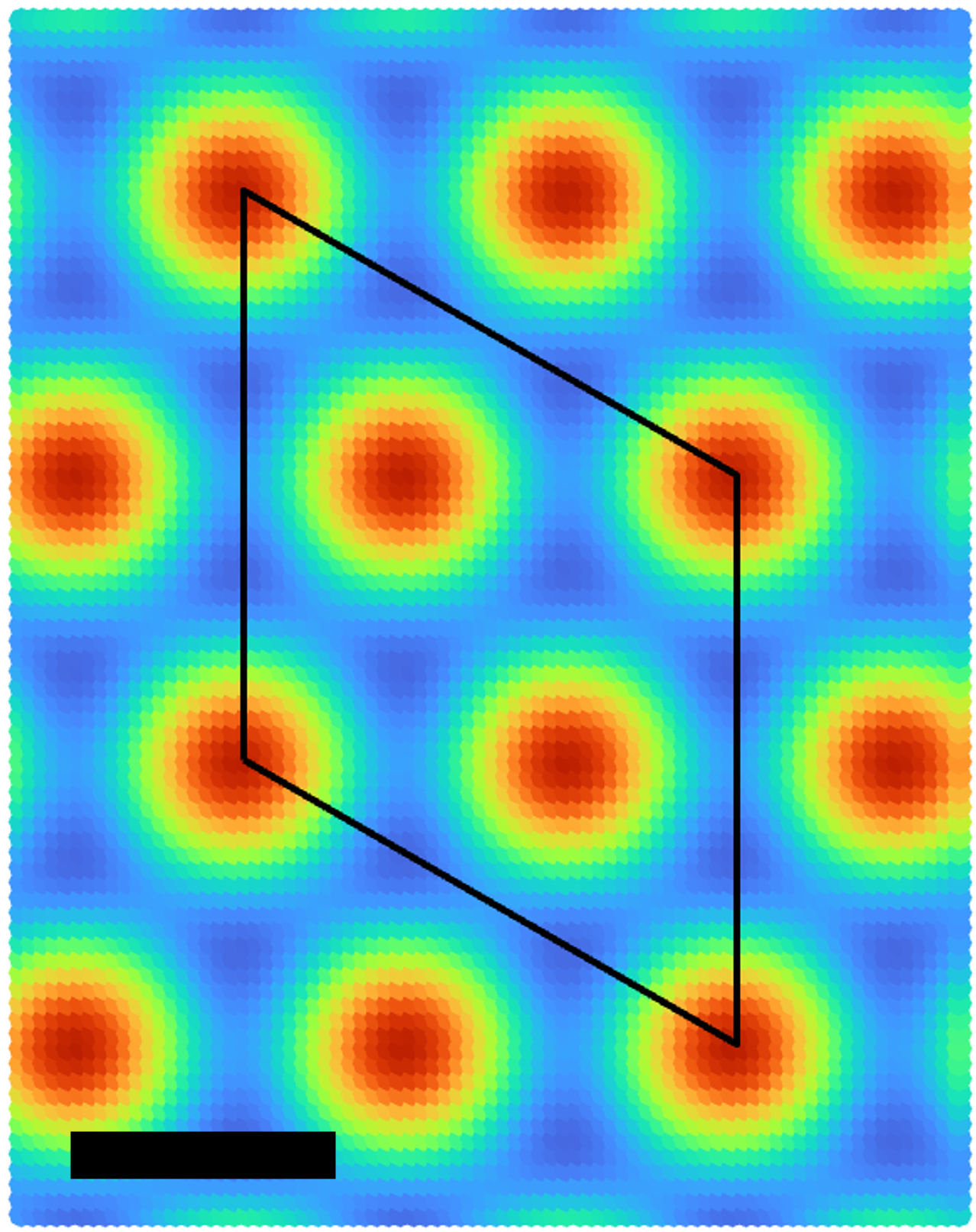}
         \caption{}
         \label{fig:out_of_plane_mode:2}
     \end{subfigure}
         \begin{subfigure}[b]{0.15\textwidth}
         \centering
         \includegraphics[trim = 0mm 0mm 0mm 0mm, clip=true,width=\textwidth]{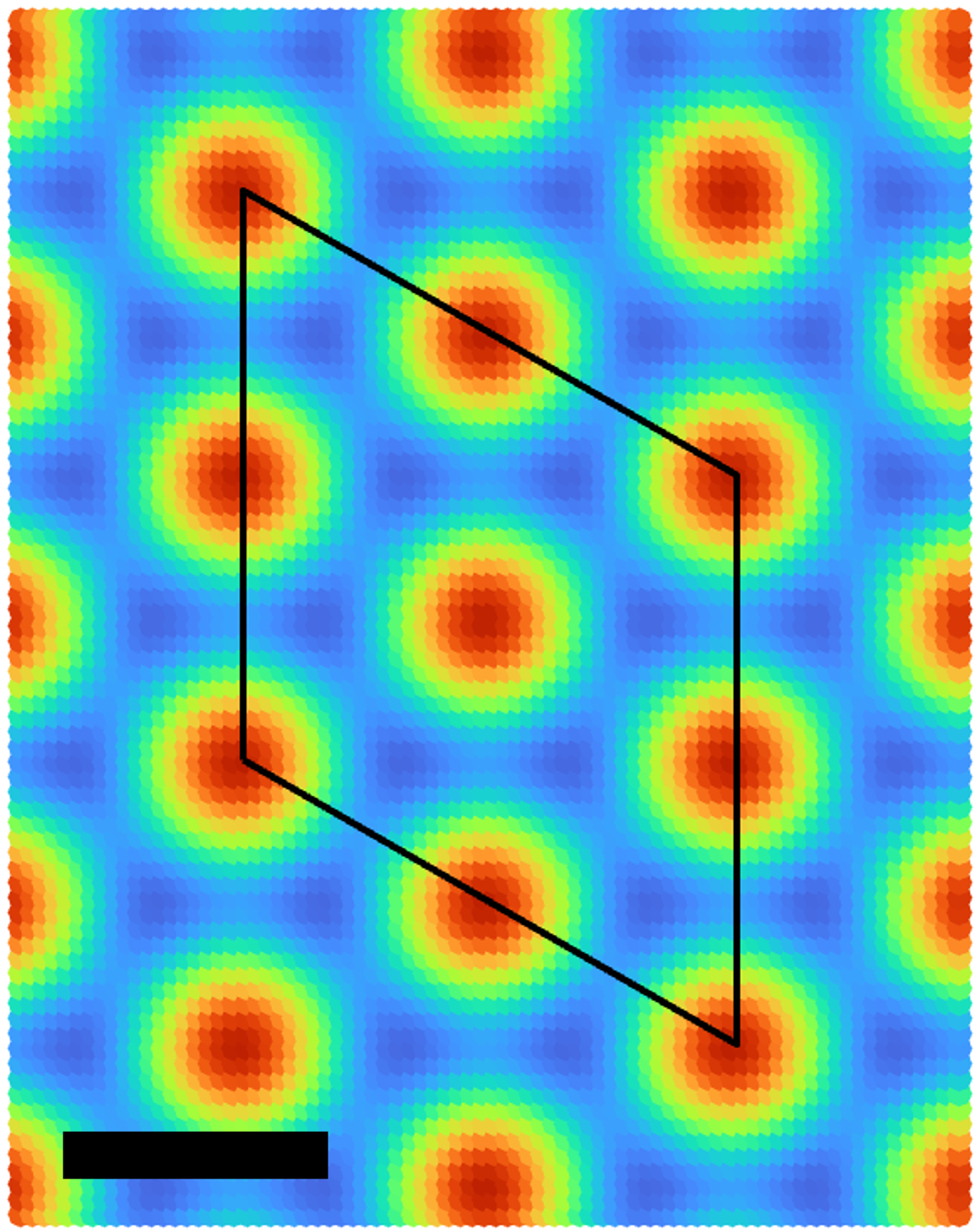}
         \caption{}
         \label{fig:out_of_plane_mode:3}
     \end{subfigure}
     \begin{subfigure}[b]{0.15\textwidth}
         \centering
         \includegraphics[trim = 0mm 0mm 0mm 0mm, clip=true,width=\textwidth]{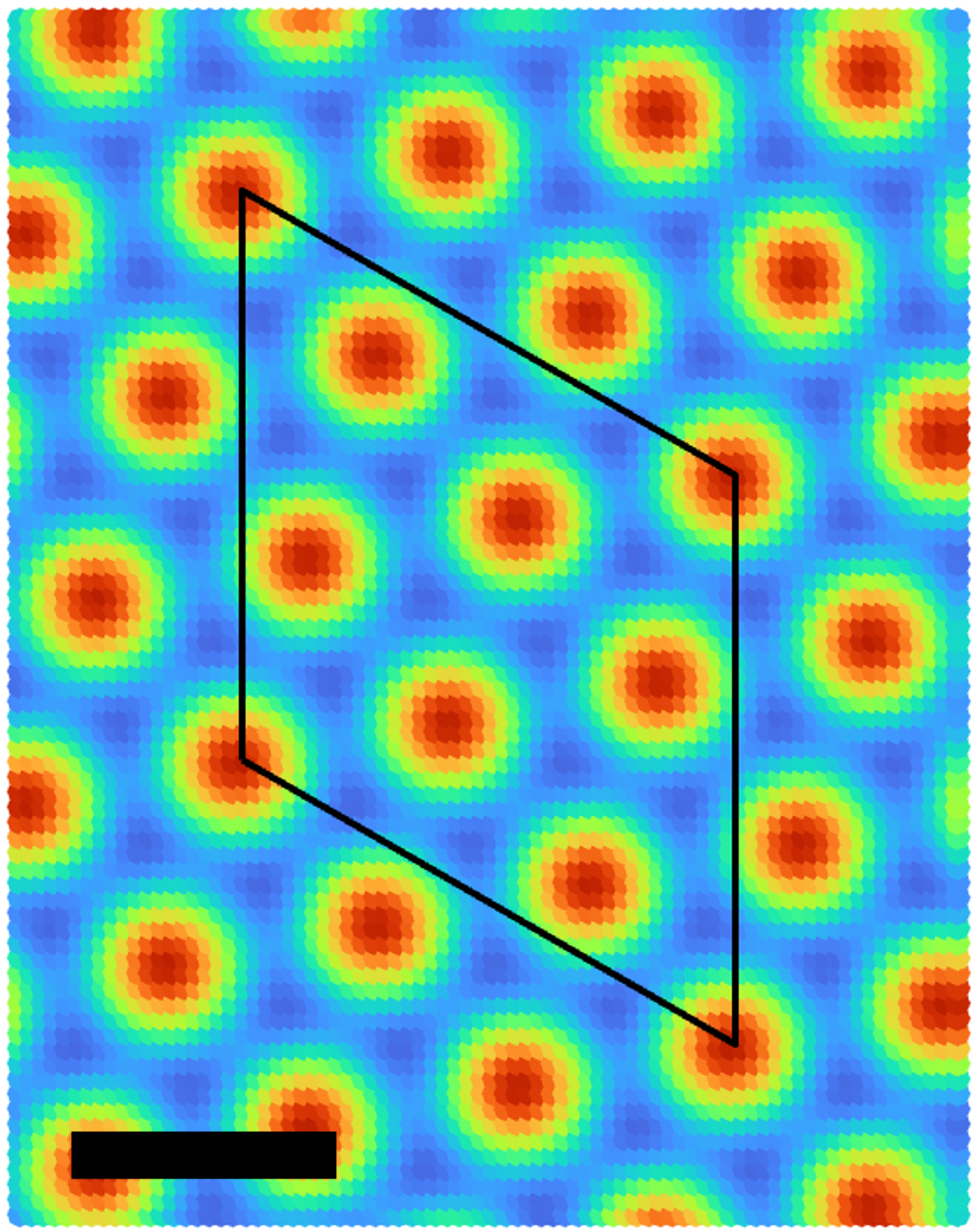}
         \caption{}
         \label{fig:out_of_plane_mode:4}
     \end{subfigure}
     \begin{subfigure}[b]{0.15\textwidth}
         \centering
         \includegraphics[trim = 0mm 0mm 0mm 0mm, clip=true,width=\textwidth]{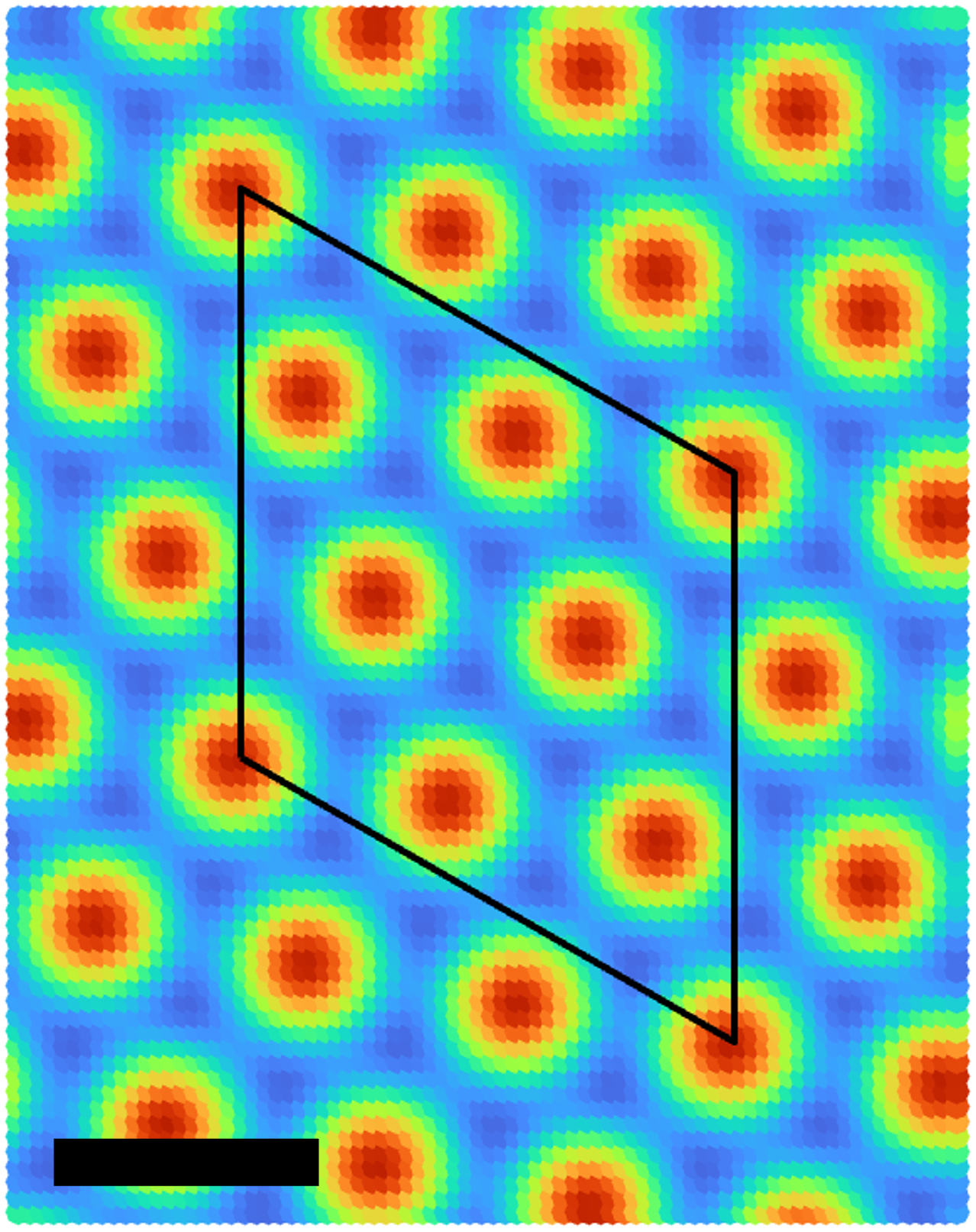}
         \caption{}
         \label{fig:out_of_plane_mode:5}
     \end{subfigure}
\captionsetup{justification=raggedright, singlelinecheck=false}
\caption{Contour plots of the combined out-of-plane displacement field for sets of modes with the same frequency. Displacements are normalized with color ranging from red for a positive displacement to blue for negative displacement along the $\khat$ direction ($z$ Cartesian coordinate). 
The solid black line is the boundary of the TBG supercell ($\theta = 2.28\degree$), coinciding with $\mathbf{a}_1$ and $\mathbf{a}_2$. (a) $\bth{1}^{\rm out}(\br)$ $+$ $\bth{2}^{\rm out}(\br)+\bth{3}^{\rm out}(\br)$, (b) $\bth{4}^{\rm out}(\br)$ $+$ $\bth{5}^{\rm out}(\br)$ $+$ $\bth{6}^{\rm out}(\br)$, (c) $\bth{7}^{\rm out}(\br)+\bth{8}^{\rm out}(\br)+\bth{9}^{\rm out}(\br)$, (d) $\bth{10}^{\rm out}(\br)$ $+$ $\bth{11}^{\rm out}(\br)$ $+$ $\bth{12}^{\rm out}(\br)$, and (e) $\bth{13}^{\rm in}(\br)$ $+$ $\bth{14}^{\rm in}$ $+$ $\bth{15}^{\rm in}(\br)$.}
\label{fig:out_of_plane_mode}
\end{figure}

The displacement fields associated with the out-of-plane basis vectors are shown in \fig{fig:out_of_plane_mode} summed over the sets of modes with the same frequency. The displacement field associated with $\bth{1}^{\rm out}(\br)+\bth{2}^{\rm out}(\br)+\bth{3}^{\rm out}(\br)$ in \fig{fig:out_of_plane_mode:1} consists of circular dome-shaped displacements centered on the AA domains located at the corners of the supercell. This field resembles the out-of-plane deformation observed in multiscale simulations of TBG relaxation (Fig.~1 in \cite{zhang2018}), which indicates that the out-of-plane basis can capture the main features of TBG layer deformation. Next, the displacement field of $\bth{4}^{\rm out}(\br)+\bth{5}^{\rm out}(\br)+\bth{6}^{\rm out}(\br)$ in \fig{fig:out_of_plane_mode:2} has smaller dome-shaped displacements centered on the AA, AB, and BA domains. The other modes include more localized deformations since the $\hat{\bG}_l$ vectors are increasing in magnitude. Similar to $\bth{k}^{\rm in}$, $\bth{l}^{\rm out}$ exhibits a more global displacement field for low $l$, and constitutes an orthogonal basis. We select the first 15 out-of-plane modes (listed in \tab{tab:out_of_plane_mode}) as the out-of-plane basis for quantifying the out-of-plane deformation of TBG layers as discussed in \sect{sec:relaxation}.

\section{Quantification of TBG relaxation}\label{sec:relaxation}
Graphene layers brought into contact to form a TBG undergo a complex deformation pattern involving both in-plane and out-of-plane displacements of the layer atoms \cite{zhang2018,guinea2019,yoo2019,sung2022}. Various models have been proposed to approximate the in-plane deformation that involves prominent localized twisting. The model of Zhang and Tadmor \cite{zhang2018} is based on an exponentially-decaying (Gaussian) rotation field centered on the AA domains. This approach is generalized in the model of Sung et al.~\cite{sung2022} in terms of three non-orthogonal, transverse periodic lattice distortions (PLDs). Here we show how both the in-plane \emph{and} out-of-plane deformation can be systematically described using the orthogonal in-plane and out-of-plane bases for an elastic plate derived in \sect{sec:normal_mode_basis}. We apply this approach to characterize the deformation of TBG computed in an atomistic simulation.

\subsection{General formulation}\label{sec:relaxation:general}
The position of atom $j$ in a graphene layer in the relaxed TBG is given by
\begin{equation}\label{eqn:position}
    \br^j = \br_0^j + \mathbf{\Delta}(\br_0^j),
\end{equation}
where $\br^j$ is the current position, $\br_0^j$ is the reference position, and $\mathbf{\Delta}(\br_0^j)$ is the displacement. $\mathbf{\Delta}(\br_0^j)$ is separated into the in-plane displacement ($\mathbf{\Delta}_{\rm in}(\br_0^j)$) and out-of-plane displacement ($\mathbf{\Delta}_{\rm out}(\br_0^j)$).
We represent the in-plane displacement $\mathbf{\Delta}_{\rm in}(\br_0)$ as a sum over the the in-plane mode vectors in \eqn{eqn:in_plane_mode} where each in-plane mode has a coefficient $A_k$ and phase $\phi_{\bar{\bG}_k}$,
\begin{equation}\label{eqn:in_plane_disp_basis_approx}
\mathbf{\Delta}_{\rm in}(\br_0)\approx
\sum_{k} A_k \bd_k\exp{(i(\bar{\bG}_k\cdot\br_0+\phi_{\bar{\bG}_k}))}.
\end{equation}
Similarly, the out-of-plane displacement $\mathbf{\Delta}_{\rm out}(\br_0)$ is taken to be a sum over the out-of-plane mode vectors in \eqn{eqn:out_of_plane_mode} with coefficients $B_l$ and phases $\phi_{\hat{\bG}_l}$,
\begin{equation}\label{eqn:out_of_plane_disp_basis_approx}
\mathbf{\Delta}_{\rm out}(\br_0)\approx
\sum_{l} B_l \khat \exp{(i(\hat{\bG}_l\cdot\br_0+\phi_{\hat{\bG}_l}))}).
\end{equation}
The real parts of \eqns {eqn:in_plane_disp_basis_approx} and \eqnx{eqn:out_of_plane_disp_basis_approx} are
\begin{align}\label{eqn:in_plane_disp_basis_approx2}
\mathbf{\Delta}_{\rm in}(\br_0) &\approx
\sum_{k} [A_k^{\rm c}\bd_k\cos{(\bar{\bG}_k\cdot\br_0)} + A_k^{\rm s}\bd_k\sin{(\bar{\bG}_k\cdot\br_0)}], \\
\mathbf{\Delta}_{\rm out}(\br_0) &\approx
\sum_{k} [B_k^{\rm c}\khat\cos{(\hat{\bG}_l\cdot\br_0)} + B_k^{\rm s}\khat\sin{(\hat{\bG}_l\cdot\br_0)}].
\end{align}
where $A_k^{\rm c}=A_k\cos{(\phi_{\bar{\bG}_k})}$, $A_k^{\rm s}=A_k\sin{(\phi_{\bar{\bG}_k})}$, $B_l^{\rm c}=B_l\cos{(\phi_{\hat{\bG}_l})}$, and $B_l^{\rm s}=B_l\sin{(\phi_{\hat{\bG}_l})}$.
To compute $A_{k^*}$ and $\phi_{\bar{\bG}_{k^*}}$ in $\mathbf{\Delta}_{\rm in}(\br_0)$ for the in-plane mode $k^*$, $\mathbf{\Delta}_{\rm in}(\br_0)$ is projected onto $\bd_{k^*}\cos{(\bar{\bG}_{k^*}\cdot\br_0)}$ and $\bd_{k^*}\sin{(\bar{\bG}_{k^*}\cdot\br_0)}$ and integrated over the supercell domain $S$. These integrations are expressed as
\begin{multline}
    \int_S\mathbf{\Delta}_{\rm in}(\br_0)\cos{(\bar{\bG}_{k^*}\cdot\br_0)}dS = \\
    \int_S \sum_{k}\bd_k\cdot\bd_{k^*}[A_k^{\rm c}\cos{(\bar{\bG}_k\cdot\br_0)}\cos{(\bar{\bG}_{k^*}\cdot\br_0)} + \\ A_k^{\rm s}\sin{(\bar{\bG}_k\cdot\br_0)}\cos{(\bar{\bG}_{k^*}\cdot\br_0)}]dS, \label{eqn:in_plane_int1}
\end{multline}
and
\begin{multline}
    \int_S\mathbf{\Delta}_{\rm in}(\br_0)\sin{(\bar{\bG}_{k^*}\cdot\br_0)}dS  = \\
    \int_S \sum_{k}\bd_k\cdot\bd_{k^*}[A_k^{\rm c}\cos{(\bar{\bG}_k\cdot\br_0)}\sin{(\bar{\bG}_{k^*}\cdot\br_0)} + \\ A_k^{\rm s}\sin{(\bar{\bG}_k\cdot\br_0)}\sin{(\bar{\bG}_{k^*}\cdot\br_0)}]dS. \label{eqn:in_plane_int2}
\end{multline}
Using orthogonality between different in-plane modes and
\begin{multline}
\int_{S}\bd_{k^*}\cdot\bd_{k^*}\sin{(\bar{\bG}_{k^*}\cdot\br_0)}\cos{(\bar{\bG}_{k^*}\cdot\br_0)}dS= \\ \bar{I}_{\rm in}(-\pi/2,0)=0
\end{multline}
explained in SM~\cite{SM}, \eqns{eqn:in_plane_int1} and \eqnx{eqn:in_plane_int2} are expressed as
\begin{multline}
    A_{k^*}^{\rm c} \int_S \cos{(\bar{\bG}_{k^*}\cdot\br_0)} \cos{(\bar{\bG}_{k^*}\cdot\br_0)}dS =  \\
    A_{k^*}^{\rm c}\bar{I}_{\rm in}(0,0), \label{eqn:in_plane_int3}
\end{multline}
and 
\begin{multline}
    A_{k^*}^{\rm s} \int_S \sin{(\bar{\bG}_{k^*}\cdot\br_0)} \sin{(\bar{\bG}_{k^*}\cdot\br_0)}dS = \\
    A_{k^*}^{\rm c}\bar{I}_{\rm in}(-\pi/2,-\pi/2), \label{eqn:in_plane_int4}
\end{multline}
where $\bd_{k^*}\cdot\bd_{k^*}=1$ is omitted.
Since the carbon atoms in a graphene layer are uniformly distributed in the supercell, the left-hand sides of \eqns{eqn:in_plane_int3} and \eqnx{eqn:in_plane_int4} are approximated as
\begin{multline}\label{eqn:in_plane_approx1}  
\int_S\mathbf{\Delta}_{\rm in}(\br_0)\cos{(\bar{\bG}_{k^*}\cdot\br_0)}dS
 \approx \\ \sum_{j=1}^{\Nlay}\mathbf{\Delta}_{\rm in}(\br_0^j)\cos{(\bar{\bG}_{k^*}\cdot\br_0^j)}\Omega, 
\end{multline}
\begin{multline}\label{eqn:in_plane_approx2}
\int_S\mathbf{\Delta}_{\rm in}(\br_0)\sin{(\bar{\bG}_{k^*}\cdot\br_0)}dS
 \approx \\ \sum_{j=1}^{\Nlay}\mathbf{\Delta}_{\rm in}(\br_0^j)\sin{(\bar{\bG}_{k^*}\cdot\br_0^j)}\Omega, 
\end{multline}
where $\Nlay$ is the number of atoms in a layer, and $\Omega=\sqrt{3}L^2/(2\Nlay)$ is the volume occupied by a carbon atom in the supercell domain \footnote{To check whether approximations in \eqns{eqn:in_plane_approx1} and \eqns{eqn:in_plane_approx2} are valid, we compute $\sum_{j=0}^{\Nlay}\cos{(\bar{\bG}_{k}\cdot\br_0^j)}\cos{(\bar{\bG}_{k}\cdot\br_0^j)}\Omega$ and compared with the closed form solution of $\int_S\cos{(\bar{\bG}_{k}\cdot\br_0)}\cos{(\bar{\bG}_{k}\cdot\br_0)}dS$, $\bar{I}_{\rm in}(0,0)$. The relative error between two values for all $\bar{\bG}_{k}$ listed in \tab{tab:in_plane_mode} and $L \in[57.5,338.7]$~\AA\ (i.e., $\theta \in [0.41,2.45]$\degree) for the supercell is less than $10^{-5}$, indicating that the approximations are valid}.
Using \eqns{eqn:in_plane_approx1} and \eqnx{eqn:in_plane_approx2}, $A^{\rm c}_{k^*}$ and $A^{\rm s}_{k^*}$ in \eqns{eqn:in_plane_int3} and \eqnx{eqn:in_plane_int4} are
\begin{multline}\label{eqn:A_k_c}
    A^{\rm c}_{k^*} = \frac{\Omega}{\bar{I}_{\rm in}(0,0)}\sum_{j=1}^{\Nlay}\mathbf{\Delta}_{\rm in}(\br_0^j)\cos{(\bar{\bG}_{k^*}\cdot\br_0^j)} = \\ \frac{2}{\Nlay}\sum_{j=1}^{\Nlay}\mathbf{\Delta}_{\rm in}(\br_0^j)\cos{(\bar{\bG}_{k^*}\cdot\br_0^j)}, 
\end{multline}
and 
\begin{multline}\label{eqn:A_k_s}
    A^{\rm s}_{k^*} = \frac{\Omega}{\bar{I}_{\rm in}(-\pi/2,-\pi/2)}\sum_{j=1}^{\Nlay}\mathbf{\Delta}_{\rm in}(\br_0^j)\sin{(\bar{\bG}_{k^*}\cdot\br_0^j)} = \\ \frac{2}{\Nlay}\sum_{j=1}^{\Nlay}\mathbf{\Delta}_{\rm in}(\br_0^j)\sin{(\bar{\bG}_{k^*}\cdot\br_0^j)}. 
\end{multline}
$A_{k^*}$ and $\phi_{\bar{\bG}_{k^*}}$ are computed via
$\sqrt{(A^{\rm c}_{k^*})^2+(A^{\rm s}_{k^*})^2}$ and $\tan^{-1}{(A^{\rm s}_{k^*}/A^{\rm c}_{k^*})}$, respectively.

Expressions for $B_{k^*}^{\rm c}$ and $B_{l^*}^{\rm s}$ in $\mathbf{\Delta}_{\rm out}(\br_0)$ for the out-of-plane mode $l^*$ follow similarly. Using the orthogonality of the out-of-plane modes (see SM\cite{SM}) and the approximations in \eqns{eqn:in_plane_approx1} and \eqnx{eqn:in_plane_approx2}, $B_{l^*}^{\rm c}$ and $B_{l^*}^{\rm s}$ are
\begin{multline} \label{eqn:B_l_c}
    B^{\rm c}_{l^*} = \frac{\Omega}{\bar{I}_{\rm out}(0,0)}\sum_{j=1}^{\Nlay}\mathbf{\Delta}_{\rm out}(\br_0^j)\cos{(\hat{\bG}_{l^*}\cdot\br_0^j)} = \\ \frac{2}{\Nlay}\sum_{j=1}^{\Nlay}\mathbf{\Delta}_{\rm out}(\br_0^j)\cos{(\hat{\bG}_{l^*}\cdot\br_0^j)}, 
\end{multline}
and
\begin{multline} \label{eqn:B_l_s}
    B^{\rm s}_{l^*} = 
    \frac{\Omega}{\bar{I}_{\rm out}(-\frac{\pi}{2},-\frac{\pi}{2})}\sum_{j=1}^{\Nlay}\mathbf{\Delta}_{\rm out}(\br_0^j)\sin{(\hat{\bG}_{l^*}\cdot\br_0^j)} = \\ \frac{2}{\Nlay}\sum_{j=1}^{\Nlay}\mathbf{\Delta}_{\rm out}(\br_0^j)\sin{(\hat{\bG}_{l^*}\cdot\br_0^j)}.
\end{multline}
$B_{l^*}$ and $\phi_{\hat{\bG}_{l^*}}$ are computed via
$\sqrt{(B^{\rm c}_{l^*})^2+(B^{\rm s}_{l^*})^2}$ and $\tan^{-1}{(B^{\rm s}_{l^*}/B^{\rm c}_{l^*})}$, respectively.

In performing the projection for the displacements $\mathbf{\Delta}_{\rm in}$ and $\mathbf{\Delta}_{\rm out}$ of a graphene layer in a relaxed TBG obtained from atomistic simulations, we find that $A_k^{\rm c}$ and $B_l^{\rm s}$ are negligible compared with $A_k^{\rm s}$ and $B_l^{\rm c}$, respectively. 
This indicates that $\phi_{\bar{\bG}_{k}}\approx(n+1/2)\pi$ and $\phi_{\hat{\bG}_{l}}\approx n\pi$ where $n$ is an integer 
according to $\tan^{-1}(A_{k^*}^{\rm s}/A_{k^*}^{\rm c})$ and $\tan^{-1}(B_{l^*}^{\rm s}/B_{l^*}^{\rm c})$ (as stated in \sect{sec:normal_mode_basis} and derived in \app{app:phasederiv})\footnote{Note that although we consider summed bases in \sect{sec:normal_mode_basis}, each basis is independently used to quantify the TBG deformation and confirm that $\phi_{\bar{\bG}_{k}}=(n+\frac{1}{2})\pi$ and $\phi_{\hat{\bG}_{l}}=n\pi$ as derived in \app{app:phasederiv}.}.
We therefore drop the cosine terms for the in-plane displacement, and the sine terms for the out-of-plane displacements, so that the total displacement is given by:
\begin{multline}\label{eqn:disp_bases}
     \mathbf{\Delta}(\br_0)= \sum_{k=1}^{12}A_k\bth{k}^{\rm in}(\br_0) \\
     + \sum_{l=1}^{15}B_l\bth{l}^{\rm out}(\br_0) + \bm{\epsilon}_{\rm in}(\br_0) + \bm{\epsilon}_{\rm out}(\br_0),
\end{multline}
where $\bth{k}^{\rm in}$ and $\bth{l}^{\rm out}$ are given in \eqns{eqn:in_plane_mode} and \eqnx{eqn:out_of_plane_mode}, respectively, and we have $A_k=A_k^{\rm s}$ and $B_l=B_l^{\rm c}$, We have retained 12 in-plane modes and 15 out-of-plane modes as discussed in \sect{sec:normal_mode_basis}. The termination of the series and discrete summation instead of integration introduce errors in the in-plane and out-of-plane displacements that are represented by $\bm{\epsilon}_{\rm in}(\br_0)$ and $\bm{\epsilon}_{\rm out}(\br_0)$. 

\subsection{Numerical results from atomistic simulations}\label{sec:relaxation2}

\begin{table}\centering
\begin{tabular}{c}
\hline
$A_{1,2,3}^{\rm top} = -A_{1,2,3}^{\rm bot}$ \\
\hline
$A_{4,5,6}^{\rm top} = A_{4,5,6}^{\rm bot}$ \\
\hline
$A_{7,8,9}^{\rm top} = -A_{7,8,9}^{\rm bot}$ \\
\hline
$A_{10,11,12}^{\rm top} = -A_{10,11,12}^{\rm bot}$ \\
\hline
\end{tabular}
\captionsetup{justification=raggedright, singlelinecheck=false}    
\caption{Relation between the in-plane coefficients $A_k^{\rm top}$ and $A_k^{\rm bot}$ for the top and bottom graphene layers in a TBG with and without a substrate. This relation reflects mirror symmetry due to the interlayer interaction between the graphene layers in a TBG.}
\label{tab:Atop_Abot}
\end{table}

In this section, we present results for the $A_k$ and $B_l$ coefficients in \eqn{eqn:disp_bases} for the top and bottom graphene layers in a TBG from the displacement fields obtained by molecular statics energy relaxation using the interatomic potentials IP$_1$ and IP$_2$ specified in \sect{sec:tbg_model}. Several twist angles are considered for both a free-standing TBG and a TBG supported on a substrate.
We confirm that basis vectors that have the same frequency (see Tables~\ref{tab:in_plane_mode} and \ref{tab:out_of_plane_mode}), share the same coefficient for both in-plane and out-of-plane bases (i.e., $A_{1,2,3}$, $A_{4,5,6}$, $A_{7,8,9}$, $A_{10,11,12}$, and $B_{1,2,3}$, $B_{4,5,6}$, $B_{7,8,9}$, $B_{10,11,12}$, $B_{13,14,15}$), even though the coefficients are computed for each basis independently. Also, we find that the in-plane coefficients for the top and bottom layers, $A_k^{\rm top}$ and $A_k^{\rm bot}$, obey the relations given in \tab{tab:Atop_Abot}. For transverse waves, $A_{1,2,3}$, $A_{7,8,9}$, and $A_{10,11,12}$ have opposite signs for the top and bottom graphene layers, indicating that they have opposite rotations (see \figs{fig:in_plane_mode:1}, \figx{fig:in_plane_mode:3}, and \figx{fig:in_plane_mode:4} for the displacement fields corresponding to these modes). 
In particular, the opposite signs of $A_{1,2,3}^{\rm top}$ and $A_{1,2,3}^{\rm bot}$ induce opposite localized twisting centered on the high-energy AA domains which reduce their size during relaxation \cite{zhang2018}.
In contrast, for a longitudinal wave, $A_{4,5,6}^{\rm top}$ and $A_{4,5,6}^{\rm bot}$ have the same sign in the mirror symmetry, which indicates that top and bottom layers have the same radial and longitudinal displacement at the AA domain (see \fig{fig:in_plane_mode:2} for the displacement field).
Modes $A_{4,5,6}^{\rm top}$ and $A_{4,5,6}^{\rm bot}$ have the same sign, which indicates that top and bottom layers have the same radial and longitudinal displacement at the AA domain (see \fig{fig:in_plane_mode:2} for the displacement field).
These findings hold for all simulation conditions, including all twist angles, in the presence or absence of an underlying substrate and for both IPs used in the present study.

The out-of-plane coefficients of the top and bottom layers, $B_l^{\rm top}$ and $B_l^{\rm bot}$, are also correlated with the same relations for all twist angles and for both IPs.  For a free-standing TBG, we find that the top and bottom layers move in opposite directions with the same magnitude, i.e., $B_l^{\rm bot}= -B_l^{\rm top}$, whereas for a TBG on a substrate, the bottom layer adheres to the substrate with negligible out-of-plane displacement, $B_l^{\rm bot}\approx 0$, so that the entire out-of-plane displacement is carried by $B_l^{\rm top}$.
\begin{figure*}\centering
     \begin{subfigure}{0.48\textwidth}
         \centering
         \includegraphics[trim = 0mm 0mm 0mm 0mm,clip=yes,width=\textwidth]{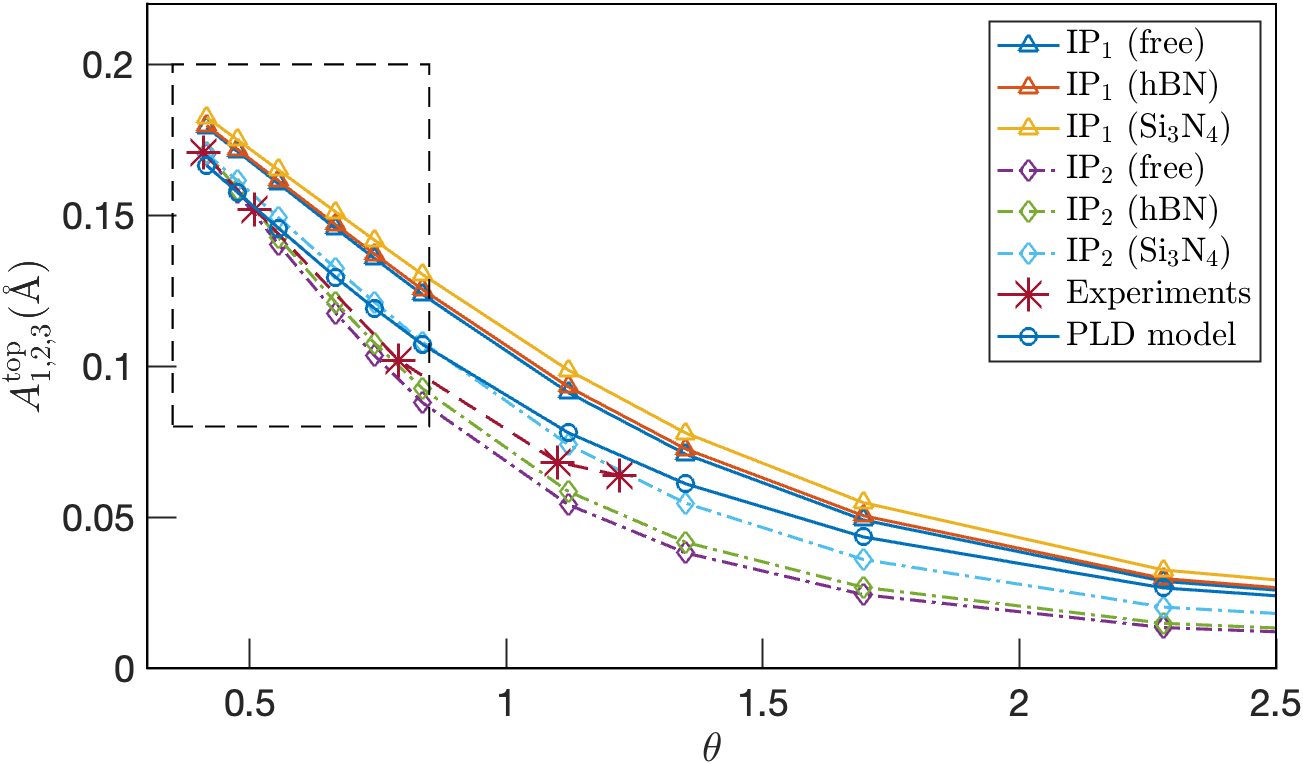}
         \caption{}
         \label{fig:relaxed_A_top:1}
     \end{subfigure}
    \begin{subfigure}{0.48\textwidth}
        \centering
         \includegraphics[trim = 0mm 0mm 0mm 0mm,clip=yes,width=\textwidth]{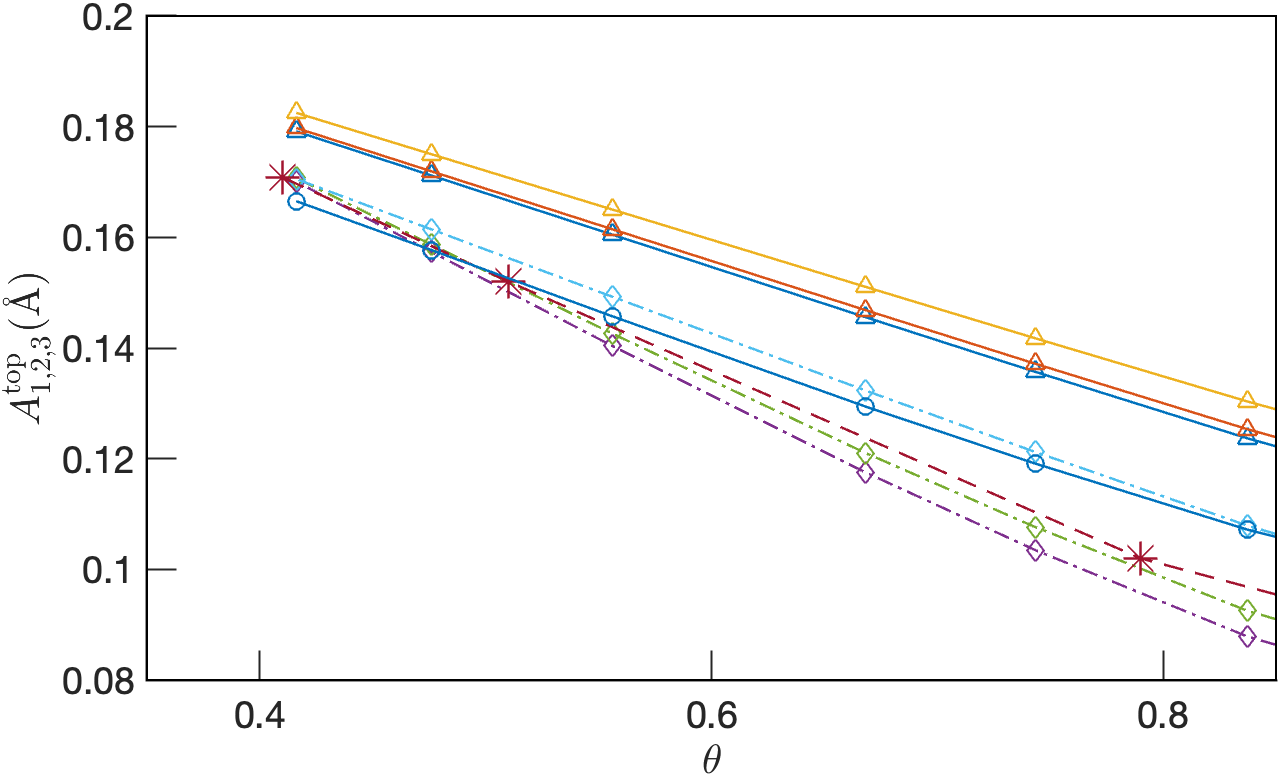}
         \caption{}
         \label{fig:relaxed_A_top:2}
     \end{subfigure}
     \\[8pt]
     \begin{subfigure}{0.48\textwidth}
        \centering
         \includegraphics[trim = 0mm 0mm 0mm 0mm,clip=yes,width=\textwidth]{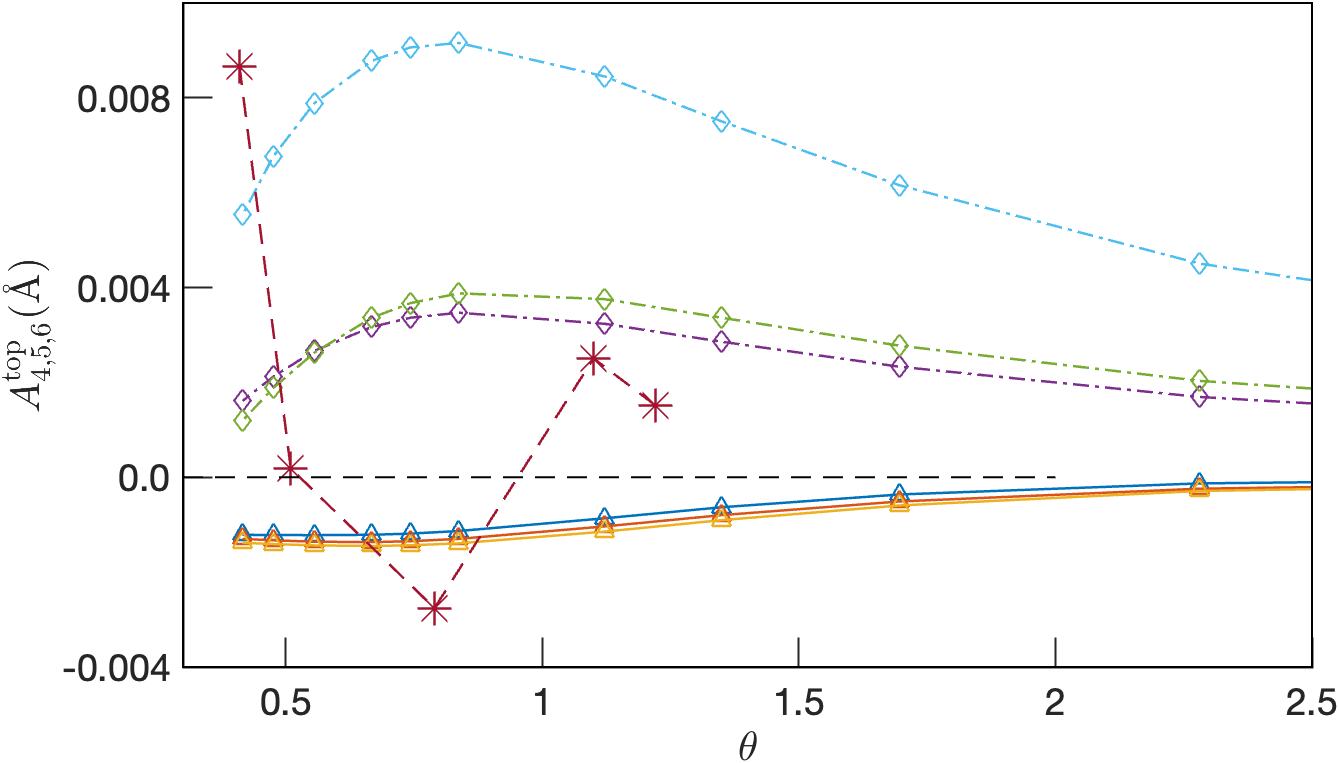}
         \caption{}
         \label{fig:relaxed_A_top:3}
     \end{subfigure}
     \begin{subfigure}{0.48\textwidth}
        \centering
         \includegraphics[trim = 0mm 0mm 0mm 0mm,clip=yes,width=\textwidth]{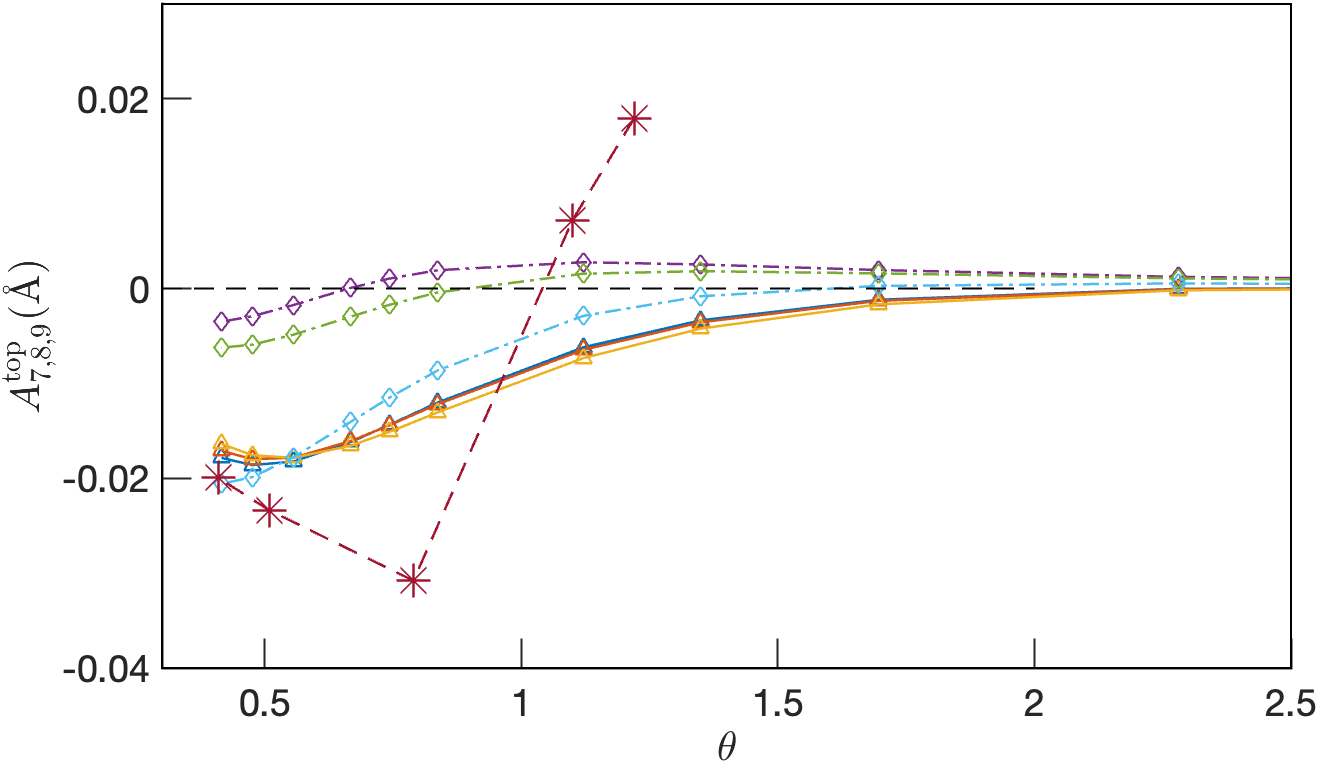}
         \caption{}
         \label{fig:relaxed_A_top:4}
     \end{subfigure}
     \\[8pt]
     \begin{subfigure}{0.48\textwidth}
     \centering
         \includegraphics[trim = 0mm 0mm 0mm 0mm,clip=yes,width=\textwidth]{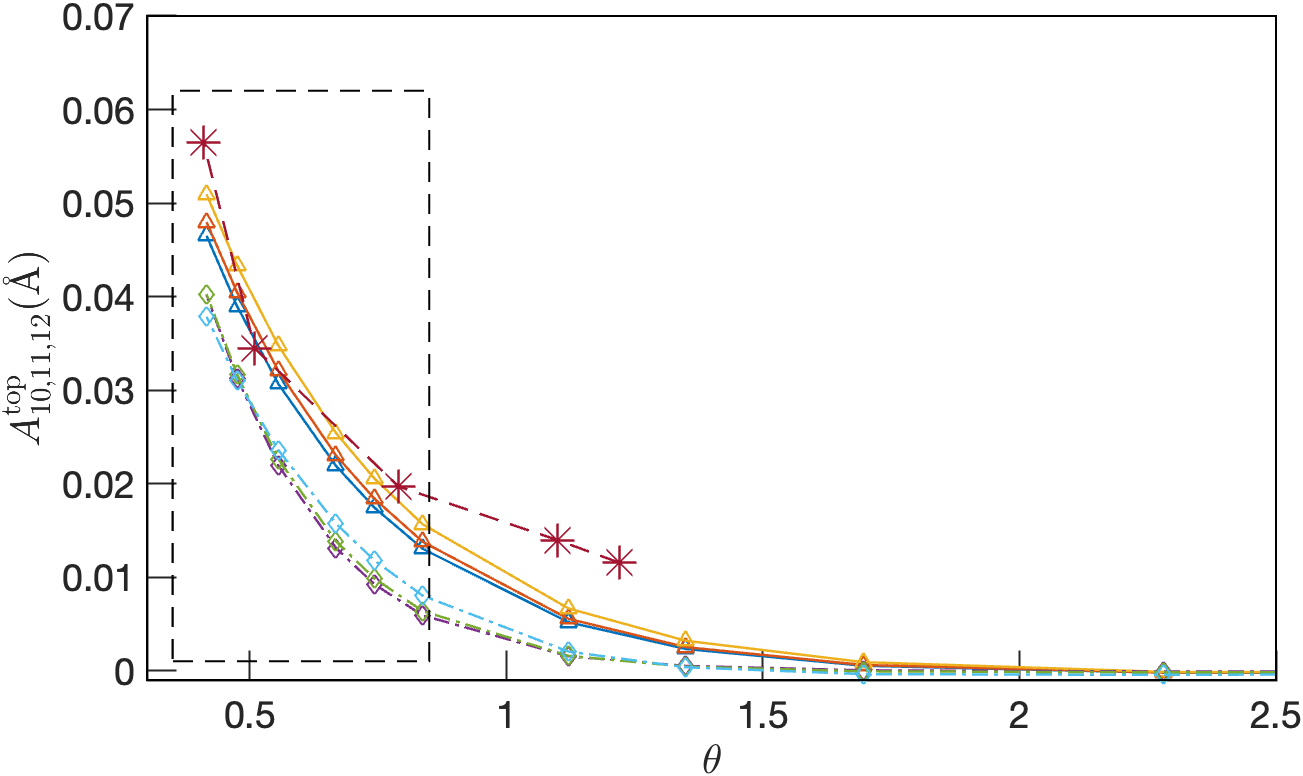}
         \caption{}
         \label{fig:relaxed_A_top:5}
     \end{subfigure}
          \begin{subfigure}{0.48\textwidth}
          \centering
         \includegraphics[trim = 0mm 0mm 0mm 0mm,clip=yes,width=\textwidth]{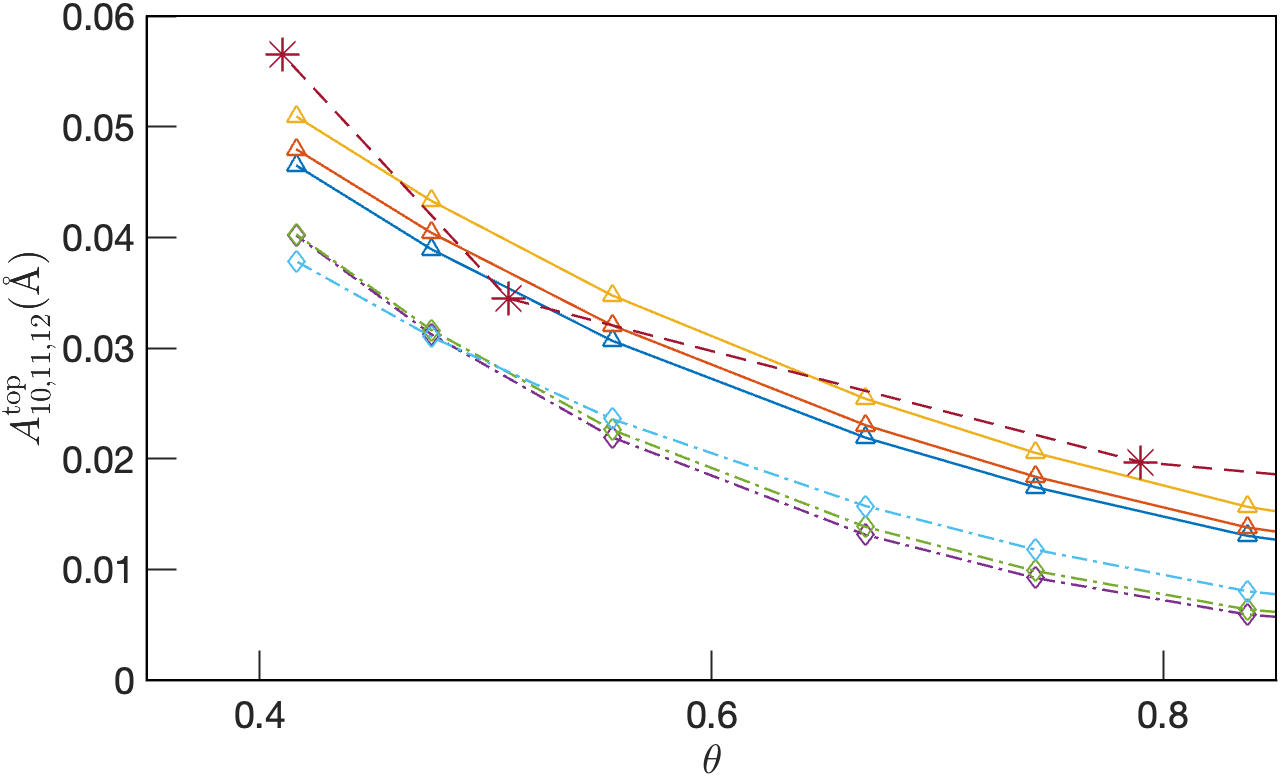}
         \caption{}
         \label{fig:relaxed_A_top:6}
     \end{subfigure}
\captionsetup{justification=raggedright, singlelinecheck=false}     
\caption {The in-plane $A_{k}^{\rm top}$ coefficients for a relaxed TBG as a function of twist angle. (a)  Results are shown for both a free-standing TBG and a TBG supported on an hBN or Si$_3$N$_4$ substrate relaxed using IP$_1$ or IP$_2$. Optimized $A_{k}$ values extracted from experimental electron diffraction images \cite{yoo2019} are also plotted as asterisks (see \sect{sec:ed:static} for details). $A_{k}$ values obtained from the modified PLD model are added for comparison. (a) $A_{1,2,3}^{\rm top}$, (b) Zoomed-in-view of the rectangular dashed region in (a), (c) $A_{4,5,6}^{\rm top}$ (d) $A_{7,8,9}^{\rm top}$, (e) $A_{10,11,12}^{\rm top}$, and (f) Zoomed-in-view of the rectangular dashed region in (e).
}
\label{fig:relaxed_A_top}
\end{figure*}

\fig{fig:relaxed_A_top} shows $A_{1,2,3}^{\rm top}$, $A_{4,5,6}^{\rm top}$, $A_{7,8,9}^{\rm top}$, and $A_{10,11,12}^{\rm top}$ as a function of twist angle for a free-standing TBG and a TBG supported on an hBN or Si$_3$N$_4$ subsrate relaxed using IP$_1$ or IP$_2$. The PLD model in \cite{sung2022} only considers fundamental transverse modes (i.e., $\bth{1,2,3}$ and  $\bth{10,11,12}$). In this study, the PLD model is extended to consider other modes, including $\bth{4,5,6}$, $\bth{7,8,9}$.
For all simulation conditions, $A_{1,2,3}^{\rm top}$ is the dominant deformation mode with a magnitude 3--25 times larger than that of $A_{4,5,6}^{\rm top}$, $A_{7,8,9}^{\rm top}$ and $A_{10,11,12}^{\rm top}$. We see tha $A_{1,2,3}^{\rm top}$ and  $A_{10,11,12}^{\rm top}$ (the second largest mode) increase with decreasing twist angle, whereas the other modes exhibit a more complex dependence on twist angle, but are quite small. Because the displacement of $\bth{1}^{\rm in}$+$\bth{2}^{\rm in}$+$\bth{3}^{\rm in}$ induces localized twisting centered on the AA domains (as shown in \fig{fig:in_plane_mode:1}), the increase of $A_{1,2,3}^{\rm top}$ with decreasing twist angle corresponds to larger localized twisting at lower twist angles. This is consistent with the results in \cite{zhang2018}.

\begin{figure}
    \centering
    \includegraphics[width=0.45\textwidth]{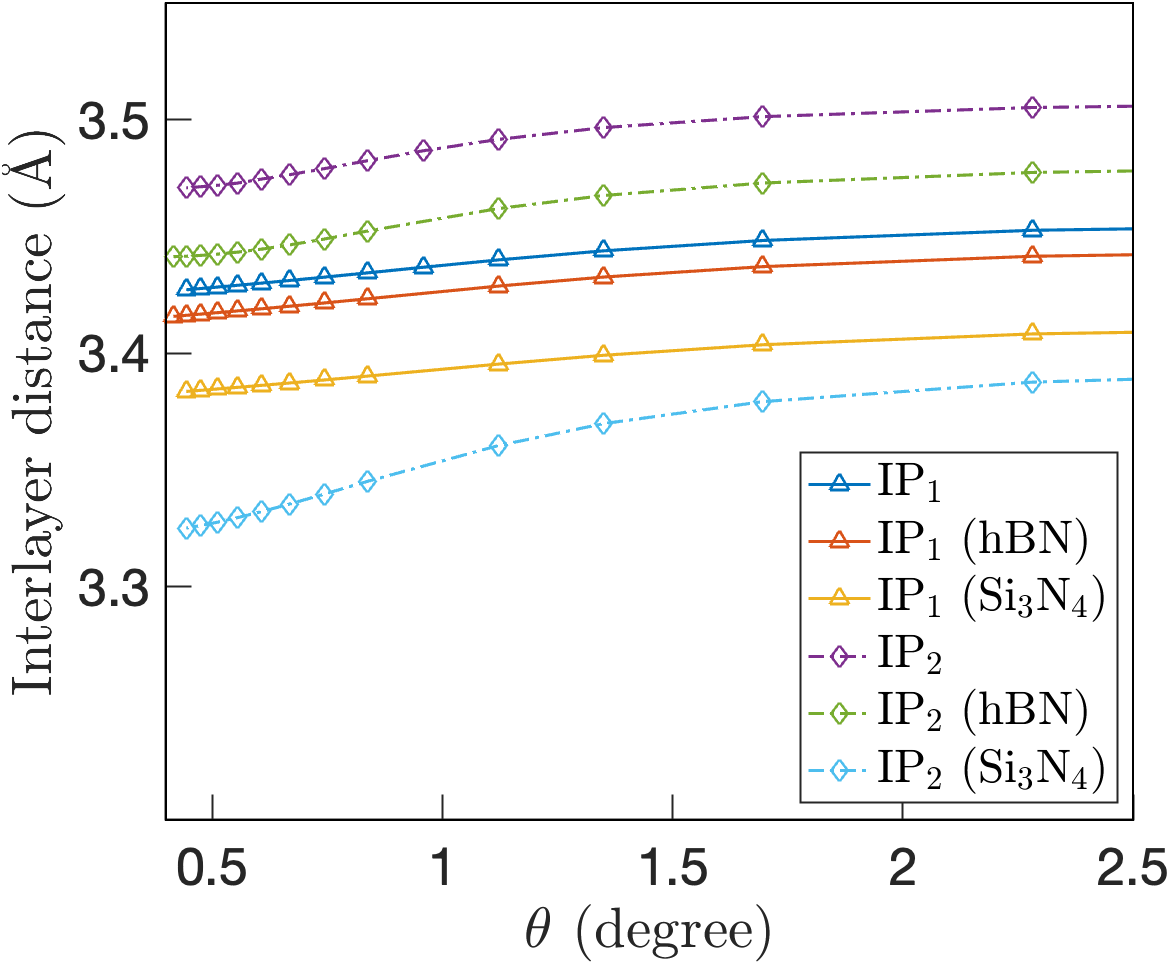}
    \caption{Average distance between the graphene layers in a relaxed TBG.}
    \label{fig:inter_dis}
\end{figure}

As shown in \fig{fig:relaxed_A_top:2}, the presence of a substrate increases the in-plane coefficient $A_{1,2,3}^{\rm top}$ compared with a free-standing TBG (more so for Si$_3$N$_4$ than hBN). The increase in $A_{1,2,3}^{\rm top}$ is attributed to the LJ interaction between the TBG and the substrate. The LJ interaction leads to a decrease in the interlayer distance (as shown in \fig{fig:inter_dis}), which increases the interlayer energy, especially at the high energy AA domain. $A_{1,2,3}^{\rm top}$ (and $A_{1,2,3}^{\rm bot}$) increase to reduce the size of the high energy AA domain at the expense of in-plane strain energy in the graphene layer \cite{zhang2018}. The effect is more pronounced for the Si$_3$N$_4$ substrate than hBN due to the larger LJ coefficients for TBG--Si$_3$N$_4$ interactions (see \tab{tab:LJ_parameters}) that lead to a smaller interlayer separation.

Comparing $A_{1,2,3}^{\rm top}$ with the experimental values plotted in \fig{fig:relaxed_A_top}, we see that IP$_2$ (hNN) provides a more accurate representation of TBG relaxation than IP$_1$ (AIREBO and DRIP).
We note that in the experiments, the TBG is sandwiched between a Si$_3$N$_4$ substrate and an hBN capping layer \cite{yoo2019}, whereas the atomistic simulations only consider a substrate.

The next largest contribution to the displacement comes from $A_{10,11,12}^{\rm top}$. \fig{fig:relaxed_A_top:5} and \figx{fig:relaxed_A_top:6} show that $A_{10,11,12}^{\rm top}$ has a similar trend to $A_{1,2,3}^{\rm top}$. i.e., $A_{10,11,12}^{\rm top}$ increases with decreasing twist angle. The magnitude of $A_{10,11,12}^{\rm top}$ grows when the substrate is supported. Here, unlike the case of $A_{1,2,3}^{\rm top}$, IP$_1$ is in better agreement with the experimental results. This shows that neither potential fully captures the TBG deformation, however since the contributions of the $A_{1,2,3}^{\rm top}$ modes to the displacement are about three times larger than $A_{10,11,12}^{\rm top}$, IP$_2$ remains the better choice for simulating TBG deformation.

Finally for the in-plane displacement, Figs.~\ref{fig:relaxed_A_top:3} and \ref{fig:relaxed_A_top:4} show the contributions of the $A_{4,5,6}^{\rm top}$ and $A_{7,8,9}^{\rm top}$ modes as a function of twist angle. As shown in \fig{fig:in_plane_mode:2}, the displacement field of $\bth{4}^{\rm in}+\bth{5}^{\rm in}+\bth{6}^{\rm in}$ includes radial and longitudinal displacements at the AA domain. Compared to other modes, the magnitude of $A_{4,5,6}^{\rm top}$ is quite small, on the order of $10^{-3}$ \AA. $A_{7,8,9}^{\rm top}$ decrease with the twist angle as shown in \fig{fig:relaxed_A_top:4}. The substrate effect increases the magnitude of $A_{7,8,9}^{\rm top}$ for IP$_2$, but less so for IP$_1$.
$A_{7,8,9}^{\rm top}$ represent the magnitude of localized twistings at AA, AB, and BA domains with the same direction, as shown in the displacement field of $\bth{7}^{\rm in}+\bth{8}^{\rm in}+\bth{9}^{\rm in}$ in \fig{fig:in_plane_mode:3}.
The functional dependence of $A_{4,5,6}^{\rm top}$ and $A_{7,8,9}^{\rm top}$ on the twist angle for both IP$_1$ and IP$_2$ are not in agreement with the experiment.  However, we note that it is difficult to extract experimental values for $A_{4,5,6}$, $A_{7,8,9}$ and $A_{10,11,12}$ because the intensity of the electron diffraction image is primrily affected by $A_{1,2,3}$, which is responsible for the majority of the deformation of graphene layers. Computation of A$_k$ values from experimental electron diffraction images is discussed in \sect{sec:ed:static}.

\begin{figure*}\centering
     \begin{subfigure}[b]{0.48\textwidth}
         \centering
         \includegraphics[trim = 0mm 0mm 0mm 0mm,clip,width=\textwidth]{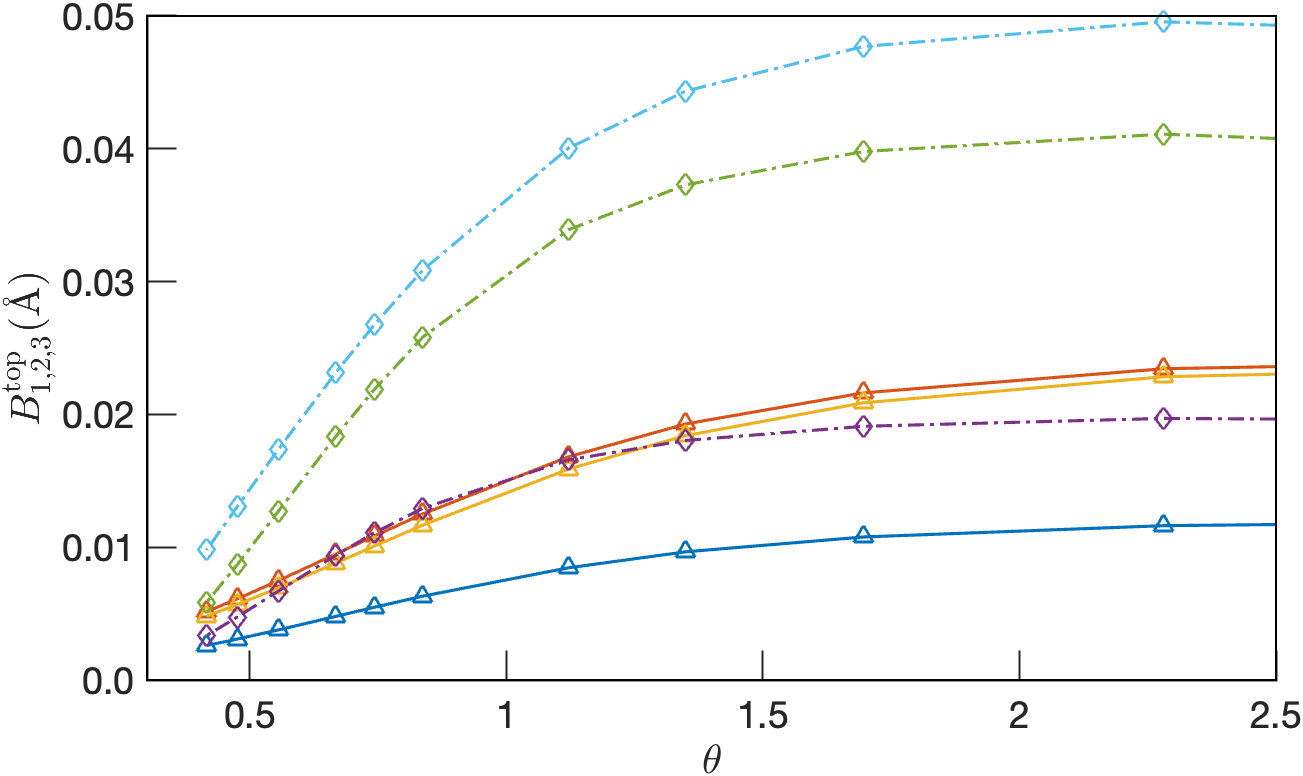}
         \caption{$B_{1,2,3}^{\rm top}$}
         \label{fig:relaxed_B_top:1}
     \end{subfigure}
     \begin{subfigure}[b]{0.48\textwidth}
         \centering
         \includegraphics[trim = 0mm 0mm 0mm 0mm,clip,width=\textwidth]{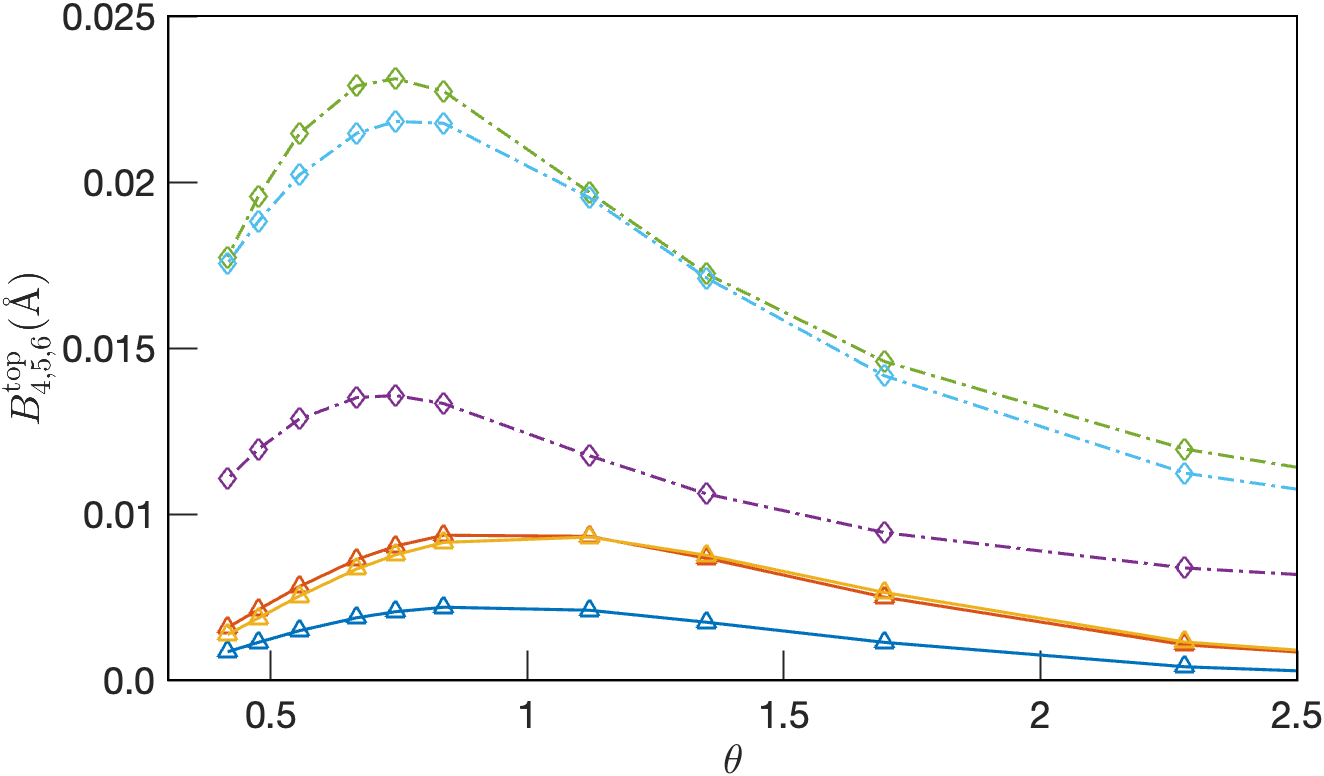}
         \caption{$B_{4,5,6}^{\rm top}$}
         \label{fig:relaxed_B_top:2}
     \end{subfigure}
     \\[8pt]
     \begin{subfigure}[b]{0.48\textwidth}
         \centering
         \includegraphics[trim = 0mm 0mm 0mm 0mm,clip,width=\textwidth]{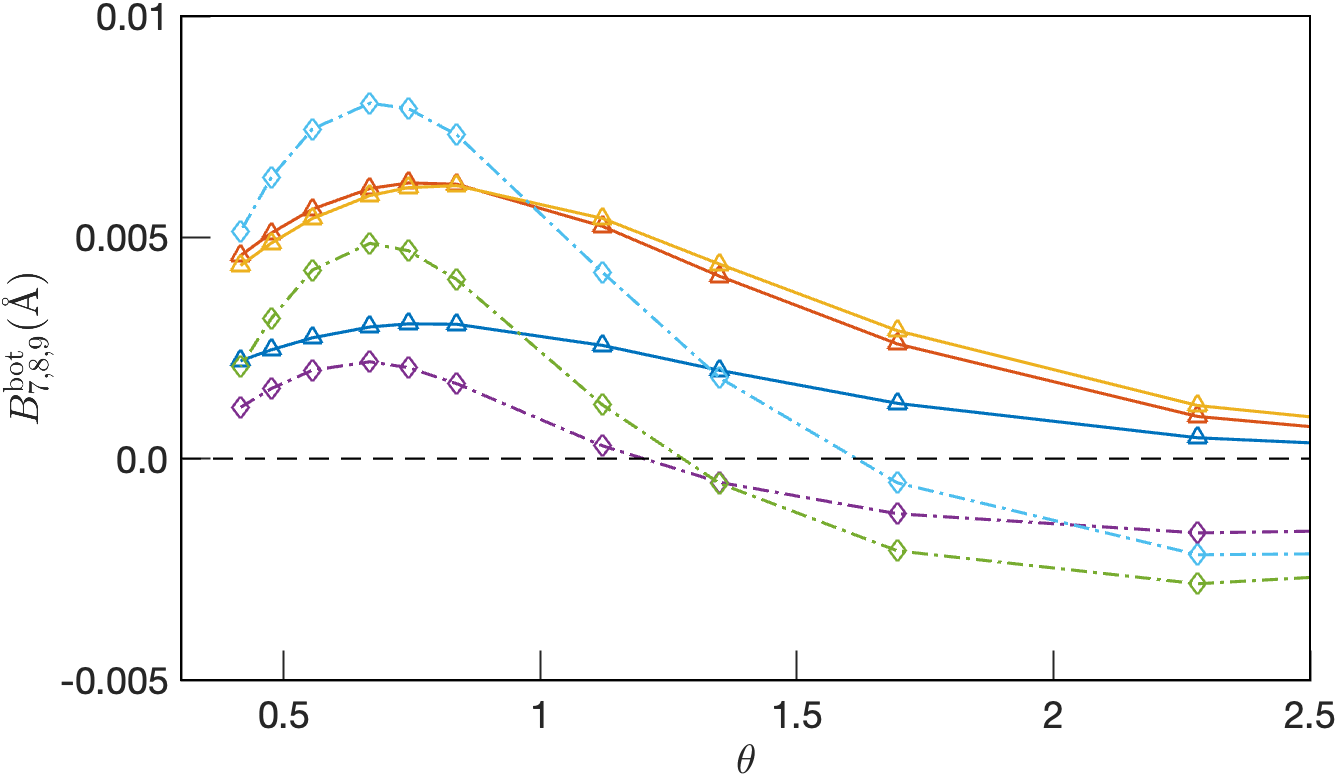}
         \caption{$B_{7,8,9}^{\rm top}$}
         \label{fig:relaxed_B_top:3}
     \end{subfigure}
     \begin{subfigure}[b]{0.48\textwidth}
         \centering
         \includegraphics[trim = 0mm 0mm 0mm 0mm,clip,width=\textwidth]{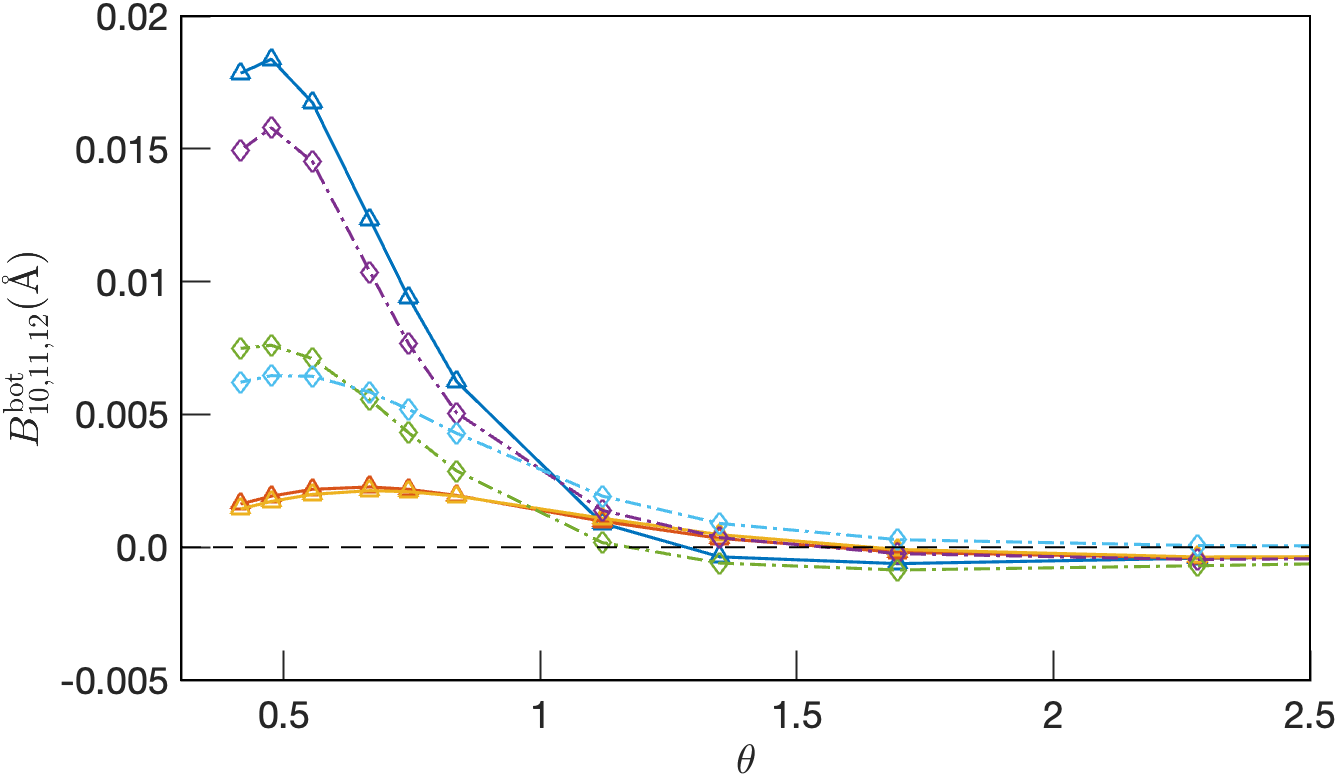}
         \caption{$B_{10,11,12}^{\rm top}$}
         \label{fig:relaxed_B_top:4}
     \end{subfigure}
     \\[8pt]
     \begin{subfigure}[b]{0.48\textwidth}
         \centering
         \includegraphics[trim = 0mm 0mm 0mm 0mm,clip,width=\textwidth]{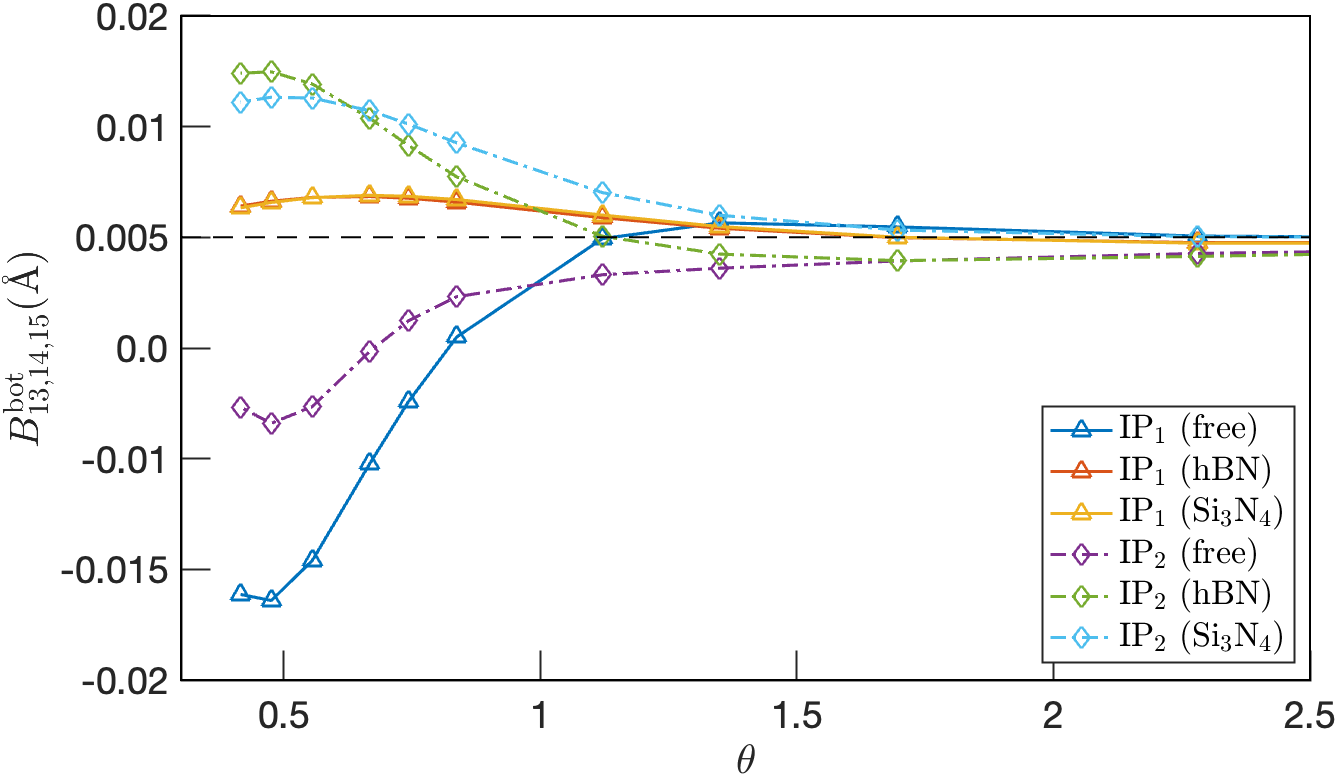}
         \caption{$B_{13,14,15}^{\rm top}$}
         \label{fig:relaxed_B_top:5}
     \end{subfigure}
\captionsetup{justification=raggedright, singlelinecheck=false} 
\caption{The out-of-plane $B_l^{\rm top}$ coefficients for a relaxed TBG as a function of twist angle. Results are shown for both a free-standing TBG and a TBG supported on an hBN or Si$_3$N$_4$ substrate relaxed using IP$_1$ or IP$_2$.}
\label{fig:relaxed_B_top}
\end{figure*}

\begin{figure*}\centering
     \begin{subfigure}[b]{0.48\textwidth}
         \centering
         \includegraphics[trim = 0mm 0mm 0mm 0mm,clip,width=\textwidth]{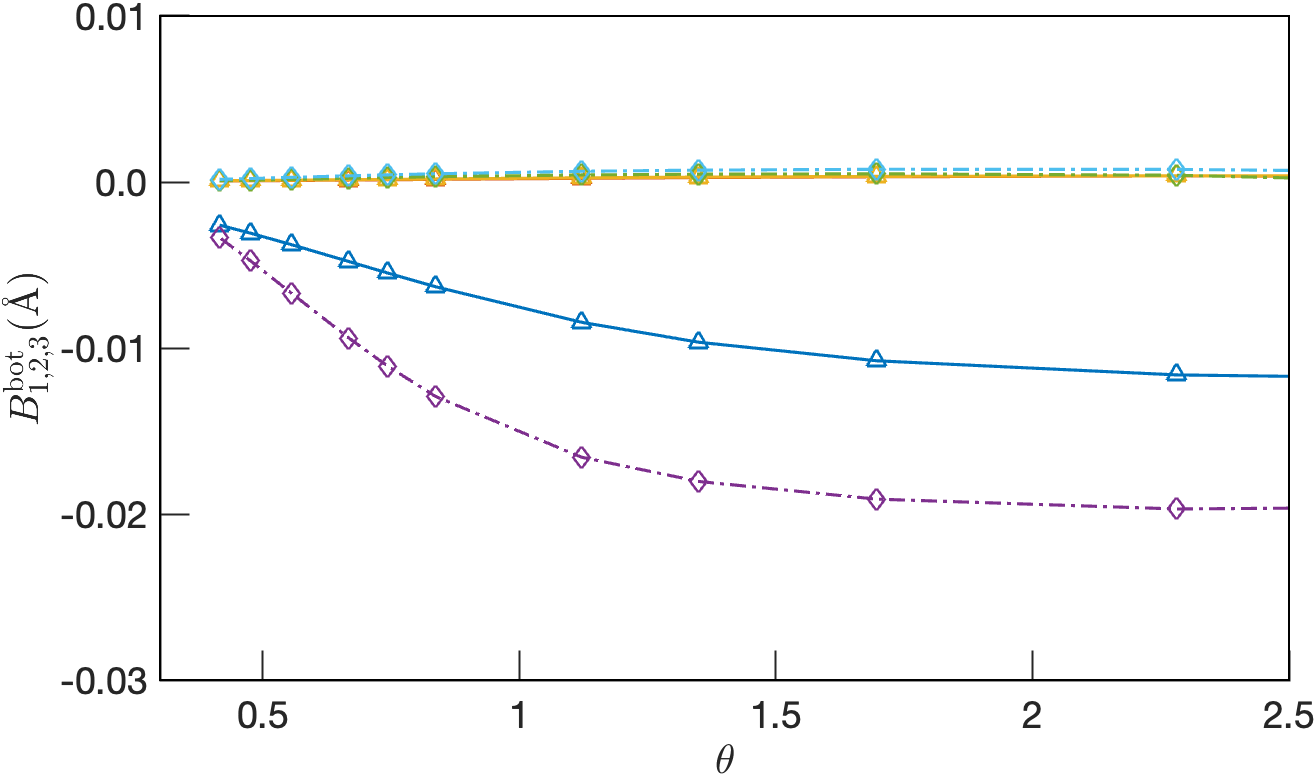}
         \caption{$B_{1,2,3}^{\rm bot}$}
         \label{fig:relaxed_B_bot:1}
     \end{subfigure}
     \begin{subfigure}[b]{0.48\textwidth}
         \centering
         \includegraphics[trim = 0mm 0mm 0mm 0mm,clip,width=\textwidth]{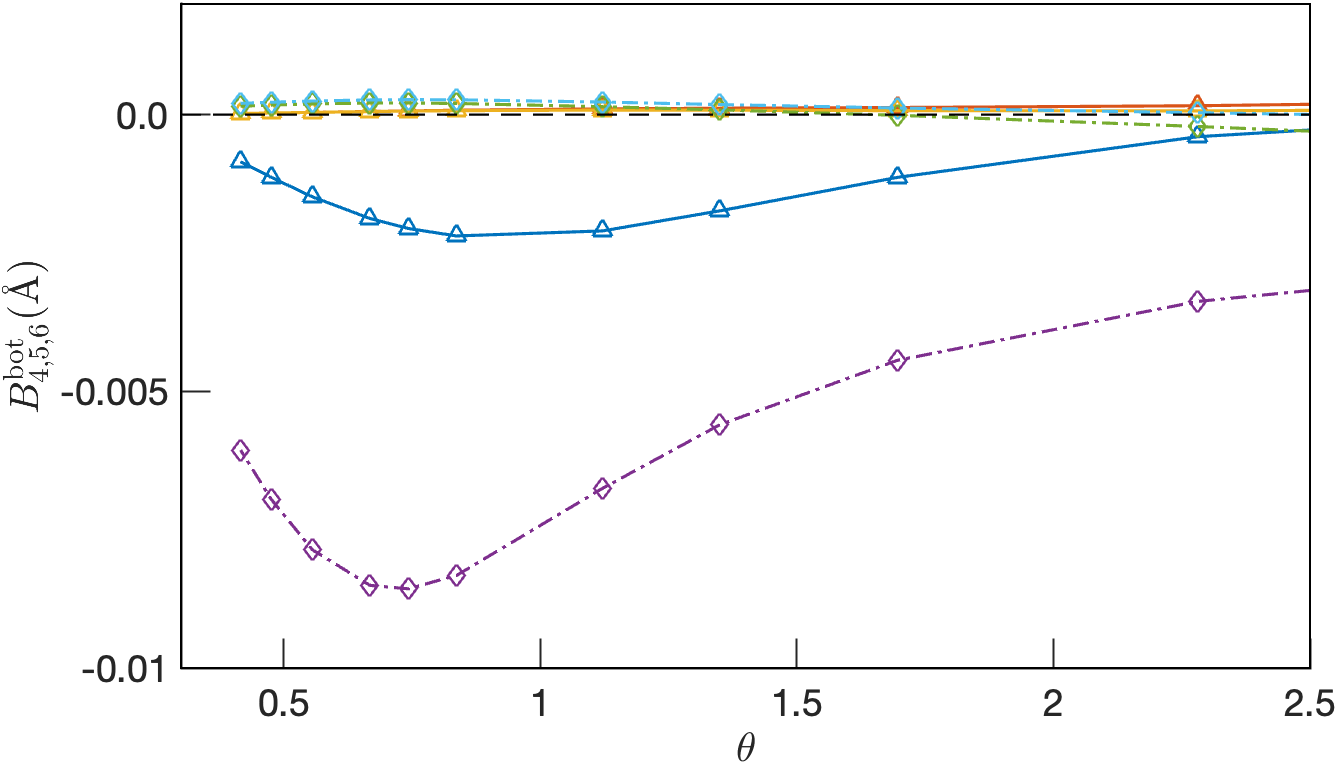}
         \caption{$B_{4,5,6}^{\rm bot}$}
         \label{fig:relaxed_B_bot:2}
     \end{subfigure}

    \begin{subfigure}[b]{0.48\textwidth}
         \centering
         \includegraphics[trim = 0mm 0mm 0mm 0mm,clip,width=\textwidth]{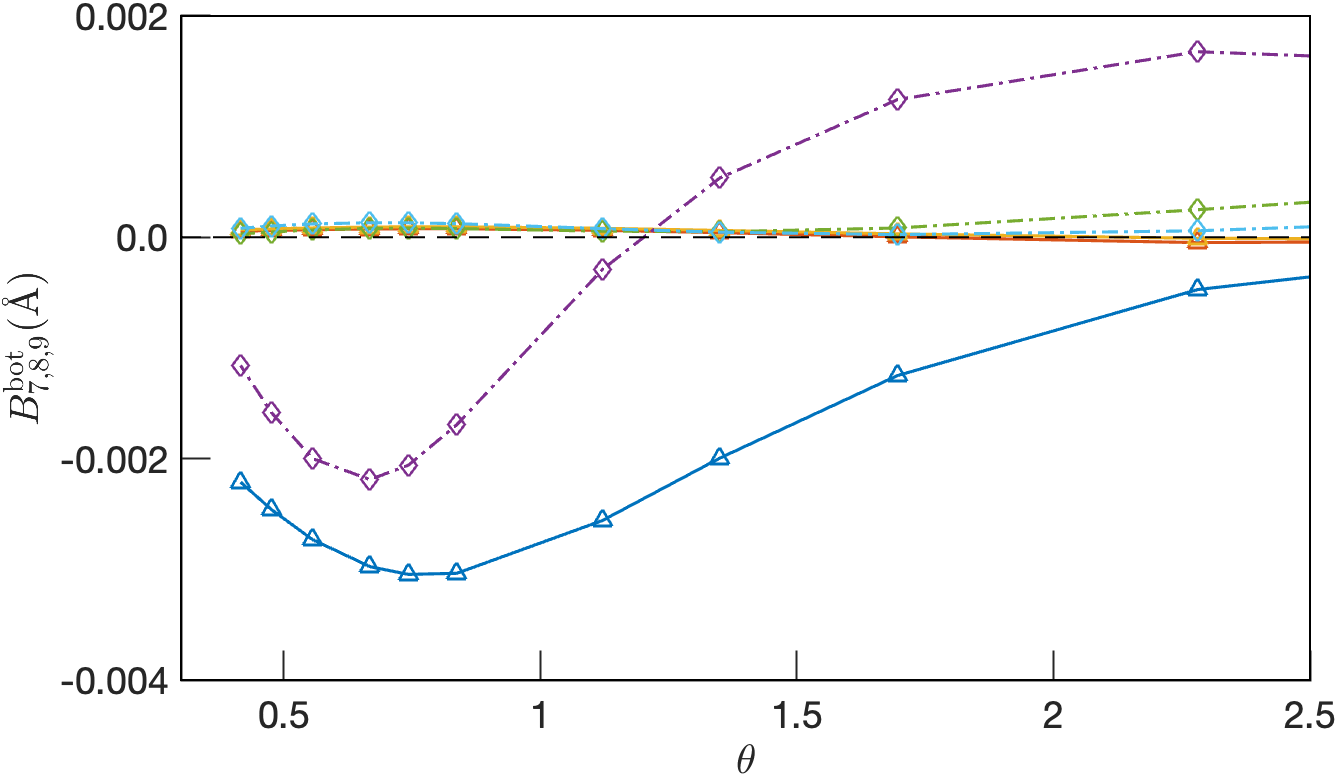}
         \caption{$B_{7,8,9}^{\rm bot}$}
         \label{fig:relaxed_B_bot:3}
     \end{subfigure}
    \begin{subfigure}[b]{0.48\textwidth}
         \centering
         \includegraphics[trim = 0mm 0mm 0mm 0mm,clip,width=\textwidth]{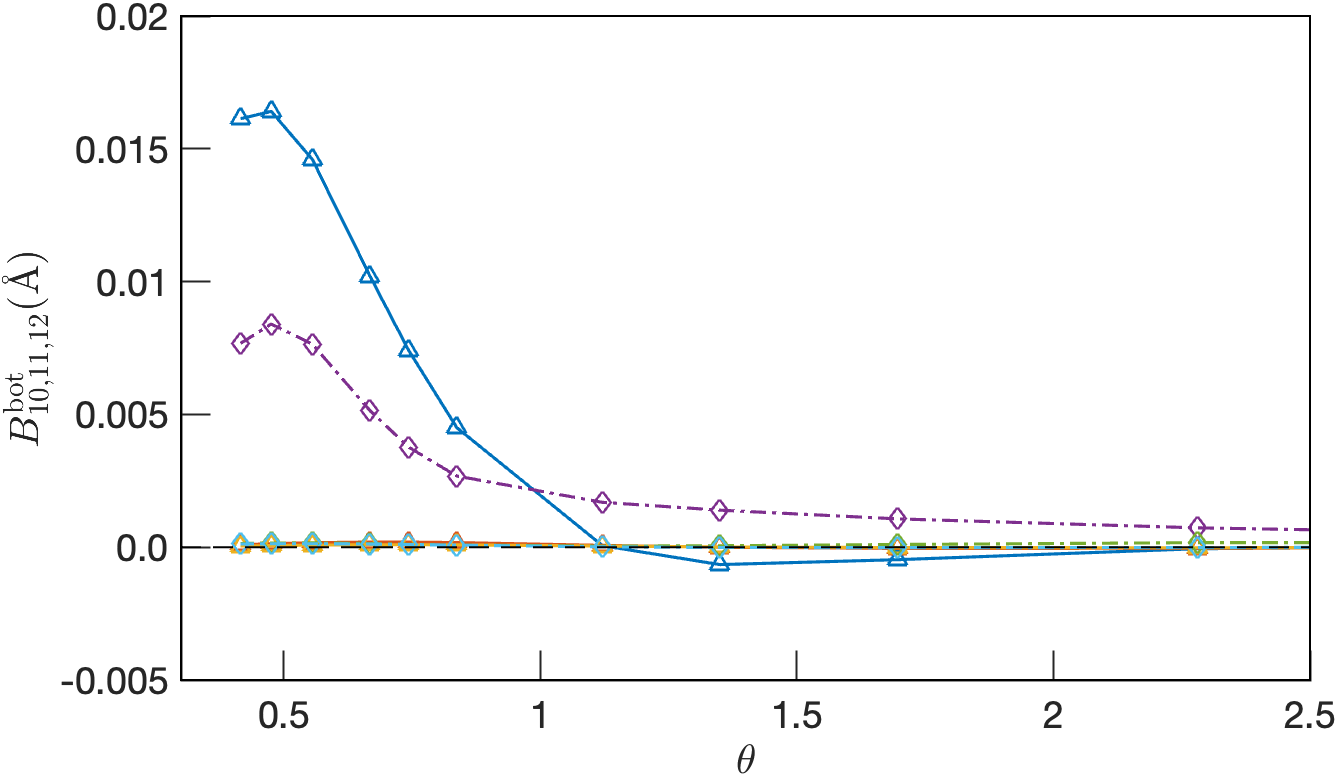}
         \caption{$B_{10,11,12}^{\rm bot}$}
         \label{fig:relaxed_B_bot:4}
     \end{subfigure}

     \begin{subfigure}[b]{0.48\textwidth}
         \centering
         \includegraphics[trim = 0mm 0mm 0mm 0mm,clip,width=\textwidth]{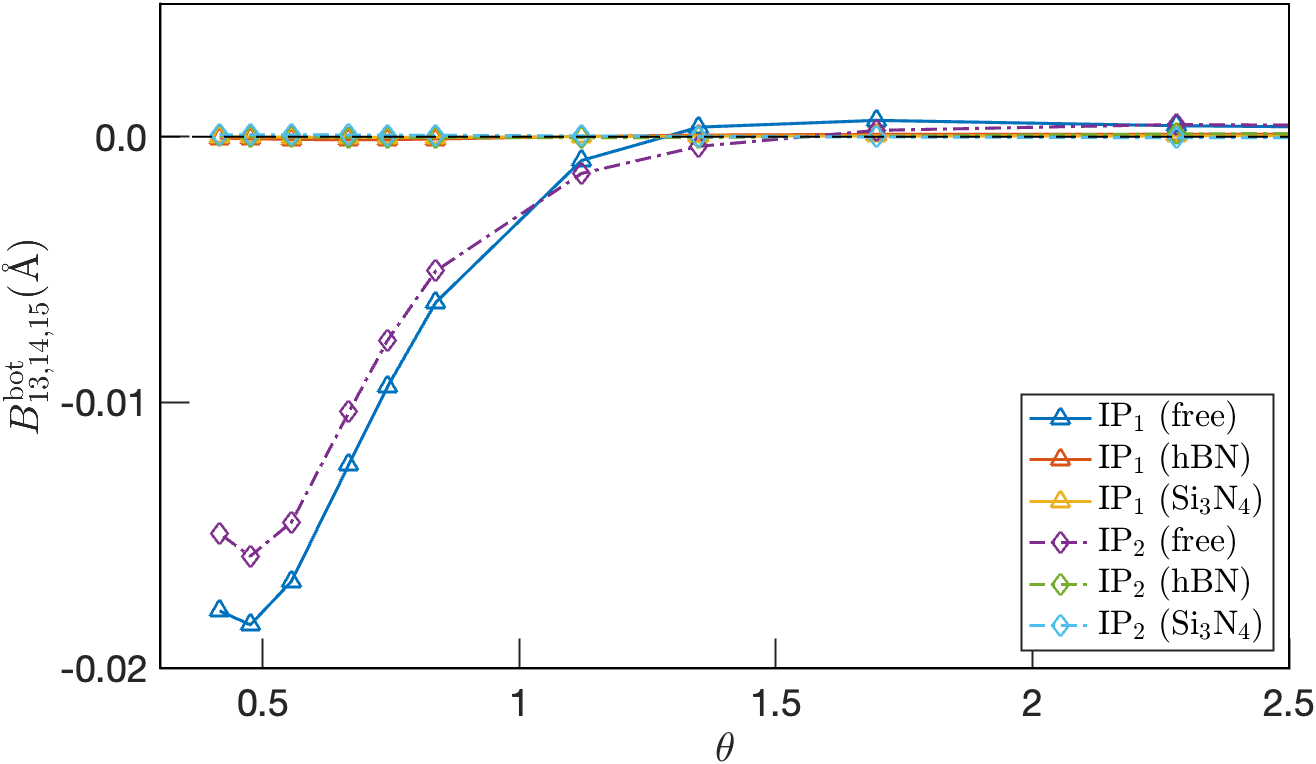}
         \caption{$B_{13,14,15}^{\rm bot}$}
         \label{fig:relaxed_B_bot:5}
     \end{subfigure}
\captionsetup{justification=raggedright, singlelinecheck=false} 
\caption{The out-of-plane $B_l^{\rm bot}$ coefficients for a relaxed TBG as a function of twist angle. Results are shown for both a free-standing TBG and a TBG supported on an hBN or Si$_3$N$_4$ substrate relaxed using IP$_1$ or IP$_2$.}
\label{fig:relaxed_B_bot}
\end{figure*}

The out-of-plane coefficients $B_l^{\rm top}$ are shown in \fig{fig:relaxed_B_top} (see \fig{fig:relaxed_B_bot} for $B_l^{\rm bot}$).
$B_{1,2,3}$ carries most of the out-of-plane displacement at large twist angles ($\theta>1.5\degree$). However it decreases rapidly with a decreasing twist angle so that higher-order modes ($l > 3$) are significant for the out-of-plane displacement at small twist angles ($\theta < 1.5\degree$).
As the twist angle decreases, all $B_l$ values become small, on the order of 10$^{-2}$~\AA, whereas the in-plane $A_{1,2,3}$ and $A_{10,11,12}$ modes increase to a displacement of order of 10$^{-1}$~\AA. This indicates that the in-plane modes (1,2,3) and (10,11,12) are most dominant in a relaxed TBG at small twist angles.

\begin{figure*}[!htbp]\centering
     \begin{subfigure}[b]{0.305\textwidth}
         \centering
         \includegraphics[trim = 0mm 0mm 0mm 0mm,clip,width=\textwidth]{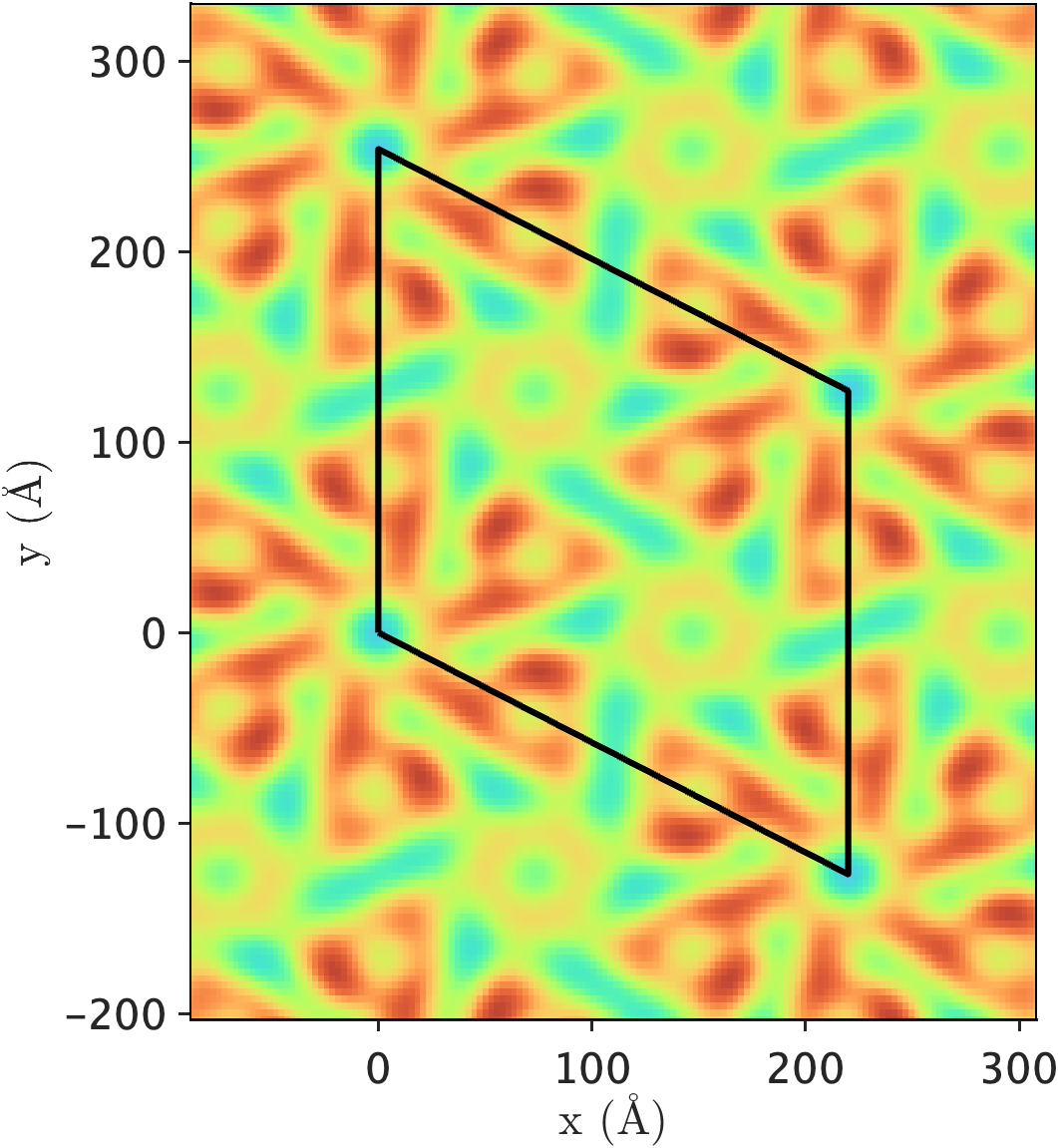}
         \caption{$\vert\bm{\epsilon}_{\rm in}\vert$ ($\theta = 0.55 \degree$)}
         \label{fig:relaxed_error:1}
     \end{subfigure}
     \begin{subfigure}[b]{0.305\textwidth}
         \centering
         \includegraphics[trim = 0mm 0mm 0mm 0mm,clip,width=\textwidth]{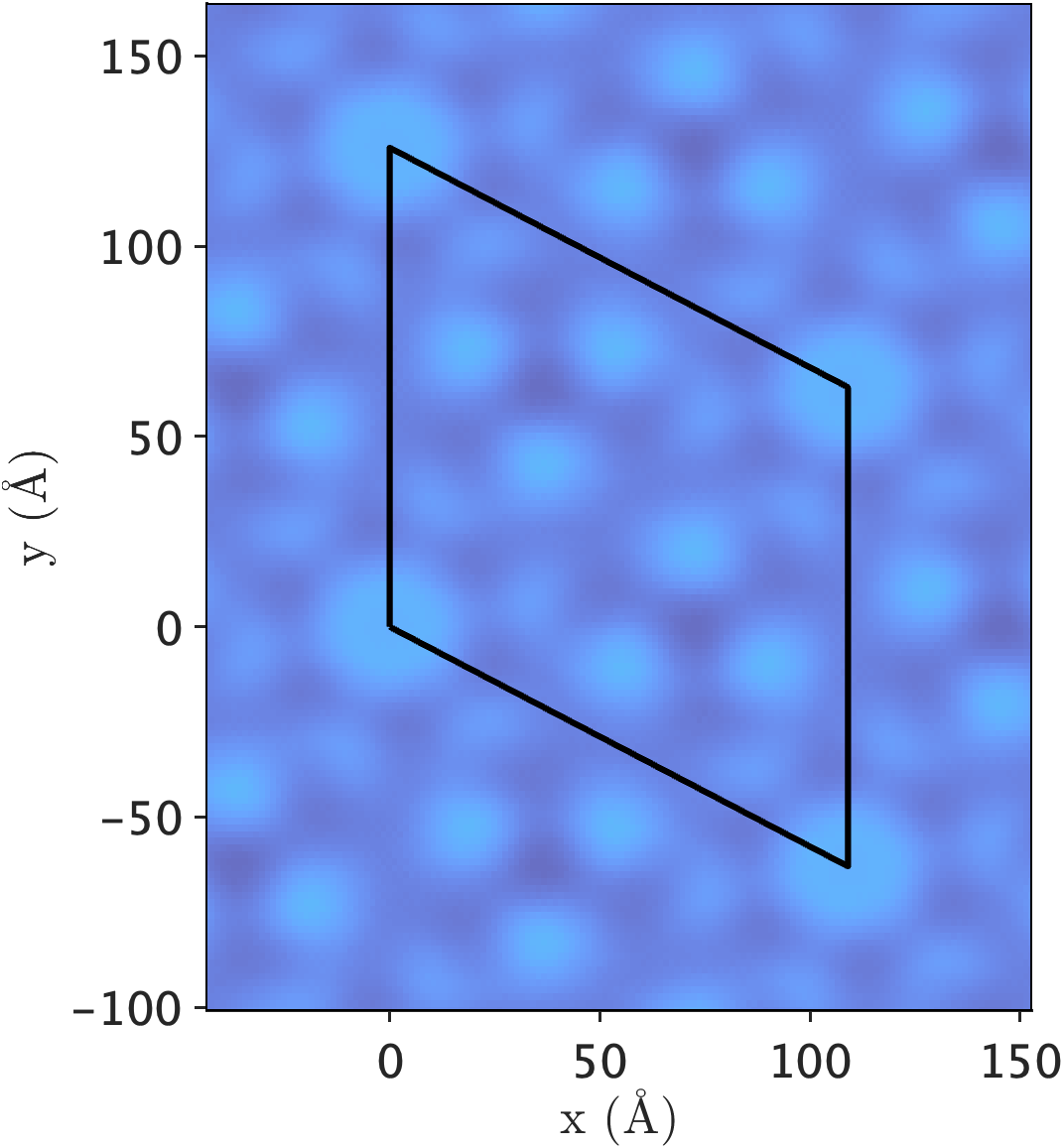}
         \caption{$\vert\bm{\epsilon}_{\rm in}\vert$ ($\theta = 1.12 \degree$)}
         \label{fig:relaxed_error:2}
     \end{subfigure}
     \begin{subfigure}[b]{0.32\textwidth}
         \centering
         \includegraphics[trim = 0mm 0mm 0mm 0mm,clip,width=\textwidth]{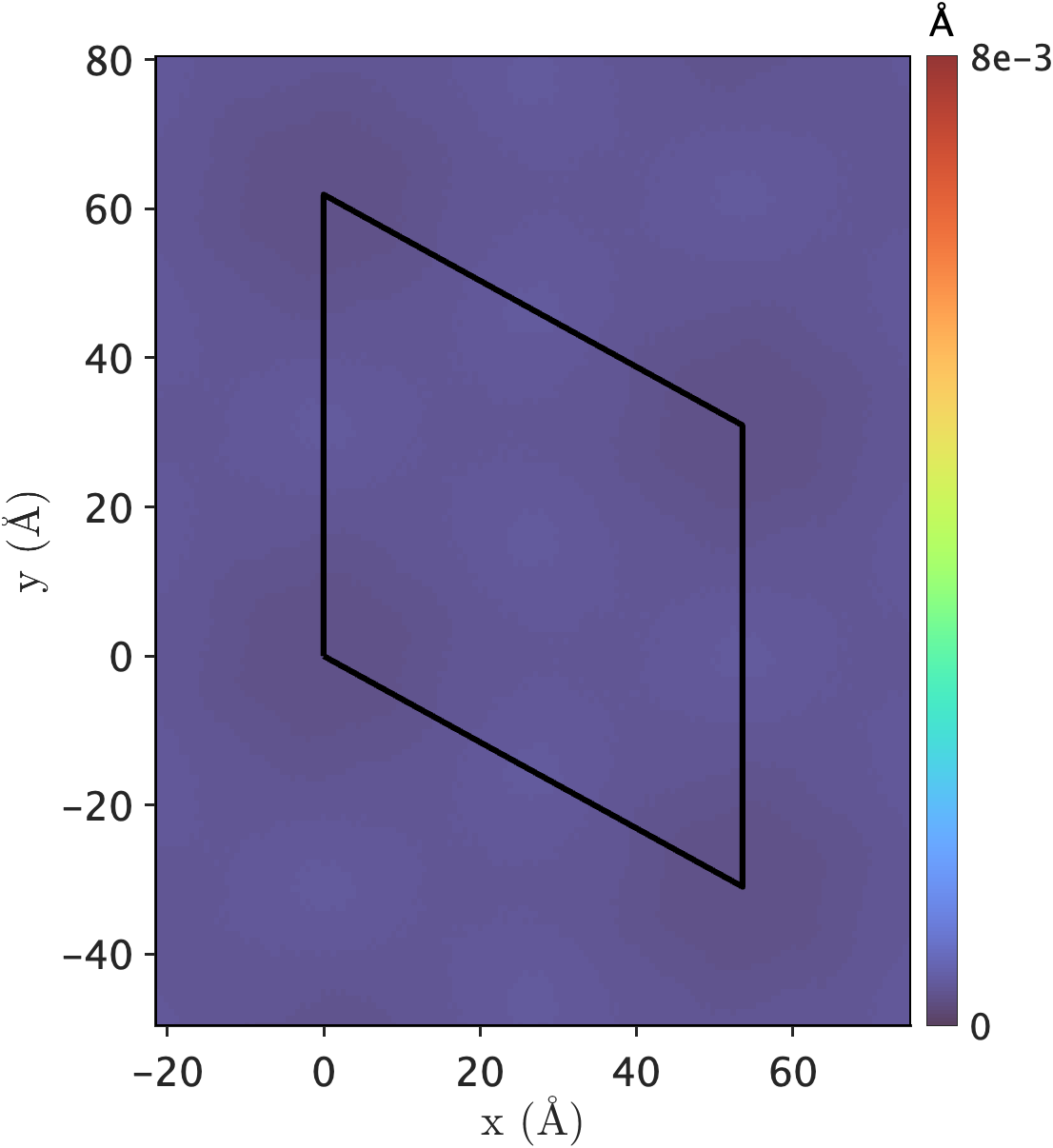}
         \caption{$\vert\bm{\epsilon}_{\rm in}\vert$ ($\theta = 2.28 \degree$)}
         \label{fig:relaxed_error:3}
     \end{subfigure}
     \\[8pt]
     \begin{subfigure}[b]{0.305\textwidth}
         \centering
         \includegraphics[trim = 0mm 0mm 0mm 0mm,clip,width=\textwidth]{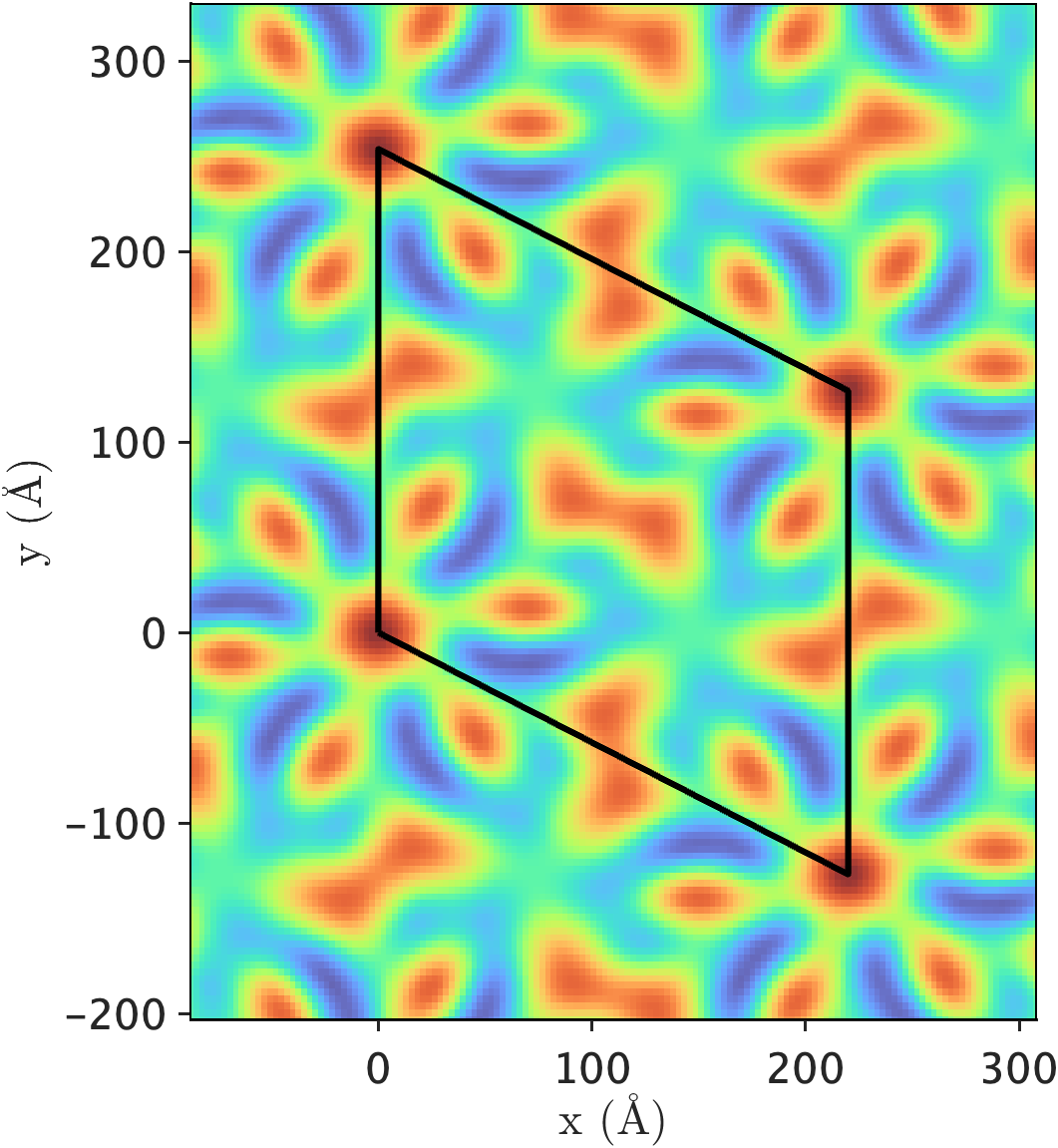}
         \caption{$\bm{\epsilon}_{\rm out}$ ($\theta = 0.55 \degree$)}
         \label{fig:relaxed_error:4}
     \end{subfigure}
     \begin{subfigure}[b]{0.305\textwidth}
         \centering
         \includegraphics[trim = 0mm 0mm 0mm 0mm,clip,width=\textwidth]{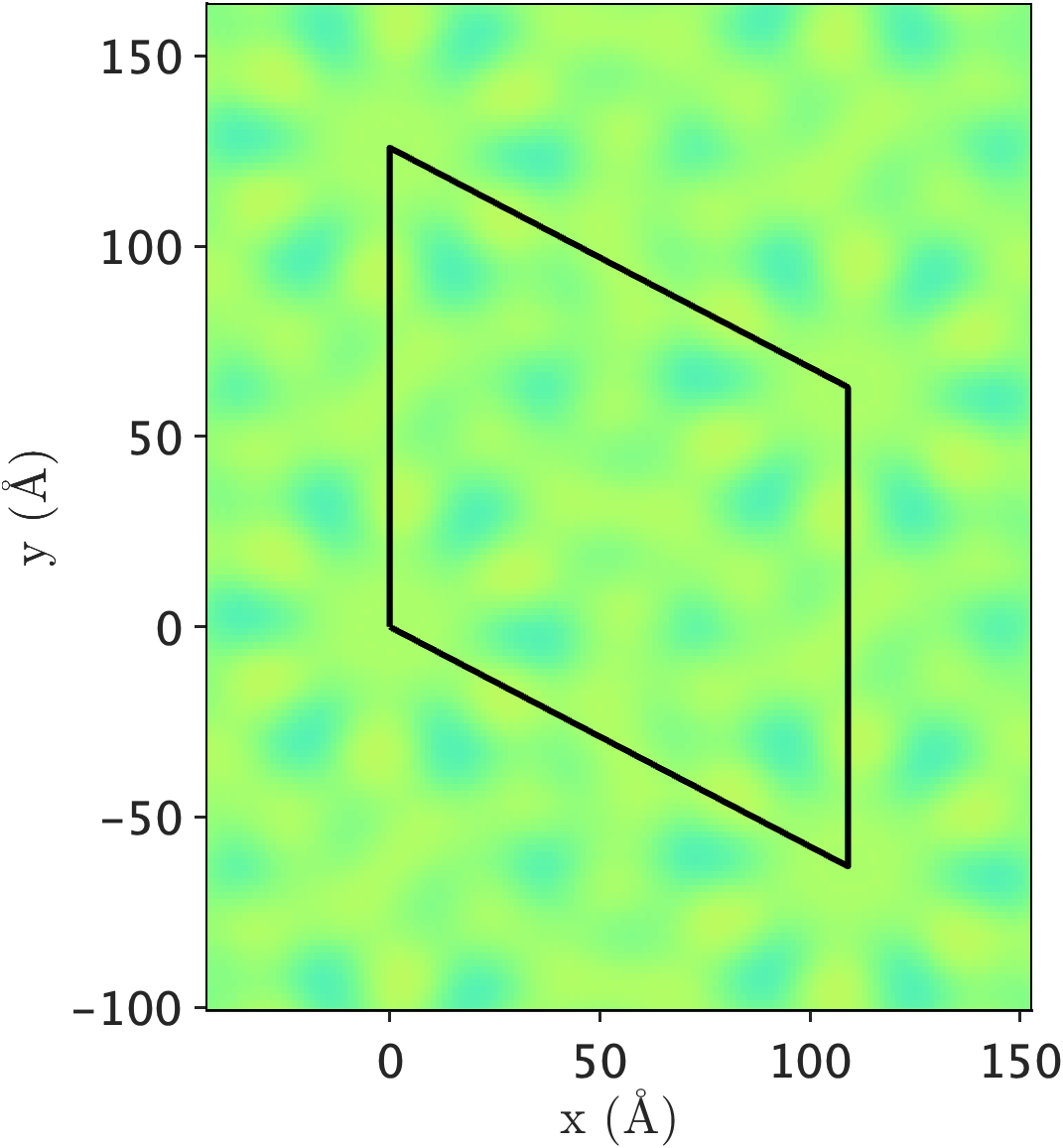}
         \caption{$\bm{\epsilon}_{\rm out}$ ($\theta = 1.12 \degree$)}
         \label{fig:relaxed_error:5}
     \end{subfigure}
     \begin{subfigure}[b]{0.32\textwidth}
         \centering
         \includegraphics[trim = 0mm 0mm 0mm 0mm,clip,width=\textwidth]{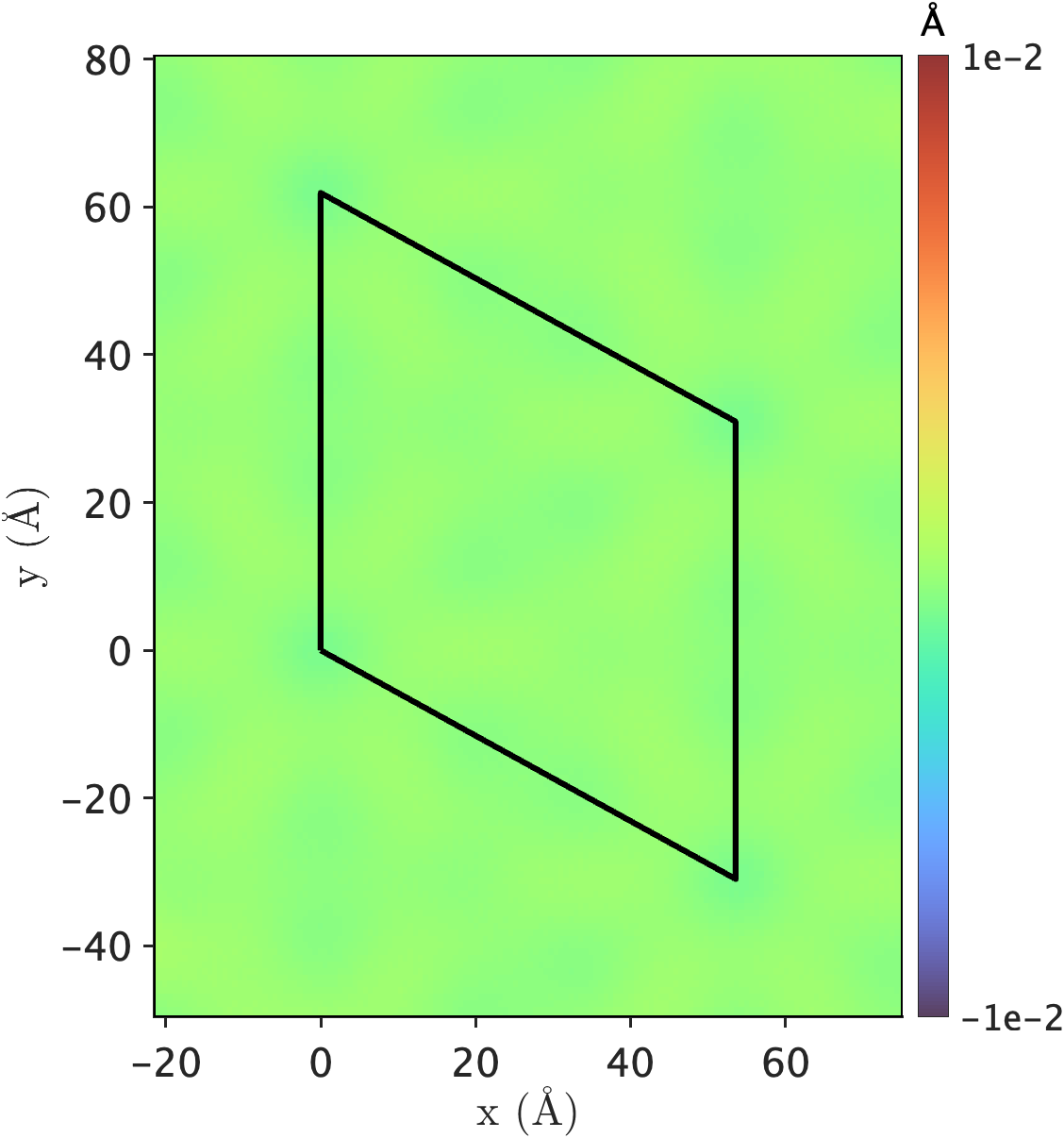}
         \caption{$\bm{\epsilon}_{\rm out}$ ($\theta = 2.28 \degree$)}
         \label{fig:relaxed_error:6}
     \end{subfigure}
\captionsetup{justification=raggedright, singlelinecheck=false} 
\caption{Contour plots of the displacement errors $\vert\bm{\epsilon}_{\rm in}(\br_0)\vert$ (absolute value) and $\bm{\epsilon}_{\rm out}(\br_0)$ of the top layer in a relaxed free-standing TBG for different values of the twist angle computed using potential IP$_2$. The solid black line is the boundary of the supercell.
}
\label{fig:relaxed_error}
\end{figure*}

Finally, the in-plane and out-of-plane displacement errors in \eqn{eqn:disp_bases}, $\bm{\epsilon}_{\rm in}(\br_0)$ and $\bm{\epsilon}_{\rm out}(\br_0)$, are shown in \fig{fig:relaxed_error} for the top layer in a relaxed free-standing TBG, computed using potential IP$_2$, for twist angles 0.55, 1.22, and 2.28$\degree$. These errors represent localized displacement fields that are not captured by the truncated elastic plate basis representation. As the twist angle decreases (for which the size of the supercell increases) the magnitude of error increases for both $\bm{\epsilon}_{\rm in}(\br_0)$ and $\bm{\epsilon}_{\rm out}(\br_0)$. This indicates that high-order modes become more important at low twist angles. However, the selected truncation in the present study (12 in-plane bases and 15 out-of-plane bases) capture the majority of the displacement for the range twist angles considered with a relative error of $\vert\bm{\epsilon}_{\rm in}+\bm{\epsilon}_{\rm out}\vert/\vert\mathbf{\Delta}\vert<0.02\ (2\%)$ for all atoms in the top layer and for all twist angles.

\section{Electron diffraction simulation}\label{sec:ed:static}
Direct experimental validation of TBG deformation patterns using atomic resolution electron microscopy is challenging due to the complex geometry of the TBG samples. Instead TBG deformation is typically investigated through electron diffraction images \cite{zhang2018,yoo2019,sung2022}.

For a thin 2D material, the electron diffraction image intensity is successfully computed by a Fourier transform of the atomic structure \cite{sung2022,hovden2016,zhang2018}. The intensity ($J(\mathbf{k})$) is given by: \begin{equation}\label{eqn:intensity}
J(\mathbf{k}) \propto \vert V^{\rm top}(\mathbf{k})+V^{\rm bot}(\mathbf{k})\vert^2,
\end{equation}
where $\mathbf{k}$ is the 2D position in reciprocal space, and $V^{\rm top}$ and $V^{\rm bot}$ are the discrete 2D Fourier transforms of the top and bottom TBG layer structures:
\begin{multline}\label{eqn:electron_diffraction_exp}
    V(\mathbf{k})= \\ \frac{1}{MN}\sum_{m=0}^{M-1}\sum_{n=0}^{N-1}\sum_{j=1}^{\Nlay}\exp{(i\mathbf{k}\cdot(\br^{j}+m\mathbf{a}_1+n\mathbf{a}_2))},
\end{multline}
where $\br^{j}$ is the real-space position of atom $j$,
$M$ and $N$ are the number of periodic copies of the supercell along the $\mathbf{a}_1$ and $\mathbf{a}_2$ directions, and $m$ and $n$ are the supercell indices.
Note that the $z$ component of $\br^j$ does not impact \eqn{eqn:electron_diffraction_exp} since the $z$ component of $\mathbf{k}$ is zero.
Due to periodicity, for any integers $m$ and $n$,
\begin{equation}
\exp{(i\mathbf{k}\cdot(\br^j+m\mathbf{a}_1+n\mathbf{a}_2))}
=
\exp{(i\mathbf{k}\cdot\br^{j})}.
\end{equation}
Therefore, \eqn{eqn:electron_diffraction_exp} simplifies to
\begin{equation}\label{eqn:electron_diffraction_exp_2}
    V(\mathbf{k})=\sum_{j=1}^{\Nlay}\exp{(i\mathbf{k}\cdot\br_{\rm}^{j})}.
\end{equation}

In \sect{sec:relaxation:general}, the positions of the atoms in a relaxed TBG are expressed via selected in-plane and out-of-plane bases, $\bth{k}^{\rm in}$ and $\bth{l}^{\rm out}$, with associated coefficients $A_k$ and $B_l$ (see \eqn{eqn:disp_bases}). Assuming that accelerated electrons are transmitted through a TBG along its normal direction and interact with layers independently, the out-of-plane positions of atoms can be omitted. Thus, \eqn{eqn:electron_diffraction_exp_2} becomes
\begin{align}
    V(\mathbf{k})&=\sum_{j=1}^{\Nlay}\exp{(i\mathbf{k}\cdot(\br_{0}^j+\mathbf{\Delta}_{\rm in}(\br_0^j))}, \nonumber \\
    &=\sum_{j=1}^{\Nlay}\exp{(i\mathbf{k}\cdot(\br_{0}^j+\sum_{k=1}^{12}A_k\bth{k}^{\rm in}(\br_0^j))}, \label{eqn:electron_diffraction_exp_4}
\end{align}
where $\br_0^j$ is the reference position of atom $j$ (in either the top or bottom layer) of an unrelaxed TBG.
Using \eqn{eqn:electron_diffraction_exp_4}, the reciprocal structures for the top and bottom layers follow as
\begin{multline}
     \Hat{V}^{\rm top}\left(\mathbf{k},A_{1,2,3}^{\rm top},A_{4,5,6}^{\rm top},A_{7,8,9}^{\rm top},A_{10,11,12}^{\rm top}\right) = \\
     \sum_{j=1}^{\Nlay}\exp{(i\mathbf{k}\cdot(\br_{0}^j+\sum_{k=1}^{12}A_k^{\rm top}\bth{k}^{\rm in}(\br_0^j))} \label{eqn:v_hat} 
\end{multline}
\begin{multline}
     \Hat{V}^{\rm bot}\left(\mathbf{k},A_{1,2,3}^{\rm bot},A_{4,5,6}^{\rm bot},A_{7,8,9}^{\rm bot},A_{10,11,12}^{\rm bot}\right) = \\ 
     \sum_{j=1}^{\Nlay}\exp{(i\mathbf{k}\cdot(\br_{0}^j+\sum_{k=1}^{12}A_k^{\rm bot}\bth{k}^{\rm in}(\br_0^j))}. \label{eqn:v_hat_bot}
\end{multline}
Using the relations between $A_k^{\rm top}$ and $A_k^{\rm bot}$ (\tab{tab:Atop_Abot}), the intensity in \eqn{eqn:intensity} is
\begin{multline}\label{eqn:intensity2}
\hat{J}(\mathbf{k},A_{1,2,3}^{\rm top},A_{4,5,6}^{\rm top},A_{7,8,9}^{\rm top},A_{10,11,12}^{\rm top}) \propto \\
\vert\hat{V}^{\rm top}(\mathbf{k},A_{1,2,3}^{\rm top},A_{4,5,6}^{\rm top},A_{7,8,9}^{\rm top},A_{10,11,12}^{\rm top})\\
+\hat{V}^{\rm bot}(\mathbf{k},-A_{1,2,3}^{\rm top},A_{4,5,6}^{\rm top},-A_{7,8,9}^{\rm top},-A_{10,11,12}^{\rm top})\vert^2.
\end{multline}

\begin{figure*}\centering
\includegraphics[width=1.0\textwidth]{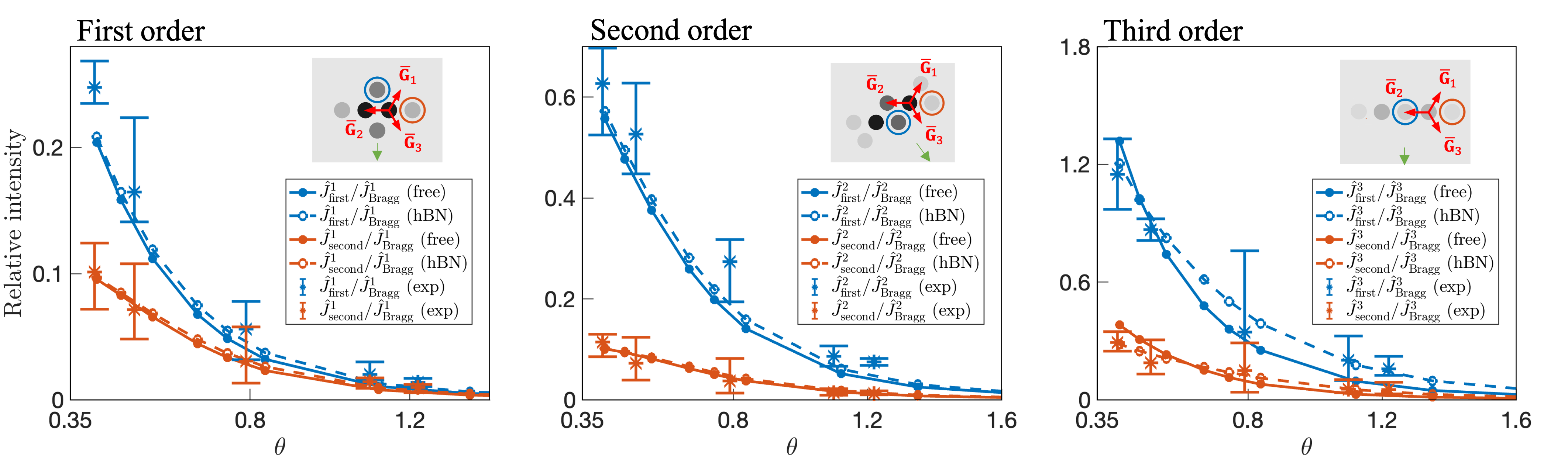}
\captionsetup{justification=raggedright, singlelinecheck=false} 
\caption{The relative intensity between the first and second superlattice peaks and Bragg peaks for the first-, second-, and third-order Bragg spots.
The blue solid and dashed lines represent the relative intensity between the first superlattice and Bragg peaks ($\hat{J}_{\rm first}^i/\hat{J}_{\rm Bragg}^i$), and the orange solid and dashed lines represent the relative intensity between the second superlattice and Bragg peaks ($\hat{J}_{\rm second}^i/\hat{J}_{\rm Bragg}^i$) for a free-standing TBG and a TBG supported on an hBN substrate relaxed with IP$_2$ (hNN). The relative position between the superlattice and Bragg peaks are represented by $\bar{\bG}_1$, $\bar{\bG}_2$, and $\bar{\bG}_3$ in the inset for each Bragg spot. The green arrow represents the direction for the origin of reciprocal space. Averaged relative intensities from experimental electron diffraction images are plotted as asterisks with error bars.
}
\label{fig:rel_I_total}
\end{figure*}

Using \eqn{eqn:intensity2} and the numerical values of $A_k$ computed in \sect{sec:relaxation2}, intensities for the superlattice and Bragg peaks for the first-, second-, and third-order Bragg spots are computed for both a free-standing TBG and a TBG supported on an hBN substrate. The intensity of the Bragg peak for the $i$-th order Bragg spot is denoted by $\hat{J}_{\rm Bragg}^i$, and the intensity of the first and second superlattice peaks in the $i$-th Bragg spot are denoted by $\hat{J}_{\rm first}^i$ and $\hat{J}_{\rm second}^i$. For a given $A_k^{\rm top}$, the intensities $\hat{J}_{\rm Bragg}^i$, $\hat{J}_{\rm second}^i$, and $\hat{J}_{\rm second}^i$ are computed using different $\mathbf{k}$ values that correspond to the position of each peak in reciprocal space as listed in \tab{tab:position_reciprocal}.

\begin{table*}\centering
\begin{tabular}{c|c|c|c}
\hline
 & Bragg peak & First superlattice peak & Second superlattice peak \\
\hline
First Bragg spot & $\mathbf{b}_2^{\rm bot}$ & $\mathbf{b}_2^{\rm bot}-\bar{\bG}_3$ & $\mathbf{b}_2^{\rm bot}-\bar{\bG}_2$\\
Second Bragg spot & $\mathbf{b}_1^{\rm bot}-\mathbf{b}_2^{\rm bot}$ & $\mathbf{b}_1^{\rm bot}-\mathbf{b}_2^{\rm bot}-\bar{\bG}_1$ & $\mathbf{b}_1^{\rm bot}-\mathbf{b}_2^{\rm bot}+\bar{\bG}_2$\\
Third Bragg spot & $2\mathbf{b}_2^{\rm bot}$ & $2\mathbf{b}_2^{\rm bot}+\bar{\bG}_2$ & $2\mathbf{b}_2^{\rm bot}-\bar{\bG}_2$\\
\hline
\end{tabular}
\captionsetup{justification=raggedright, singlelinecheck=false} 
\caption{The reciprocal position $\mathbf{k}$ for the Bragg, first and second superlattice peaks associated with the first-, second-, and third- order Bragg spots. $\mathbf{b}_1^{\rm bot}=\mathbf{R}\mathbf{b}_1$ and $\mathbf{b}_2^{\rm bot}=\mathbf{R}\mathbf{b}_2$ are the reciprocal vectors for the TBG bottom layer unit cell where $\mathbf{b}_1 = 4\pi/(3a)(\sqrt{3}/2\ihat-1/2\jhat)$ and $\mathbf{b}_2 = 4\pi/(3a)\jhat$ are reciprocal vectors of an (untwisted) graphene layer unit cell, and the rotation matrix for twisting is  $\mathbf{R}=\begin{bmatrix}
\cos\theta/2 & -\sin\theta/2 \\ \sin\theta/2  &
\cos\theta/2
\end{bmatrix}$.}
\label{tab:position_reciprocal}
\end{table*}

The intensities of the first and second superlattice peaks relative to the Bragg peak (i.e., $\hat{J}_{\rm first}^i/\hat{J}_{\rm Bragg}^i$ and $\hat{J}_{\rm second}^i/\hat{J}_{\rm Bragg}^i$) for the relaxed free-standing TBG and a TBG supported on an hBN substrate for a range twist angles is plotted in \fig{fig:rel_I_total}. For comparison, relative intensities obtained from experimental electron diffraction images \cite{yoo2019} are also plotted.
Experimental diffraction patterns are quantified by simultaneously fitting up to six 2D Gaussians peaks around each Bragg peak and calculating the volume under each fitted Gaussian. The non-linear least squares fit is performed with Matlab using the {\tt lsqcurvefit} function.
Sung et al.\ \cite{sung2022} reported that an increase in $A_{1,2,3}$ results in increases in the relative intensities ($\hat{J}_{\rm first}^i/\hat{J}_{\rm Bragg}^i$ and $\hat{J}_{\rm second}^i/\hat{J}_{\rm Bragg}^i$) for all Bragg spots. In \fig{fig:rel_I_total}, we see that higher values of $A_{1,2,3}$ for the relaxed TBG supported on an hBN substrate compared to that of the relaxed free-standing TBG (see \fig{fig:relaxed_A_top:1}) slightly increases the relative intensity at the same $\theta$ compared to that of relaxed free-standing TBG for the first- and second- Brag spots. However, we find that the relative intensities in the third-order Bragg spot for the hBN-supported TBG are lower than those for the free-standing TBG when $\theta<0.4\degree$. This discrepancy may be due to the different degrees of freedom of the PLD representation and the more general basis used in the present study to express the TBG deformation.

The expression for the electron diffraction intensity in \eqn{eqn:intensity2} depends on the four coefficients $A_{1,2,3}$, $A_{4,5,6}$, $A_{7,8,9}$, and $A_{10,11,12}$. These parameters can be obtained from the experiment by minimizing the difference between the measured relative intensities and those obtained from \eqn{eqn:intensity2}. The use of relative intensities cancels out the undetermined proportionality constant in \eqn{eqn:intensity2}. Specifically, we define the following cost function
\begin{multline}\label{eqn:phi}
    \phi(A_{1,2,3}^{\rm top},A_{4,5,6}^{\rm top},A_{7,8,9}^{\rm top},A_{10,11,12}^{\rm top}) = \\ \sum_{i=1}^{3}\biggl[\left(\hat{J}_{\rm first}^i/\hat{J}_{\rm Bragg}^i-J_{\rm first}^{i,\rm exp}/J_{\rm Bragg}^{i,\rm exp}\right)^2
    \\ +\left(\hat{J}_{\rm second}^i/\hat{J}_{\rm Bragg}^i-J_{\rm second}^{i,\rm exp}/J_{\rm Bragg}^{i,\rm exp}\right)^2\biggr],
\end{multline}
where $\hat{J}^i_{\rm first}$, $\hat{J}^i_{\rm second}$, $\hat{J}^i_{\rm Bragg}$ are intensities of first and second superlattice and Bragg peaks for the $i$th Bragg spot computed via \eqn{eqn:intensity2}, and $\hat{J}^{i, \rm{exp}}_{\rm first}$, $\hat{J}^{i, \rm{exp}}_{\rm second}$, $\hat{J}^{i, \rm{exp}}_{\rm Bragg}$ are the intensities of the first and second superlattice and Bragg peaks for $i$th Bragg spots obtained from an experimental electron diffraction image.
The optimal coefficients $A_k$ are obtained by minimizing $\phi$ using initial guesses for $A_{1,2,3}^{\rm top}$ and $A_{10,11,12}^{\rm top}$ computed using eqn.~(2) in \cite{sung2022}. The initial guesses for $A_{4,5,6}^{\rm top}$ and $A_{7,8,9}^{\rm top}$ are taken to be zero because their magnitudes are small compared to those of $A_{1,2,3}^{\rm top}$ and $A_{10,11,12}^{\rm top}$. The minimization is performed using the CG method implemented within a Matlab code (provided along with this article) until the norm of the gradient is less than 10$^{-6}$. (Gradients of $\phi$ with respect to the coefficients $A_k$ are given in SM\cite{SM}.) The $A_k^{\rm top}$ obtained from experimental electron diffraction images at different twist angles \cite{yoo2019} are plotted as the asterisks in \fig{fig:relaxed_A_top}.

As mentioned in \sect{sec:relaxation}, the dependence on twist angle of the coefficients  $A_{1,2,3}$ and $A_{10,11,12}$ obtained by fitting to the experiments are consistent with those obtained by fitting to the simulation results (i.e., exponential increase in $A_{1,2,3}$ and $A_{10,11,12}$ with decreasing twist angle). In contrast the dependence of the experimental $A_{4,5,6}$ and $A_{7,8,9}$ coefficients on twist angle is not in good agreement with those obtained from the atomistic simulations.
We believe that this discrepancy is due to the fact that the intensities of the first and second superlattice peaks used to determine the coefficients $A_k$ (see \eqn{eqn:phi}) are mainly determined by $A_{1,2,3}$ and $A_{10,11,12}$ \cite{sung2022,hovden2016}, and thus at the given resolution of experimental electron diffraction images, $A_{4,5,6}$ and $A_{7,8,9}$ are not computed accurately \cite{yoo2019}. More accurate values would require high-resolution electric diffraction images. Alternatively, $A_{4,5,6}$ and $A_{7,8,9}$ could be computed from intensities of superlattice peaks at $\bth{4,5,6}^{\rm in}$ and $\bth{7,8,9}^{\rm in}$ in reciprocal space, respectively \cite{hovden2016}. However, we were unable to accurately measure these peaks from experimental electron diffraction images \cite{yoo2019} because the intensities were too low.

\section{Quantification of TBG dynamics}\label{sec:phonon}
\subsection{Calculation of the TBG dynamical matrix}
The dynamics of the TBG near the relaxed equilibrium structure are investigated by computing the phonon band structure. This is obtained by computing the TBG dynamical matrix, which at wave vector $\mathbf{k}$ is given by
\begin{equation}\label{eqn:dyn_matrix}
    D^{ij}_{\alpha\beta}(\mathbf{k}) = \sum_{R=0}^{\infty}\frac{1}{m_{\rm C}}\left.\frac{\partial^2 U}{\partial r_{i\alpha}^{0} \partial r_{j\beta}^{R}}\right\rvert_{\rm {eq}}e^{i\mathbf{k}\cdot\left(\mathbf{r}_i^{0}-\mathbf{r}_j^R\right)},
\end{equation}
where $i,j=1,\dots,\Ntbg$ are atom numbers in the reference cell of a TBG ($\Ntbg=2\Nlay$ is the number of atoms in the TBG supercell), $\alpha, \beta=1,2,3$ are Cartesian components, $\mathbf{r}^R_i$ is the position vector of atom $i$ in periodic cell $R$ ($R=0$ refers to the reference cell),   $r^R_{i\alpha}$ is component $\alpha$ of $\mathbf{r}^R_i$, $m_{\rm C}$ is the mass of a carbon atom, $U$ is the energy of the relaxed structure (indicated by ``eq'').
The dynamical matrix at the $\Gamma$ point (i.e., $\mathbf{k}=\mathbf{0}$) simplifies to
\begin{equation}\label{eqn:dyn_matrix_k_0}
    D^{ij}_{\alpha\beta}(\mathbf{0}) = \sum_{R=0}^{\infty}\frac{1}{m_{\rm C}}\left.\frac{\partial^2 U}{\partial r_{i\alpha}^{0} \partial r_{j\beta}^{R}}\right\rvert_{\rm {eq}}.
\end{equation}

\begin{algorithm}[t!]
\centering
\captionsetup{justification=raggedright, singlelinecheck=false} 
\caption{Calculation of the dynamical matrix $\mathbf{D}(\mathbf{k})$ for a given $\mathbf{k}$}\label{alg:calcdyn}
\begin{algorithmic}[H]
\Require An atom in the reference cell interacting with at most one periodic image of itself.
\For{$i=1$ to $\Ntbg$}
    \For{$j=1$ to $\Ntbg$}
        \For{$R=0$ to 8}
        \If{Atom $i$ in the reference cell interacts with atom $j$ in $R$ (i.e., $R=R'$)}
        \State{Compute $D^{ij}_{\alpha\beta}(\mathbf{0})$ from \eqn{eqn:dyn_matrix_k_0_2}}
        \State{$D^{ij}_{\alpha\beta}(\mathbf{k}) := D^{ij}_{\alpha\beta}(\mathbf{0})e^{i\mathbf{k}\cdot\left(\mathbf{r}_{i}^0-\mathbf{r}_{j}^R\right)}$}
        \EndIf
        \EndFor
    \EndFor
\EndFor
\end{algorithmic}
\end{algorithm}

In practice, direct calculation of the phonon band structure for large atomistic systems (as in the case of TBGs with small twist angle and a hence large supercell) is computationally prohibitive due to the large number of energy computations required at different $\mathbf{k}$ values.
Overcoming this computational burden is key to computing the phonon band structure of TBGs with low twist angle. This can be done by approximating the TBG phonon band structure, as in the multiscale approach of Lu et al.\ \cite{lu2022}. However, in this study the phonon band sructure is computed exactly using \eqn{eqn:dyn_matrix} capitalizing on the short range of atomic interactions and the periodicity of the supercell to reduce the computational cost. Specifically, we use the fact that for sufficiently low twist angle $\theta$, the cutoff distance for carbon atom interactions ($d_{\rm cutoff}$) is less than the supercell size $L$ (see \eqn{eqn:LAA}).
This implies that an atom in the reference cell can only interact with at most a single periodic image of itself. Therefore the summations in \eqns{eqn:dyn_matrix} and \eqnx{eqn:dyn_matrix_k_0} are dropped, and we have
\begin{equation}\label{eqn:dyn_matrix2}
    D^{ij}_{\alpha\beta}(\mathbf{k}) = \frac{1}{m_{\rm C}}\left.\frac{\partial^2 U}{\partial r_{i\alpha}^{0} \partial r_{j\beta}^{R'}}\right\rvert_{\rm {eq}}e^{i\mathbf{k}\cdot\left(\mathbf{r}_i^{0}-\mathbf{r}_j^{R'}\right)},
\end{equation}
and
\begin{equation}\label{eqn:dyn_matrix_k_0_2}
    D^{ij}_{\alpha\beta}(\mathbf{0}) = \frac{1}{m_{\rm C}}\left.\frac{\partial^2 U}{\partial r_{i\alpha}^{0} \partial r_{j\beta}^{R'}}\right\rvert_{\rm {eq}},
\end{equation}
where $R'$ corresponds to the cell containing atom $j$ that interacts with atom $i$ in the reference cell.
Examining \eqns{eqn:dyn_matrix2} and \eqnx{eqn:dyn_matrix_k_0_2} we see that when $d_{\rm cutoff}<L$, the dynamical matrix for any $\mathbf{k}$ can be obtained by calculating the value at $\mathbf{k}=\mathbf{0}$ in \eqn{eqn:dyn_matrix_k_0_2} and then multiplying by $e^{i\mathbf{k}\cdot\left(\mathbf{r}_i^{0}-\mathbf{r}_j^{R'}\right)}$. This is the case in the present study where the cutoffs for IP$_1$ (AIREBO+DRIP) and IP$_2$ (hNN), $d_{\rm cutoff}^{{\rm IP}_1}=12$~\AA\ and $d_{\rm cutoff}^{{\rm IP}_2}=10$~\AA, are less than $L$ for the range of $\theta$ angles studied (e.g., $L=61.8$~\AA\ for $\theta=2.28\degree$). Thus it is sufficient to consider only the first periodic images along the $x$ and $y$ directions, for a total of 8 periodic images.

A detailed procedure for computing the dynamical matrix is given in Algorithm~\ref{alg:calcdyn}.\footnote{A Matlab code for Algorithm~\ref{alg:calcdyn} is provided in \cite{SM}.} $\mathbf{D}(\mathbf{0})$ for the relaxed TBG is computed using the \texttt{dynamical\_matrix} from option of the PHONON package in LAMMPS \cite{kong2011}. Use of this algorithm significantly reduces the cost for computing the dynamical matrix since only a single computation of $\bD(\mathbf{0})$ is required to obtain $\bD(\mathbf{k})$ at any $\mathbf{k}$ values. For example, computation of the dynamical matrix for a TBG at $\theta=1.12\degree$ ($N_{\rm TBG}= 10,444$) takes 9131 seconds on 10 AMD Opteron 6344 2.6 GHz nodes with 24 cores per node (total of 240 cores). Full calculation without taking advantage of \eqns{eqn:dyn_matrix2} and \eqnx{eqn:dyn_matrix_k_0_2} would take 300 times longer, $300\times 9131= 2.74\times 10^6$~seconds.

\begin{figure}[!htbp]\centering
         \includegraphics[trim = 0mm 0mm 0mm 0mm,clip,width=0.45\textwidth]{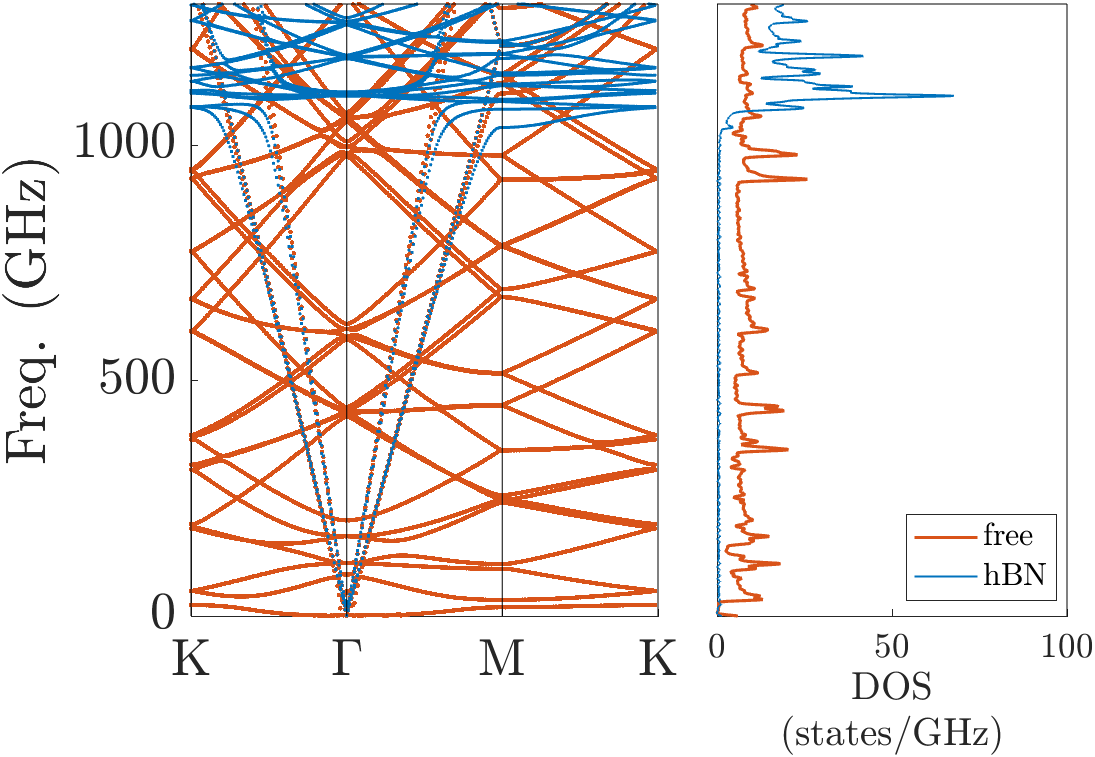}
\captionsetup{justification=raggedright, singlelinecheck=false} 
\caption{Phonon band structure of a TBG with $\theta = 2.28\degree$. The phonon frequency versus wave vectors and density of states (DOS) are shown for relaxed free-standing (orange) and hBN-supported (blue) TBGs.}
\label{fig:phonon_p1q29}
\end{figure}

The eigen decomposition of the dynamical matrix $\bD(\mathbf{k})$ yields the eigenvectors $\mathbf{v}_n$ and eigenvalues $W_n$ for phonon modes $n=1,\dots,3\Ntbg$.\footnote{The eigenvalues and eigenvectors are computed for $\mathbf{D}$ reshaped as a $3\Ntbg \times 3\Ntbg$ matrix.} The phonon frequencies follow from the eigenvalues as $f_n^{\rm TBG}=\sqrt{W_n/2\pi}$. The phonon band structure ($f_n^{\rm TBG}$ for a range of $\mathbf{k}$ values) and associated density of states (DOS) are presented in \fig{fig:phonon_p1q29} for a free-standing TBG and a TBG supported on an hBN substrate computed for the relaxed (minimum energy) structures for $\theta = 2.28 $\degree \;($\Nlay = 1,262$). All calculations in this section are performed using IP$_2$ (hNN) because it is in better agreement with experiments than IP$_1$ (AIREBO+DRIP) as discussed in \sect{sec:relaxation}.

Interestingly, the results in \fig{fig:phonon_p1q29} show that most of the phonons in the hbN-supported TBG have frequencies close to 1100~GHz, whereas those of the free-standing TBG exhibit a wide range of frequencies. This difference is attributed to the interaction between the TBG and the hBN substrate, which has a strong effect on the out-of-plane components of the dynamical matrix.

\subsection{Analysis of TBG phonon band structure using elastic plate basis}
\subsubsection{Method}\label{sec:phonon_basis_method}
To gain understanding into the TBG phonon band structure, each phonon mode eigenvector $\mathbf{v}_n(\mathbf{k})$ ($n=1,\dots,3\Ntbg$) at $\Gamma$ can be represented using the elastic plate basis derived in \sect{sec:normal_mode_basis} \footnote{Phonons at $\Gamma$ are key for understanding into dynamics of TBG since they appear in every TBG supercell regardless of the boundary condition of a TBG sample}. The first $3\Nlay$ components of $\mathbf{v}_n(\mathbf{k})$ are associated with the atoms in the top layer, and next $3\Nlay$ components with the bottom layer, so that
\begin{equation}\label{eqn:v_bot_v_top}
\mathbf{v}_n(\mathbf{k})=[\mathbf{v}^{\rm top}, \mathbf{v}^{\rm bot}],
\end{equation}
where for simplicity we have not included the dependence on $\mathbf{k}$ and $n$ in the terms on the right-hand side, but it is implied. We further denote by $\mathbf{u}_{i}^{\rm top}$, $\mathbf{w}_{i}^{\rm top}$ ($i=1,\dots,\Nlay$), and $\mathbf{u}_{j}^{\rm bot}$, $\mathbf{w}_{j}^{\rm bot}$ ($j=1,\dots,\Nlay$), the the in-plane and out-of-plane components of the atoms in $\mathbf{v}^{\rm top}$ and $\mathbf{v}^{\rm bot}$, respectively. Since the eigenvectors $\mathbf{v}_n(\mathbf{k})$ are normalized (i.e., $\vert \mathbf{v}_n(\mathbf{k})\vert^2=1$), we have that
\begin{multline}\label{eqn:ortho_u_w}
    \vert\mathbf{v}^{\rm top}\vert^2 + \vert\mathbf{v}^{\rm bot}\vert^2 = \\ 
    \sum_{i=1}^{\Nlay}(\vert\mathbf{u}_i^{\rm top}\vert^2
        + \vert\mathbf{w}_i^{\rm top}\vert^2) +
       \sum_{j=1}^{\Nlay}(
    \vert\mathbf{u}_{j}^{\rm bot}\vert^2
    + \vert\mathbf{w}_{j}^{\rm bot}\vert^2)
= 1.
\end{multline}
Next we approximate $\mathbf{u}_{i}^{\rm top}+\mathbf{w}_{i}^{\rm top}$ and $\mathbf{u}_{j}^{\rm bot}+\mathbf{w}_{j}^{\rm bot}$ using the basis expansion in \eqn{eqn:disp_bases},
\begin{align}
    \mathbf{u}_i^{\rm top} &\approx \sum_{k=1}^{12} A_k^{\rm top} \bth{k}^{\rm in}(\br_0^i), \quad 
    \mathbf{w}_{i}^{\rm top} \approx \sum_{l=1}^{15} B_l^{\rm top}\bth{l}^{\rm out}(\br_0^i), \nonumber \\
    \mathbf{u}_j^{\rm bot} &\approx  \sum_{k=1}^{12} A_k^{\rm bot} \bth{k}^{\rm in}(\br_0^j), \quad
    \mathbf{w}_{j}^{\rm bot} \approx \sum_{l=1}^{15} B_l^{\rm bot}\bth{l}^{\rm out}(\br_0^j),
\label{eqn:AB_top_bottom}
\end{align}
where $\br_0^i$ and $\br_0^j$ are reference positions of atoms $i$ and $j$ in the top and bottom layers, and the coefficients $A_k^{\rm top}$, $B_l^{\rm top}$, $A_k^{\rm bot}$, $B_l^{\rm bot}$ are computed as explained in \sect{sec:relaxation:general}.
Substituting the expansions in \eqns{eqn:AB_top_bottom} into \eqn{eqn:ortho_u_w} gives
\begin{align}\label{eqn:mag_A_B}
& \vert\mathbf{v}^{\rm top}\vert^2 \approx 
\vert\tilde{\mathbf{v}}^{\rm top}\vert^2 \equiv  \nonumber \\
        & \sum_{i=1}^{\Nlay}\biggl[\sum_{k=1}^{12}(A_k^{\rm top}\bth{k}^{\rm in}(\br_0^i))^2 + \sum_{l=1}^{15}(B_l^{\rm top}\bth{k}^{\rm out}(\br_0^i))^2\biggr] \nonumber 
\end{align}
\begin{align}
& \vert\mathbf{v}^{\rm bot}\vert^2 \approx  
\vert\tilde{\mathbf{v}}^{\rm bot}\vert^2 \equiv \nonumber \\ 
      &\sum_{j=1}^{\Nlay}\biggl[\sum_{k=1}^{12}(A_k^{\rm bot}\bth{k}^{\rm in}(\br_0^j))^2 + \sum_{l=1}^{15}(B_l^{\rm bot}\bth{l}^{\rm out}(\br_0^j))^2\biggr],
\end{align}
where now
\begin{equation} \label{eqn:ortho_approx}
\vert\tilde{\mathbf{v}}^{\rm top}\vert^2+\vert\tilde{\mathbf{v}}^{\rm bot}\vert^2=1-\epsilon,
\end{equation}
where $\epsilon$ is a small positive number that is present because the expansion is a projection onto an incomplete basis. Next, using
\begin{multline}\label{eqn:sum_Nlay/2}
    \sum_{i=1}^{\Nlay} (\bth{k}^{\rm in}(\br_0^i))^2 =  \sum_{i=1}^{\Nlay} (\bth{k}^{\rm out}(\br_0^i))^2 = \\ 
	\sum_{j=1}^{\Nlay} (\bth{k}^{\rm in}(\br_0^j))^2 = \sum_{j=1}^{\Nlay} (\bth{k}^{\rm out}(\br_0^j))^2 =
	 \frac{\Nlay}{2},
\end{multline}
from \eqns{eqn:A_k_s} and \eqnx{eqn:B_l_c} (by assuming that $\mathbf{\Delta_{\rm in}}(r_0^j) = \bth{k^*}^{\rm in}(r_0^j)=\sin{(\bar{\bG}_{k^*}\cdot\br_0^j)}$ and $\mathbf{\Delta_{\rm out}}(r_0^j) = \bth{l^*}^{\rm out}(r_0^j) = \cos{(\hat{\bG}_{l^*}\cdot\br_0^j)}$), \eqn{eqn:mag_A_B} takes the form
\begin{align}\label{eqn:mag_A_B_2}
    & \vert\tilde{\mathbf{v}}^{\rm top}\vert^2 =
    \frac{\Nlay}{2}\biggl[\sum_{k=1}^{12} (A_k^{\rm top})^2 +
                          \sum_{l=1}^{15} (B_l^{\rm top})^2\biggr] \nonumber \\ 
    & \vert\tilde{\mathbf{v}}^{\rm bot}\vert^2 =
    \frac{\Nlay}{2}\biggl[\sum_{k=1}^{12} (A_k^{\rm bot})^2 +
                          \sum_{l=1}^{15} (B_l^{\rm bot})^2\biggr].
\end{align}
Introducing the notation,
\begin{equation}
\bar{A}_k\equiv\frac{\Nlay}{2}(A_k)^2, \quad
\bar{B}_l\equiv\frac{\Nlay}{2}(B_l)^2,
\end{equation}
\eqn{eqn:mag_A_B_2} becomes
\begin{align}\label{eqn:mag_A_B_3}
    \vert\tilde{\mathbf{v}}^{\rm top}\vert^2 &=
    \sum_{k=1}^{12} \bar{A}_k^{\rm top} +
    \sum_{l=1}^{15} \bar{B}_l^{\rm top} \nonumber \\ 
    \vert\tilde{\mathbf{v}}^{\rm bot}\vert^2 &=
    \sum_{k=1}^{12} \bar{A}_k^{\rm bot} +
    \sum_{l=1}^{15} \bar{B}_l^{\rm bot}.
\end{align}

\subsubsection{Numerical results}
In \eqn{eqn:mag_A_B_3}, the relative sizes of $\vert\tilde{\mathbf{v}}^{\rm top}\vert^2$ and $\vert\tilde{\mathbf{v}}^{\rm bot}\vert^2$ indicate the contributions of the top and bottom layers to a given eigenvector, and for the top and bottom layers, the sums $ \sum_{k=1}^{12} \bar{A}_k$ and $\sum_{l=1}^{15} \bar{B}_l$ indicate the relative contributions of the in-plane and out-of-plane modes, respectively. In this section we report the results for a free-standing TBG and a TBG supported on an hBN substrate for twist angles of $1.12\degree$ and  $2.28\degree$ evaluated at the $\Gamma$ point (i.e., $\mathbf{k}=\mathbf{0}$). First an atomistic simulation is performed to minimize the energy of the structure and obtain the relaxed equilibrium configuration. Next, the dynamical matrix is computed using Algorithm~\ref{alg:calcdyn}. Finally, a Matlab script (provided with this article) is used to perform the eigendecomposion and projection.

\begin{figure*}[t]\centering
    \begin{subfigure}[b]{0.49\textwidth}
         \centering
         \includegraphics[trim = 0mm 0mm 0mm 0mm,clip=true,width=\textwidth]{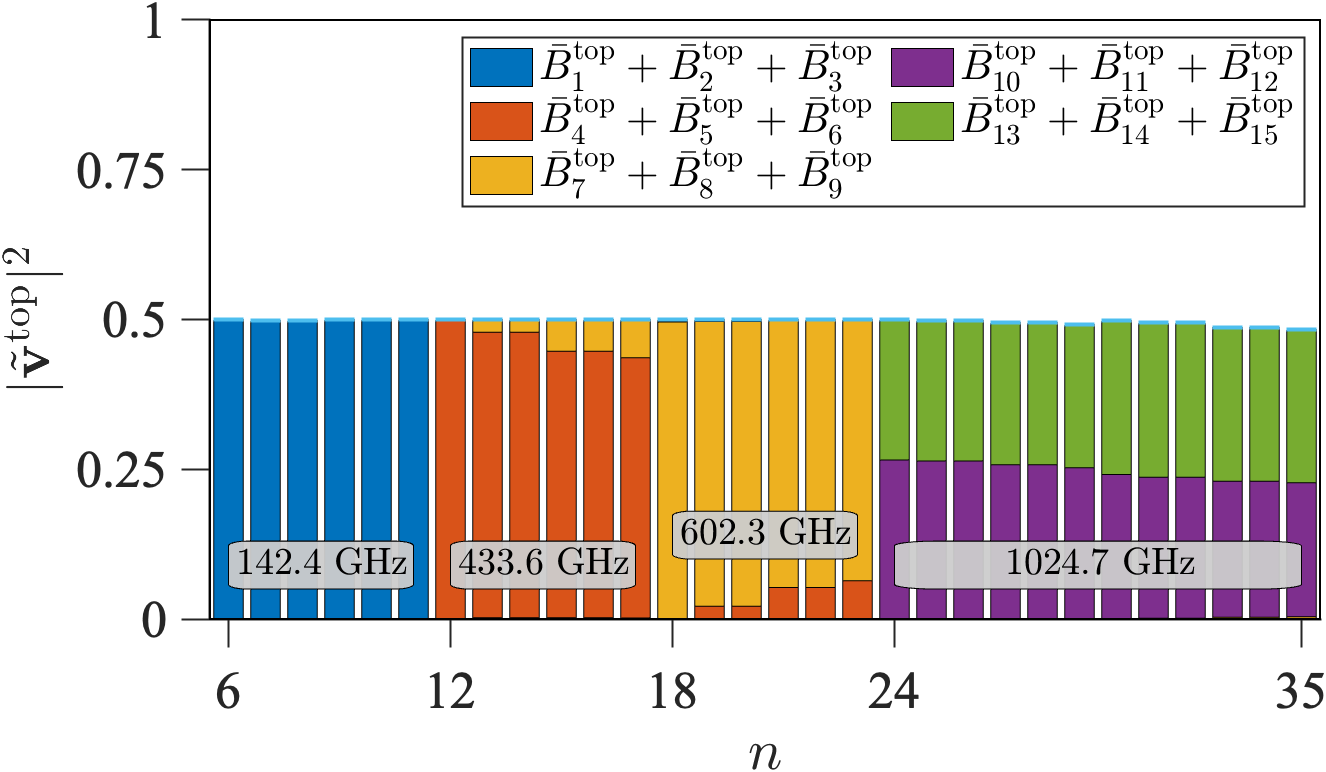}
         \caption{Modes 6 to 35 for top layer}
         \label{fig:phonon_analy_p1q29_no_sub:1}
    \end{subfigure}
    \begin{subfigure}[b]{0.49\textwidth}
         \centering
         \includegraphics[trim = 0mm 0mm 00mm 0mm,clip=true,width=\textwidth]{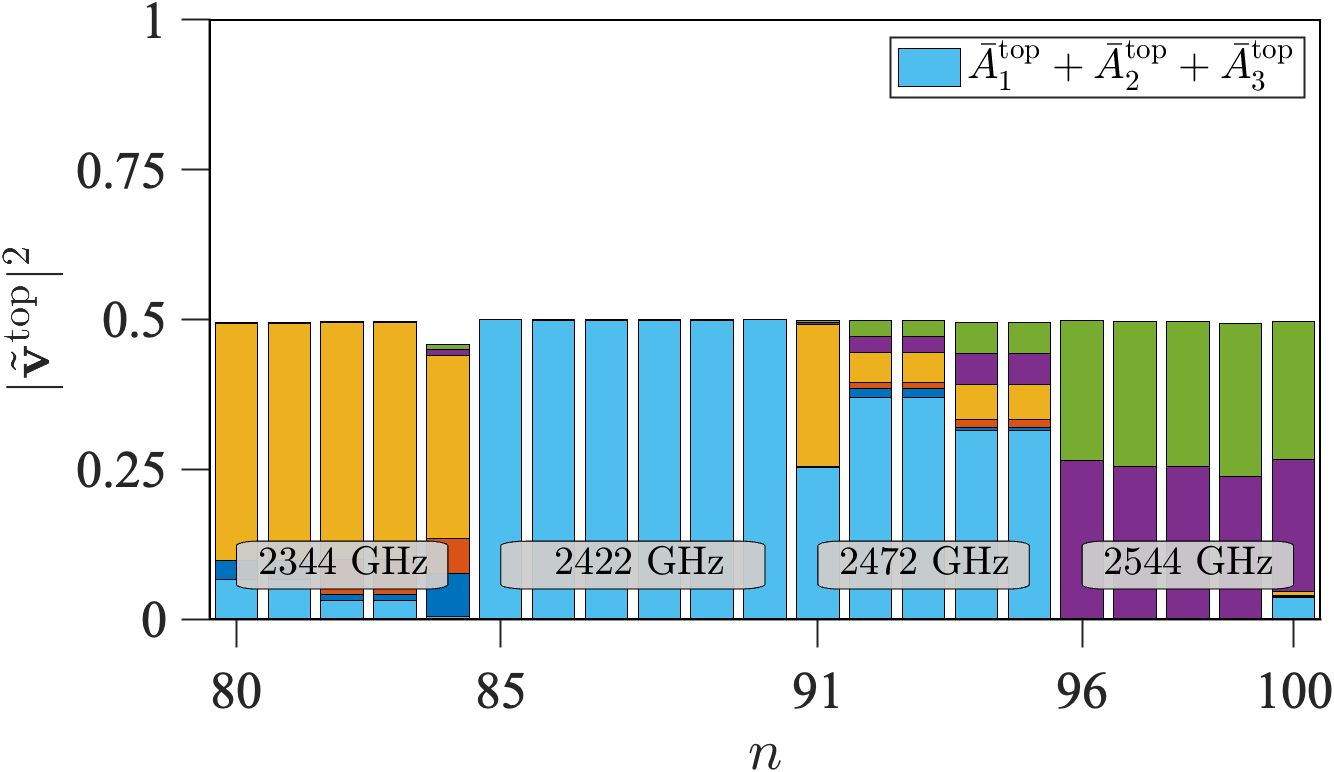}
         \caption{Modes 80 to 100 for top layer}
         \label{fig:phonon_analy_p1q29_no_sub:2}
     \end{subfigure}
\\[10pt]
     \begin{subfigure}[b]{0.49\textwidth}
         \centering
         \includegraphics[trim = 0mm 0mm 00mm 0mm,clip=true,width=\textwidth]{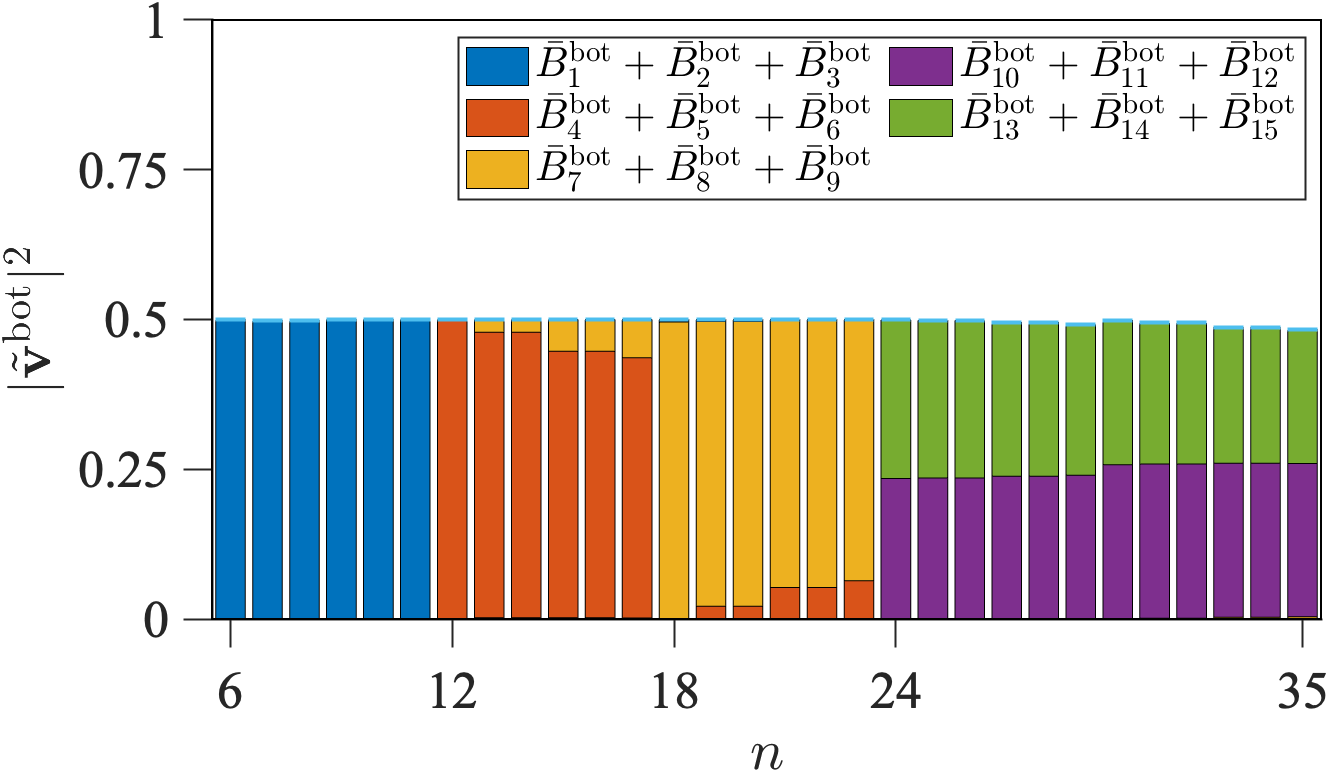}
         \caption{Modes 6 to 35 for bottom layer}
         \label{fig:phonon_analy_p1q29_no_sub:3}
     \end{subfigure}
     \begin{subfigure}[b]{0.49\textwidth}
         \centering
         \includegraphics[trim = 0mm 0mm 00mm 0mm,clip=true,width=\textwidth]{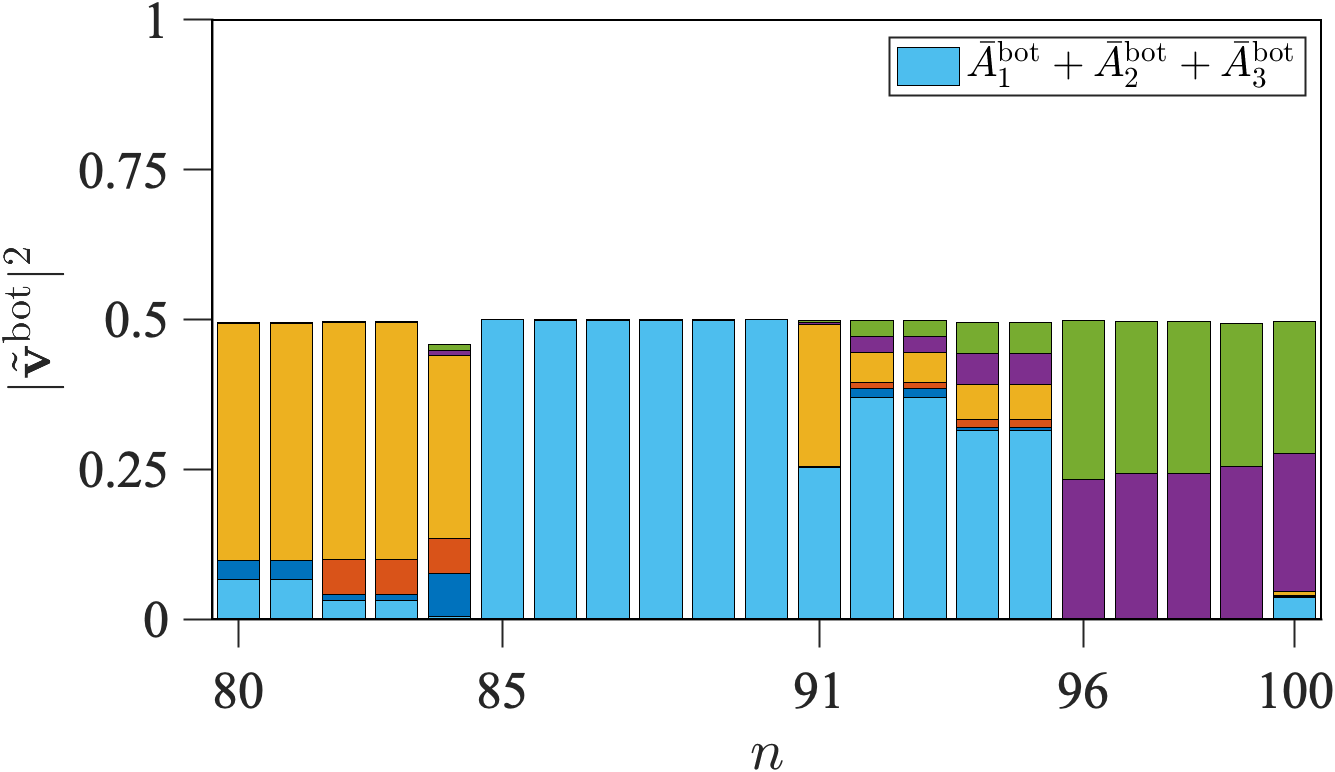}
         \caption{Modes 80 to 100 for bottom layer}
         \label{fig:phonon_analy_p1q29_no_sub:4}
     \end{subfigure}
\captionsetup{justification=raggedright, singlelinecheck=false} 
\caption{Contributions to the squared magnitude of $\mathbf{v}_n(\mathbf{0})$ for different mode numbers $n$ from the top and bottom graphene layers of a relaxed free-standing TBG at $\theta=2.28\degree$ computed from atomistic simulation results. At each mode number, the contributions from different out-of-plane and in-plane basis terms are shown as colored bars.  Terms with the same frequencies are summed together as indicated in the legend boxes.}
\label{fig:phonon_analy_p1q29_no_sub}
\end{figure*}

We begin by considering a free-standing TBG with a twist angle of $\theta=2.28\degree$. The first three phonon modes ($n=1,2,3$) associated with translations of the TBG along the two in-plane and one out-of-plane directions have zero frequency and coefficients $\bar{A_k}=0$ and $\bar{B_l}=0$. The next two phonon modes ($n=4,5$) are shearing modes in which the top and bottom layers translate in opposite in-plane directions. These shearing modes are not captured by the in-plane or out-of-plane bases (i.e., $\bar{A_k}\approx0$ and $\bar{B_l}\approx0$). The computed frequency of these modes is 22.63 GHz. Results for modes larger than 5 are plotted in \fig{fig:phonon_analy_p1q29_no_sub}. Amplitudes of basis functions that have the same frequency are summed together (see \tabs{tab:in_plane_mode} and \tabx{tab:out_of_plane_mode}), and only terms that have a sizeable contribution to the overall amplitude are shown. The range of phonon modes in the figure, $6,\ldots,35$ and $80,\ldots,100$, are selected to focus on low-frequency modes and the first phonon modes dominated by in-plane deformation (i.e., $\sum_{l=1}^{15}\bar{B}_l \ll \sum_{k=1}^{12}\bar{A}_k$), respectively.

In the low frequency range (modes $6,\ldots,35$), out-of-plane deformation is dominant (i.e., $\sum_{k=1}^{12}\bar{A}_k \ll \sum_{l=1}^{15}\bar{B}_l$) in both the top and bottom layers as shown in \figs{fig:phonon_analy_p1q29_no_sub:1} and \figx{fig:phonon_analy_p1q29_no_sub:3}. The top and bottom layers deform in the same direction and each contribute about 0.5 of the squared amplitude of $\mathbf{v}_n(\mathbf{0})$ (with the small difference captured by higher-order basis terms). This indicates that the dynamics of each layer can be represented by the out-of-plane normal mode of an elastic plate.
The frequencies of phonon mode sets $(6,\ldots,11)$, $(12,\ldots,17)$, $(18,\ldots,23)$, and $(24,\ldots,35)$ are computed as 120--203, 423--443, 587--620, and 976--1068 GHz, which is comparable to out-of-plane normal mode frequencies of the elastic plate listed in \tab{tab:out_of_plane_mode}, $f_{1,2,3}^{\rm out}$, $f_{4,5,6}^{\rm out}$, $f_{7,8,9}^{\rm out}$, and $f_{10,11,12}^{\rm out}$ (144, 432, 576, and 1009 GHz). The difference between the frequencies obtained from the phonon calculation and the out-of-plane normal mode frequencies from elastic plate theory is due to interlayer interactions and the deformation of the graphene layers in the relaxed free-standing TBG, which are not accounted for in the elastic plate theory.

The first phonon mode in which in-plane deformation begins to be noticeable is 80. Results for modes $80,\ldots,100$ are shown in \figs{fig:phonon_analy_p1q29_no_sub:2} and \figx{fig:phonon_analy_p1q29_no_sub:4}. In particular, modes 85 to 90 are almost entirely captured by $\bar{A}_1+\bar{A}_2+\bar{A}_3$ in the top and bottom layers, which contribute equally to the squared magnitude of $\mathbf{v}_n(\mathbf{0})$. The frequency range of these phonon modes (2421--2423 GHz) is comparable to the elastic normal mode frequency $f_{1,2,3}^{\rm in}$ (2436.77 GHz) in \tab{tab:in_plane_mode}. The other modes above 80 in the ranges $(80,\ldots,84)$ and $(91,\ldots,95)$ have both in-plane and out-of-plane contributions.
These modes appear since the interlayer interaction changes both in-plane and out-of-plane components of the dynamical matrix. The deformation is captured by the enumeraed elastic bases (i.e., $\sum_{k=1}^{12}(\bar{A}_k^{\rm top}+\bar{A}_k^{\rm bot}) + \sum_{l=1}^{15}(\bar{B}_l^{\rm top}+\bar{B}_l^{\rm bot}) \approx 1 $), however, their frequencies are not predicted by either the in-plane or out-of-plane normal modes of the elastic plate model due to mixed in-plane and out-of-plane deformation within a single phonon mode. Although the frequencies of these modes are not captured, we note that the deformation and frequency of phonon modes (6,$\ldots$,35) and (85,$\ldots$,90), which have a high deformation at a low frequency for in-plane and out-of-plane deformations, are successfully predicted by in-plane and out-of-plane normal modes of a simple elastic plate model.

Next, we conisder a TBG supported on an hBN substrate at $\theta = 2.28\degree$. The first two phonon modes correspond to translations of the TBG along in-plane directions. Unlike the case of a free-standing TBG, there is no phonon mode for a rigid translation of the TBG along the out-of-plane direction due to the TBG--hBN substrate interactions.
The following two modes ($n=3,4$) are shearing modes that have a frequency of 24.54~GHz, which is close to the value obtained for the free-standing TBG (22.63~GHz). The shearing mode frequencies are similar because the in-plane components of the dynamical matrix of a TBG are not directly influenced by the presence of a substrate. The substrate affects this frequency indirectly through changes to layer deformation and spacing following energy minimization. Phonon mode 5 is a translation of graphene layers along the out-of-plane direction with different magnitudes ($\sum_{i=1}^{\Nlay}(\vert\mathbf{u}_i^{\rm top}\vert^2+ \vert\mathbf{w}_i^{\rm top}\vert^2)=0.772$ and $\sum_{j=1}^{\Nlay}(\vert\mathbf{u}_j^{\rm bot}\vert^2+\vert\mathbf{w}_j^{\rm bot}\vert^2)=0.23$) that has a frequency of 1104 GHz, leading $\bar{A}_k=0$ and $\bar{B}_l=0$.

\begin{figure*}[t]\centering
    \begin{subfigure}[b]{0.49\textwidth}
         \centering
         \includegraphics[trim = 0mm 0mm 0mm 0mm,clip=true,width=\textwidth]{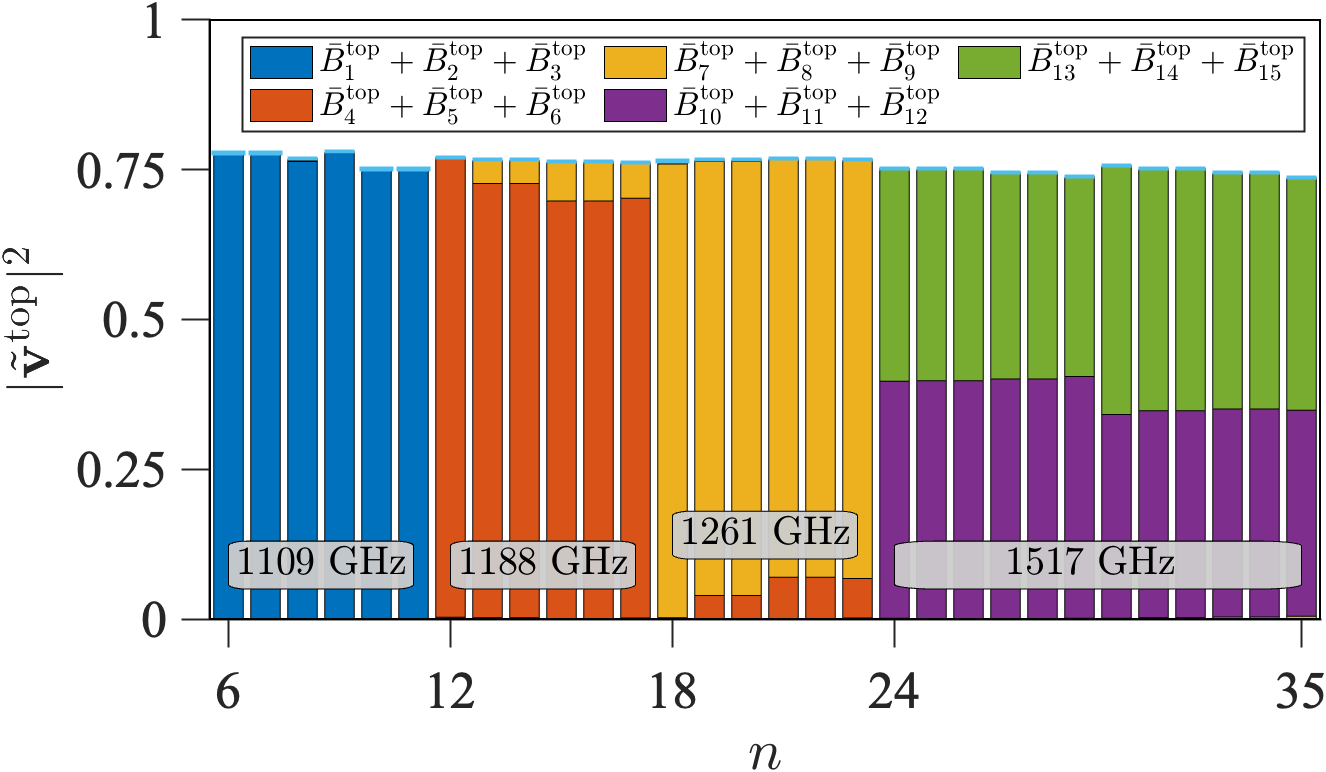}
         \caption{Modes 6 to 35 for top layer}
         \label{fig:phonon_analy_p1q29_hBN:1}
    \end{subfigure}
    \begin{subfigure}[b]{0.49\textwidth}
         \centering
         \includegraphics[trim = 0mm 0mm 00mm 0mm,clip=true,width=\textwidth]{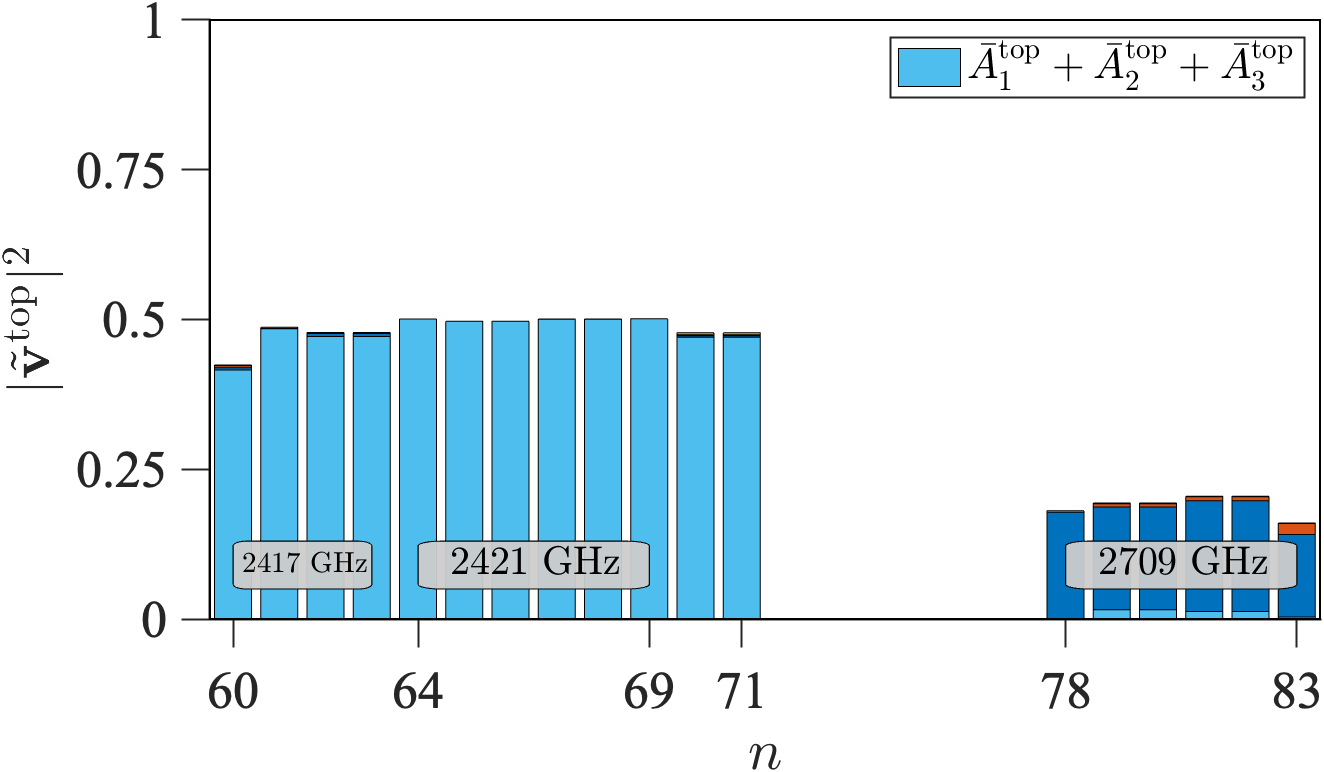}
         \caption{Modes 60 to 83 for top layer}
         \label{fig:phonon_analy_p1q29_hBN:2}
     \end{subfigure}
\\[10pt]
     \begin{subfigure}[b]{0.49\textwidth}
         \centering
         \includegraphics[trim = 0mm 0mm 00mm 0mm,clip=true,width=\textwidth]{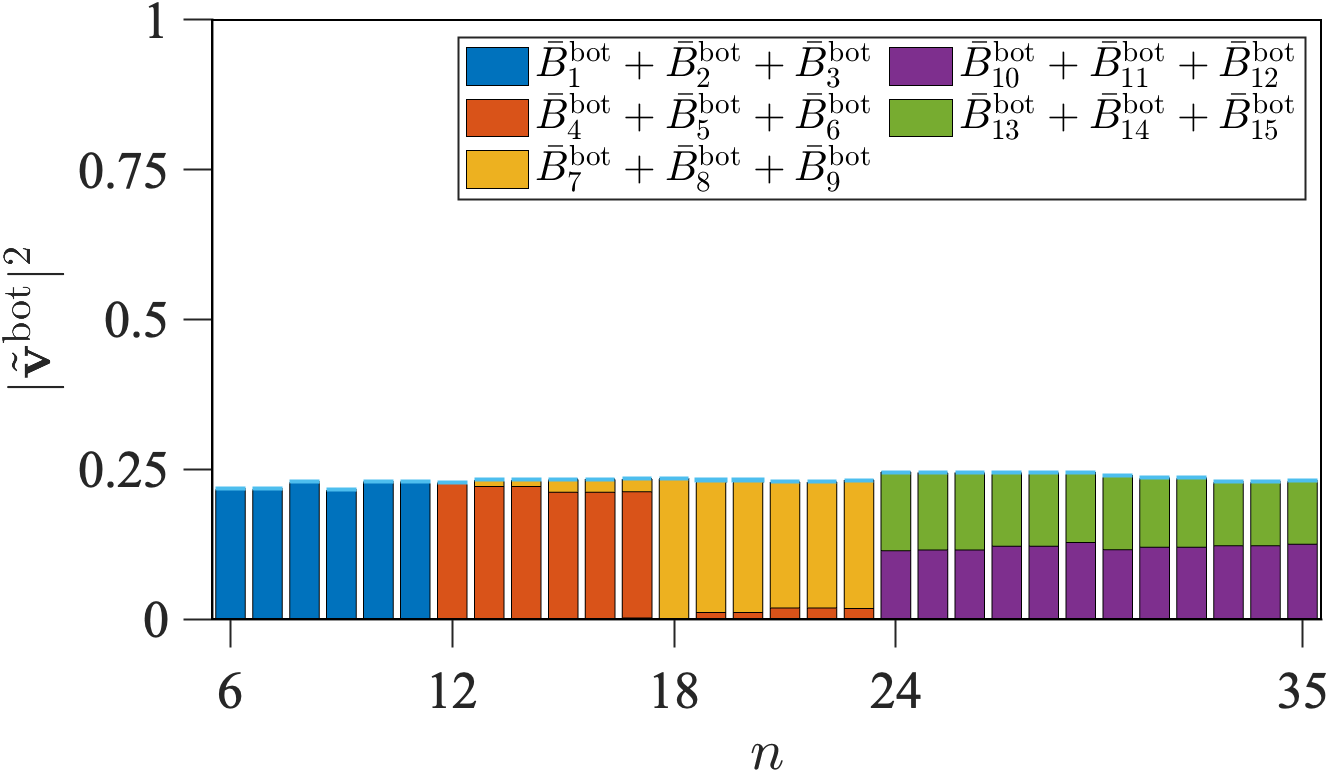}
         \caption{Modes 6 to 35 for bottom layer}
         \label{fig:phonon_analy_p1q29_hBN:3}
     \end{subfigure}
     \begin{subfigure}[b]{0.49\textwidth}
         \centering
         \includegraphics[trim = 0mm 0mm 00mm 0mm,clip=true,width=\textwidth]{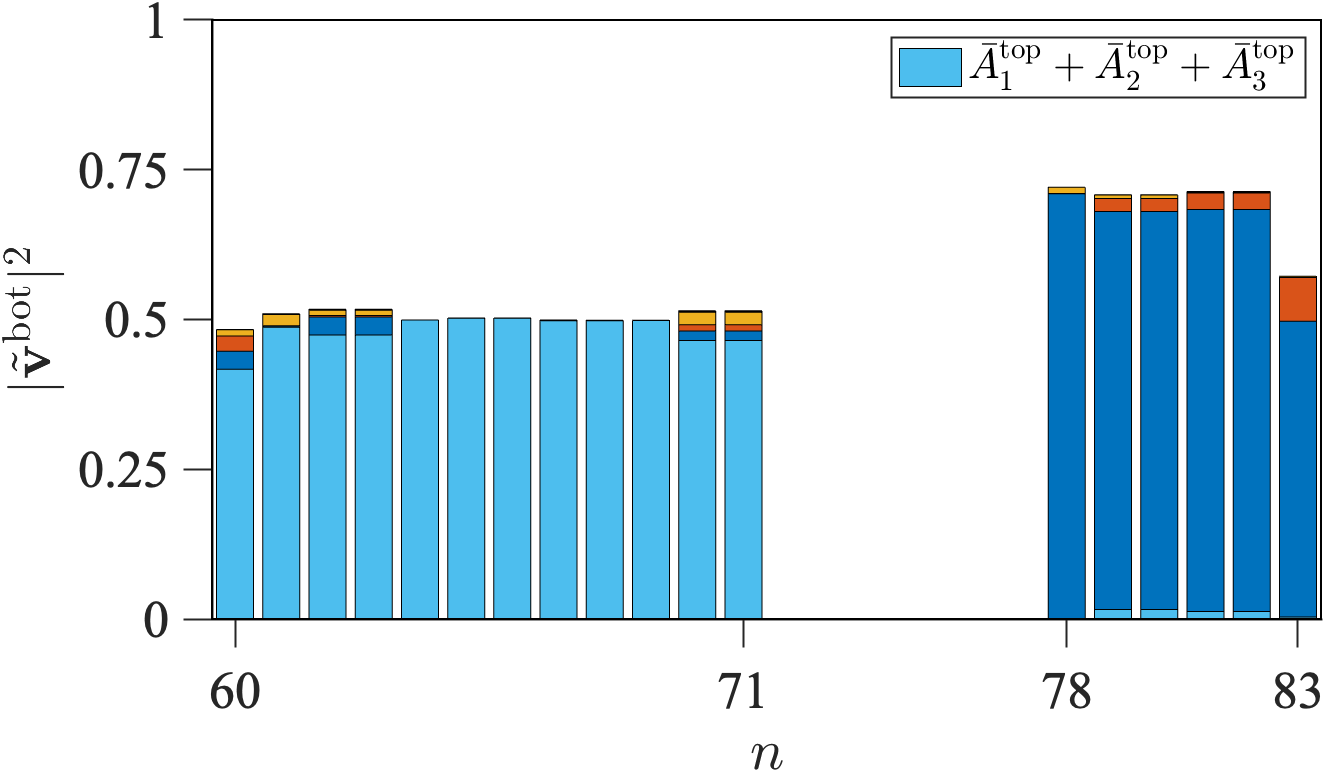}
         \caption{Modes 60 to 83 for bottom layer}
         \label{fig:phonon_analy_p1q29_hBN:4}
     \end{subfigure}
\captionsetup{justification=raggedright, singlelinecheck=false}      
\caption{Contributions to the squared magnitude of $\mathbf{v}_n(\mathbf{0})$ for different mode numbers $n$ from the top and bottom graphene layers of a relaxed TBG supported on an hBN substrate at $\theta=2.28\degree$ computed from atomistic simulation results. At each mode number, the contributions from different out-of-plane and in-plane basis terms are shown as colored bars.  Terms with the same frequencies are summed together as indicated in the legend boxes.}
\label{fig:phonon_analy_p1q29_hBN}
\end{figure*}

The out-of-plane ($\bar{B}_l$) and in-plane ($\bar{A}_k$) contributions to phonon modes (6,$\ldots$,35) and (60,$\ldots$,83) are plotted in \fig{fig:phonon_analy_p1q29_hBN}. As shown in \figs{fig:phonon_analy_p1q29_hBN:1} and \figx{fig:phonon_analy_p1q29_hBN:3}, out-of-plane deformation is dominant for phonon modes (6,$\ldots$,35). This is similar to the results for the free-standing TBG in \fig{fig:phonon_analy_p1q29_no_sub}, however whereas the top and bottom layers contributed equally to $|\mathbf{v}_n|^2$ for the free-standing TBG, for the supported TBG, the top squared magnitude is $\sim 0.77$ and the bottom squared magnitude is $\sim 0.23$. These values and associated frequencies are not predicted by the
the elastic plate model discussed in \sects{sec:in_plane_basis} and \sectx{sec:out_of_plane_basis}. To capture substrate effects, we construct a double elastic plate model in which interlayer and TBG--substrate interactions are represented by uniformly distributed harmonic springs. The normal modes of the double plate model are derived in \cite{SM}.

\begin{table*}[t]\centering
\begin{tabular}{c|c|c|c|c|c}
\hline
\multicolumn{2}{c|}{ $f^{\rm dbl}_{k}$ (GHz)} &
\multicolumn{2}{c|}{ $f^{\rm TBG}_{n}$ (GHz)} &
$\mathbf{g}_k^{\rm dbl}$ & $\tilde{\bG}_k^{\rm dbl}$  \\
\hline
\multirow{1}{*}{$f^{\rm dbl}_{0}$} & 1073.7 &
\multirow{1}{*}{$f^{\rm TBG}_{5}$} & 1109 &
\multirow{15}{5.5em}{[0.885,0.458]} & $\bG(0,0)$  \\
\cline{1-4} \cline{6-6}
\multirow{3}{*}{$f^{\rm dbl}_{1,2,3}$} & \multirow{3}{*}{1083.28} &
\multirow{3}{*}{$f^{\rm TBG}_{6,7,...,11}$} & \multirow{3}{*}{1105.96--1113.34} &
& $\bG(0,1)$  \\
\cline{6-6}
& & & & & $\bG(-1,0)$ \\
\cline{6-6}
& & & & & $\bG(1,-1)$ \\
\cline{1-4} \cline{6-6}
\multirow{3}{*}{$f^{\rm dbl}_{4,5,6}$} & \multirow{3}{*}{1157.08} &
\multirow{3}{*}{$f^{\rm TBG}_{12,13,...,17}$} & \multirow{3}{*}{1183.53--1191.6} &
& $\bG(-2,1)$  \\
\cline{6-6}
& & & & & $\bG(1,-2)$ \\
\cline{6-6}
& & & & & $\bG(1,1)$ \\
\cline{1-4} \cline{6-6}
\multirow{3}{*}{$f^{\rm dbl}_{7,8,9}$} & \multirow{3}{*}{1217.99} &
\multirow{3}{*}{$f^{\rm TBG}_{18,19,...,23}$} & \multirow{3}{*}{1255.28--1271.3} &
& $\bG(0,2)$  \\
\cline{6-6}
& & & & & $\bG(-2,0)$ \\
\cline{6-6}
& & & & & $\bG(2,-2)$ \\

\cline{1-4} \cline{6-6}
\multirow{3}{*}{$f^{\rm dbl}_{10,11,12}$} & \multirow{3}{*}{1425.6} &
\multirow{3}{*}{$f^{\rm TBG}_{24,19,...,35}$} & \multirow{3}{*}{1485.44--1544.38} &
& $\bG(1,2)$  \\
\cline{6-6}
& & & & & $\bG(-2,-3)$ \\
\cline{6-6}
& & & & & $\bG(3,-1)$ \\
\cline{1-4} \cline{6-6}
\multirow{3}{*}{$f^{\rm dbl}_{13,14,15}$} & \multirow{3}{*}{1425.6} &
\multirow{3}{*}{$f^{\rm TBG}_{24,19,...,35}$} & \multirow{3}{*}{1485.44--1544.38} &
& $\bG(1,-3)$  \\
\cline{6-6}
& & & & & $\bG(2,1)$ \\
\cline{6-6}
& & & & & $\bG(3,-2)$ \\
\hline
\end{tabular}
\captionsetup{justification=raggedright, singlelinecheck=false} 
\caption{First 15 elements of $f_{k}^{\rm dbl}$, $\mathbf{g}_{k}^{\rm dbl}$, and  $\bd_{k}^{\rm dbl}$ computed using the double plate model for the relaxed hBN-supported TBG.
The frequency of phonon mode $n$, $f_n^{\rm TBG}$ is included for comparison with $f_{k}^{\rm dbl}$ and arranged based on the similarity between the phonon wave vectors and $\bG$ from the double plate model.}
\label{tab:double_hBN}
\end{table*}

The results for $f_{k}^{\rm dbl}$, $\mathbf{g}_{k}$, and $\tilde{\bG}_k$ for the hBN-supported TBG at $\theta=2.28\degree$ are computed using the double plate model (see~\cite{SM}) and listed in \tab{tab:double_hBN}. We note that the order of $\tilde{\bG}_1$ to $\tilde{\bG}_{15}$ is the same as that of $\hat{\bG}_1$ to $\hat{\bG}_{15}$ as shown in \tabs{tab:out_of_plane_mode} and \tabx{tab:double_hBN}.
The frequency of phonon mode $n$ of the relaxed hBN-supported TBG ($f_n^{\rm hBN}$) is included for comparison with
$f_{k}^{\rm dbl}$. The double plate model successfully predicts $f_n^{\rm hBN}$ for the low-frequency phonon modes (5,$\ldots$,35) that exhibit dominant out-of-plane deformation in the TBG dynamics.
The magnitudes of the top and bottom graphene layers in terms of normal modes of the double plate model are computed by defining the deformation vectors $\bar{\mathbf{w}}^{\rm top} = [\bar{w}_1(\br_0^1),\bar{w}_1(\br_0^2),\ldots,\bar{w}_1(\br_0^{\Nlay})]$ and $\bar{\mathbf{w}}^{\rm bot} = [\bar{w}_2(\br_0^1),\bar{w}_2(\br_0^2),\ldots,\bar{w}_2(\br_0^{\Nlay})]$. Using \eqn{eqn:sum_Nlay/2}, for each $\tilde{\bG}$, magnitudes of $\bar{\mathbf{w}}^{\rm top}$ and $\bar{\mathbf{w}}^{\rm bot}$ follow as
\begin{align}\label{eqn:v_dbl}
     \vert \bar{\mathbf{w}}^{\rm top}  \vert^2 = \sum_{i=1}^{\Nlay} (\bar{w}_1(\br_0^i))^2  &= \frac{\Nlay}{2}(\eta_{\tilde{\bG}})^2, \\
     \vert \bar{\mathbf{w}}^{\rm bot} \vert^2 = \sum_{j=1}^{\Nlay} (\bar{w}_2(\br_0^j))^2 &= \frac{\Nlay}{2}(\eta_{\tilde{\bG}})^2.
\end{align}
Since $(\eta_{\tilde{\bG}}^1)^2+(\eta_{\tilde{\bG}}^2)^2=1$ (see SM\cite{SM}), the normalized magnitudes of the top and bottom graphene layers in the double plate model are  $(\eta_{\tilde{\bG}}^1)^2$ and $(\eta_{\tilde{\bG}}^2)^2$. We note that the computed  $[\eta_{\tilde{\bG}}^1,\eta_{\tilde{\bG}}^2]$ is [0.885,0.458] for all $\tilde{\bG}$'s listed in \tab{tab:double_hBN}.
For phonon modes $(5,\ldots,35)$ of the hBN-supported TBG in which out-of-plane deformation is dominant (\figs{fig:phonon_analy_p1q29_hBN:1} and \figx{fig:phonon_analy_p1q29_hBN:3}), the magnitudes ($[(\eta_\bG^1)^2,(\eta_\bG^2)^2]=[0.783,0.21]$) from the double plate model provide good agreement with the magnitudes ($[\sum_{k=1}^{15} \bar{B}_l^{\rm top},\sum_{k=1}^{15}\bar{B}_l^{\rm bot}]=[0.772,0.23]$) from the phonon modes. Therefore, the double plate model provides a reasonable representation for the frequency and relative deformation for out-of-plane dominant phonon modes of an hBN-supported TBG.

The first phonon modes that have a dominant in-plane deformation for the relaxed hBN-supported TBG are modes 64--69. The frequency range of these modes is 2402--2422 GHz. Both the top and bottom layers have magnitudes of $\bar{A}_k$ equal to 0.5. Similar to phonon modes 85--90 of the relaxed free-standing TBG, these modes are captured by the first in-plane normal modes of the elastic plate ($f_{1,2,3}^{\rm in}$ = 2436.77 GHz). This indicates that the substrate has a minor effect on the frequency and $\mathbf{v}$ of phonon modes in which $\bar{A}_1+\bar{A}_2+\bar{A}_3$ is dominant (for comparison, the averaged frequency of phonon modes 64--69 for the relaxed free-standing TBG is 2421 GHz).
Modes 60--63 and 70--71 also have significant $\bar{A}_k$ and have frequencies of 2417 and 2458 GHz. However, these modes cannot be explained by the in-plane and out-of-plane normal modes of the elastic plate because $\bar{B}_l^{\rm bot}\neq 0$. i.e., the phonon modes have both non-negligible in-plane and out-of-plane deformations.

\begin{figure}[t]\centering
         \includegraphics[trim = 0mm 0mm 0mm 0mm,clip,width=0.45\textwidth]{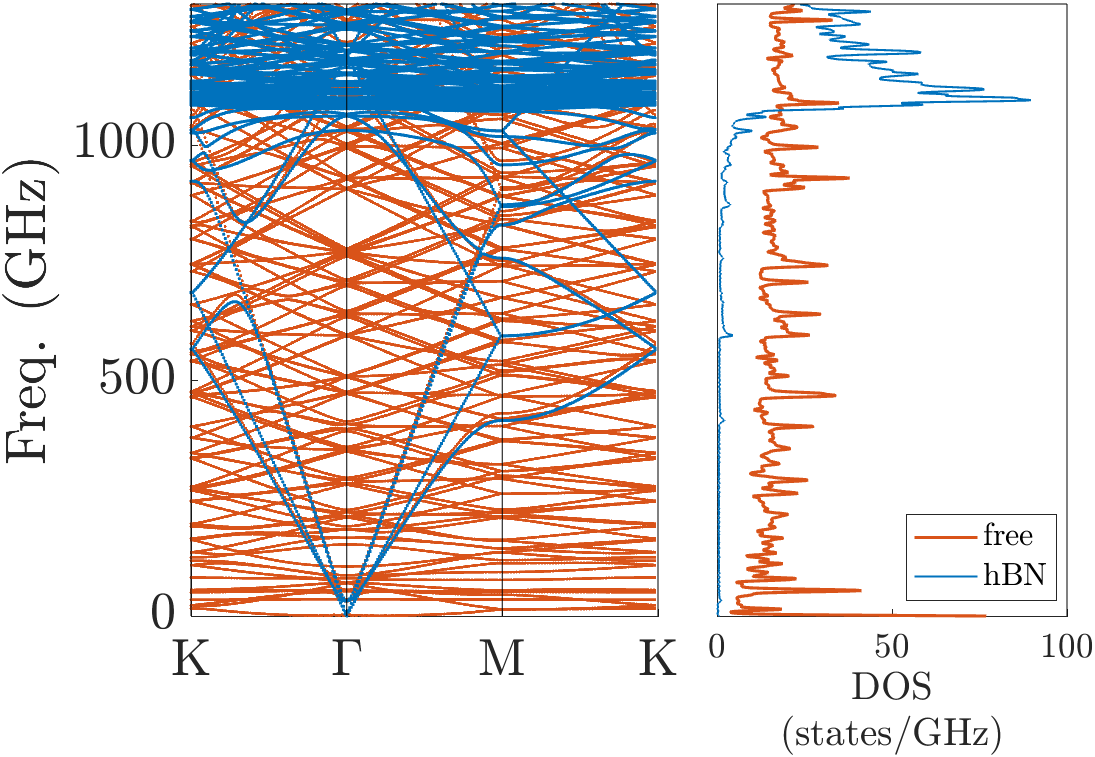}
\captionsetup{justification=raggedright, singlelinecheck=false}          
\caption{Phonon band structure of a TBG with $\theta = 1.12\degree$. The phonon frequency versus wave vectors and density of states (DOS) are shown for relaxed free-standing (orange) and hBN-supported (blue) TBGs.}
\label{fig:phonon_p1q59}
\end{figure}

To study the effect of the TBG supercell size on phonons, we compute the phonon band diagram and DOS for a relaxed free-standing TBG and hBN-supported TBG for a twist angle of $\theta=1.12\degree$ ($\Nlay = 5222$). The results are presented in \fig{fig:phonon_p1q59}. Comparing the DOS in \fig{fig:phonon_p1q59} with that in \fig{fig:phonon_p1q29} for a twist angle of $\theta=2.28\degree$, we see qualitatively similar behavior. In both cases, the free-standing TBG has a uniform distribution across frequencies, and the hBN-supported TBG exhibits a high concentration around 1100~GHz. However as explained below, the twist angle does affect the freqenecies of in-plane and out-of-plane phonons in the low frequency range.

\begin{figure*}[t]\centering
    \begin{subfigure}[b]{0.49\textwidth}
         \centering
         \includegraphics[trim = 0mm 0mm 0mm 0mm,clip=true,width=\textwidth]{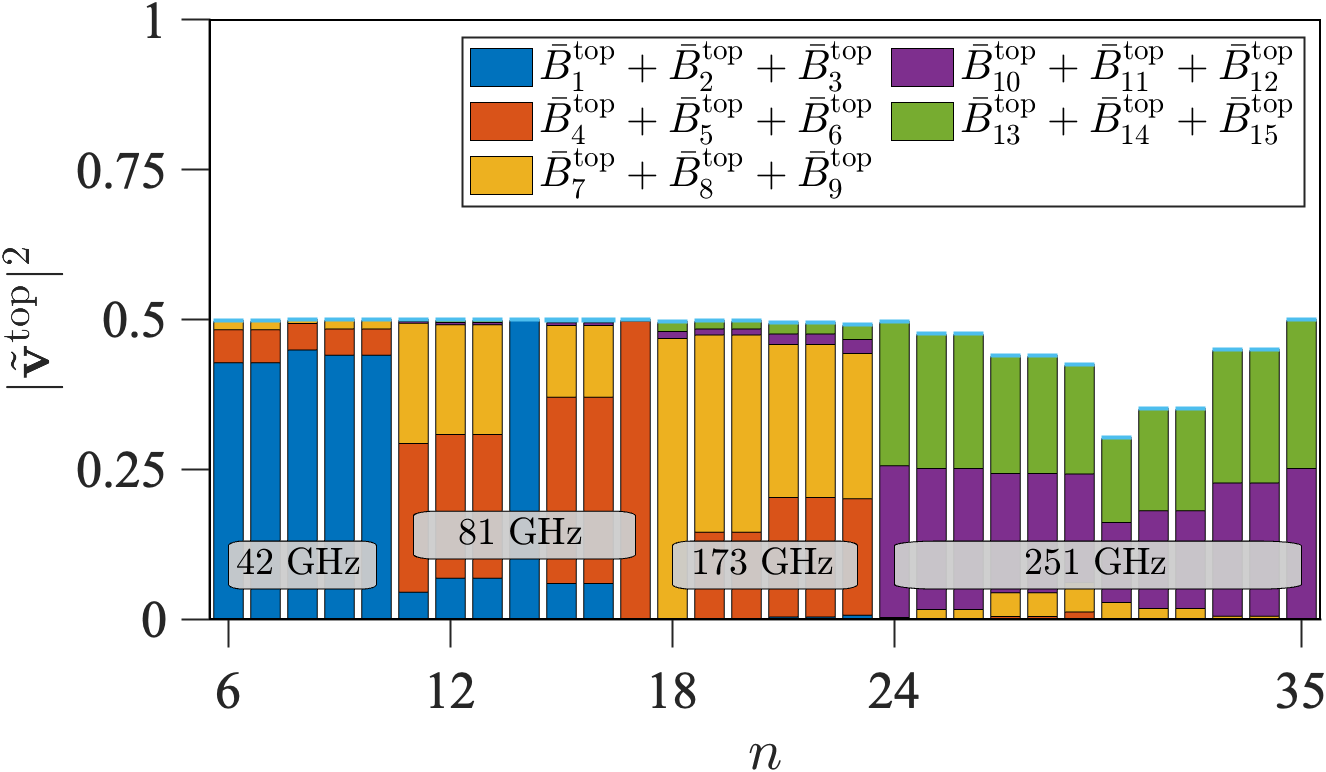}
         \caption{Modes 6 to 35 for top layer}
         \label{fig:phonon_analy_p1q59_no_sub:1}
    \end{subfigure}
    \begin{subfigure}[b]{0.49\textwidth}
         \centering
         \includegraphics[trim = 0mm 0mm 00mm 0mm,clip=true,width=\textwidth]{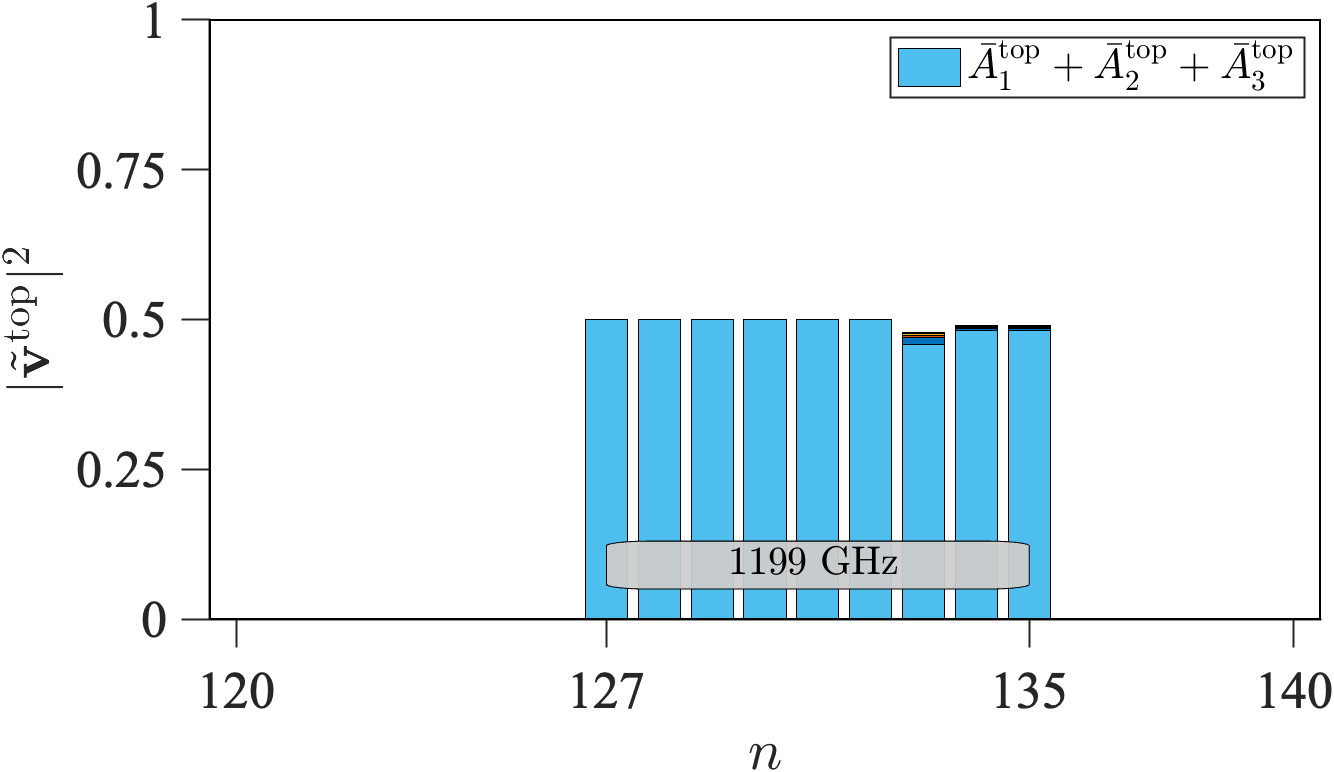}
         \caption{Modes 60 to 83 for top layer}
         \label{fig:phonon_analy_p1q59_no_sub:2}
     \end{subfigure}
\\[10pt]
     \begin{subfigure}[b]{0.49\textwidth}
         \centering
         \includegraphics[trim = 0mm 0mm 00mm 0mm,clip=true,width=\textwidth]{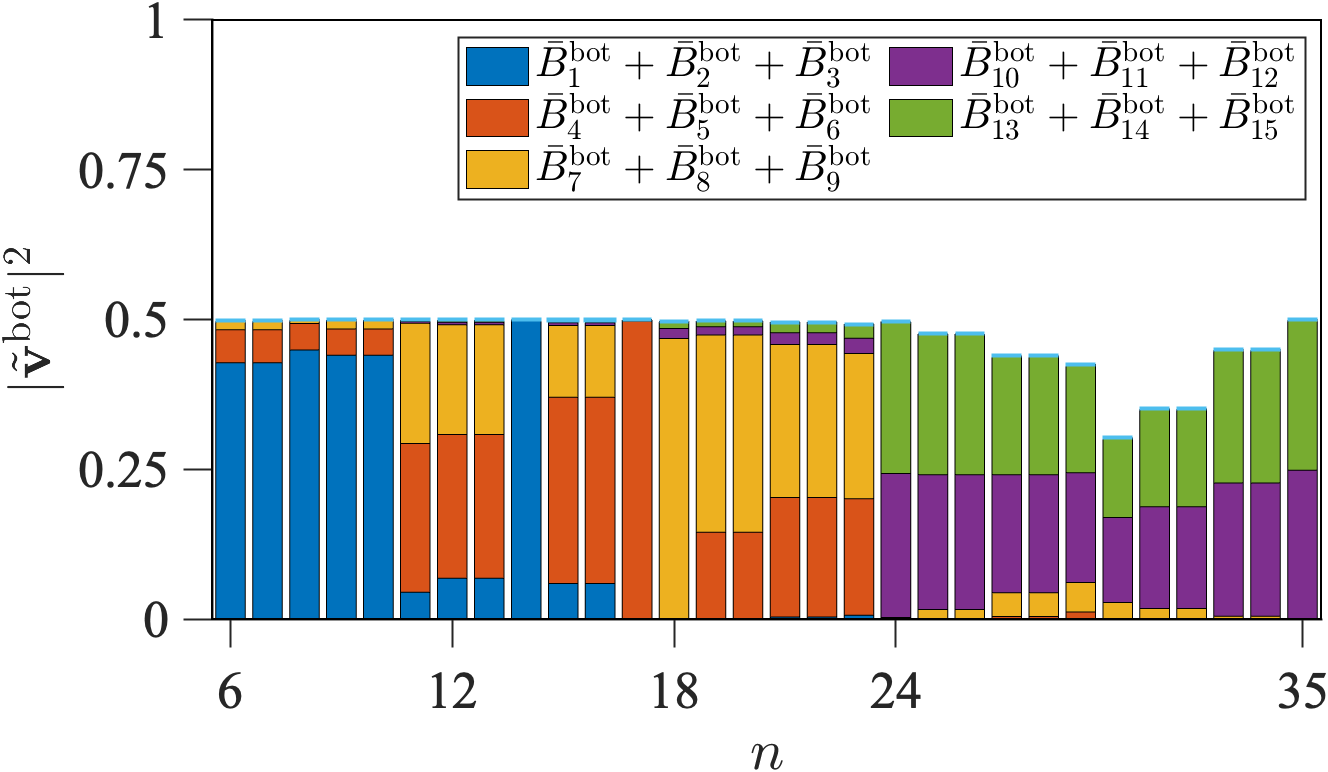}
         \caption{Modes 6 to 35 for bottom layer}
         \label{fig:phonon_analy_p1q59_no_sub:3}
     \end{subfigure}
     \begin{subfigure}[b]{0.49\textwidth}
         \centering
         \includegraphics[trim = 0mm 0mm 00mm 0mm,clip=true,width=\textwidth]{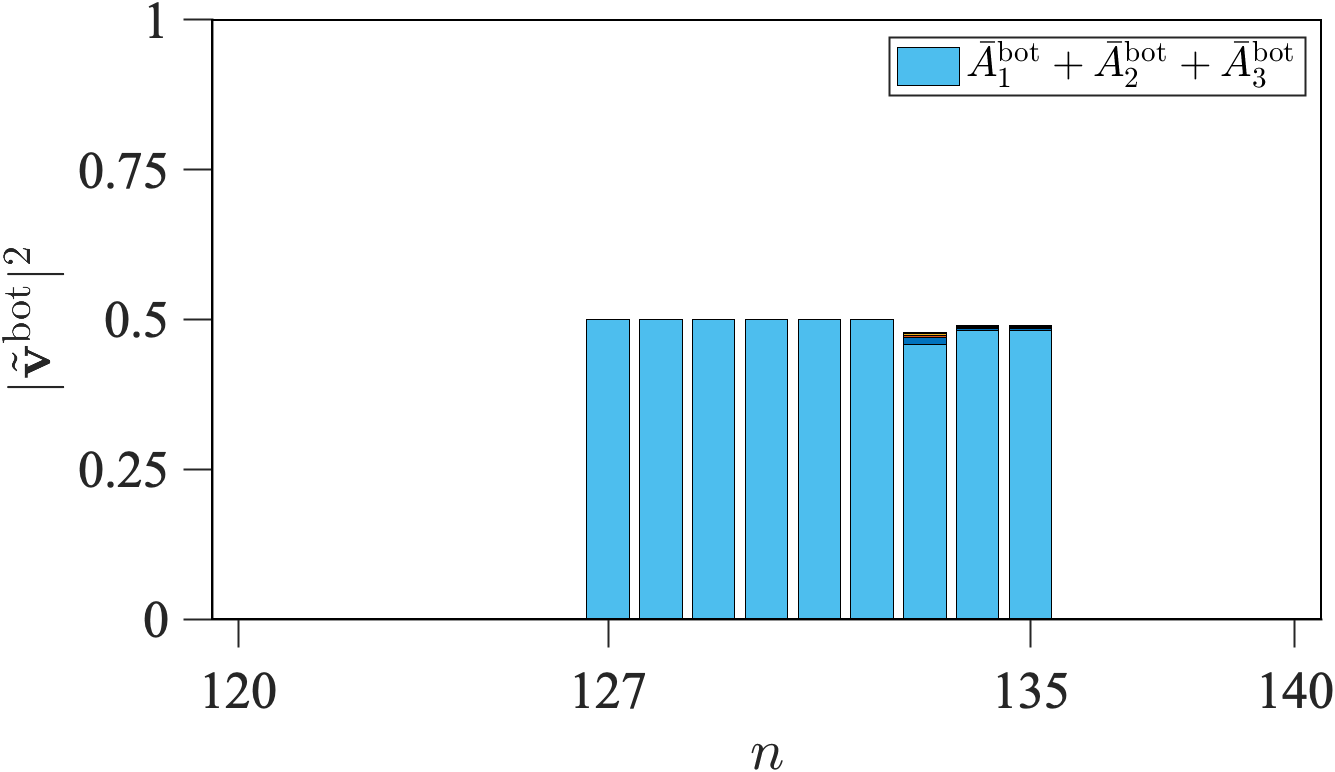}
         \caption{Modes 60 to 83 for bottom layer}
         \label{fig:phonon_analy_p1q59_no_sub:4}
     \end{subfigure}
\captionsetup{justification=raggedright, singlelinecheck=false} 
\caption{Contributions to the squared magnitude of $\mathbf{v}_n(\mathbf{0})$ for different mode numbers $n$ from the top and bottom graphene layers of a relaxed free-standing TBG at $\theta=1.12\degree$ computed from atomistic simulation results. At each mode number, the contributions from different out-of-plane and in-plane basis terms are shown as colored bars.  Terms with the same frequencies are summed together as indicated in the legend boxes.}
\label{fig:phonon_analy_p1q59_no_sub}
\end{figure*}

Similar to the analysis of low-frequency phonon modes (i.e., in-plane/out-of-plane dominant phonon modes) in \fig{fig:phonon_analy_p1q29_no_sub}, phonon modes of TBGs at $\theta=1.12\degree$ are analyzed following the approach described in \sect{sec:phonon_basis_method}. First, $\bar{A}_k$ and $\bar{B}_l$ of the relaxed free-standing TBG at $\theta=1.12$\degree\;are computed. The first three phonon modes comprise two rigid in-plane and one out-of-plane translation of the TBG, and the following two shearing modes (modes 4 and 5) have a frequency of 31.1 GHz. The computed $\bar{A}_k$ and $\bar{B}_l$ values for phonon modes (6,$\ldots$,35) and (120,$\ldots$,140) are plotted in \fig{fig:phonon_analy_p1q59_no_sub} to focus on the low-frequency phonon modes and the first phonon modes dominated by in-plane deformation.
Similar to $\bar{A}_k$ and $\bar{B}_l$ of the relaxed free-standing TBG at $\theta=2.28\degree$, phonon modes (6,$\ldots$,35) are mainly captured by $\bar{B}_l$.
However, different from the results for $\theta = 2.28\degree$ in \fig{fig:phonon_analy_p1q29_no_sub}, $\bar{B}_l$ combinations associated with different frequencies contribute to the same phonon mode. For example, there is a non-negligible contribution of $\bar{B}_4+\bar{B}_5+\bar{B}_6$ to phonon mode 6 shown in \fig{fig:phonon_analy_p1q59_no_sub:1} and \fig{fig:phonon_analy_p1q59_no_sub:3} for both layers. Furthermore, the frequency of phonon 14, where $\bar{B}_{1}+\bar{B}_{2}+\bar{B}_{3}$ is dominant, does not fall between the high $\bar{B}_{1}+\bar{B}_{2}+\bar{B}_{3}$ phonon modes (i.e., modes 6 to 10). Instead, it falls  within phonon modes where $\bar{B}_{4}+\bar{B}_{5}+\bar{B}_{6}$ is dominant (i.e., modes 11 to 17) while having similar frequencies.
Different frequencies from the same $\bar{B}$ combinations cannot occur in the double layer model because each $\bar{B}$ is associated with a specific frequency. However, in the relaxed TBG model with a small twist angle ($\theta = 1.12 \degree$), different frequencies for the same $\bar{B}$ combinations occur due to the large deformation in the graphene layers compared to those at a large angle of $\theta=2.28 \degree$, shown as the high $A_{1,2,3}$ and $B_{l}$ values at a low twist angle in \figs{fig:relaxed_A_top} and \figx{fig:relaxed_B_top}. The significant deformation of the relaxed TBG at a low twist angle makes it difficult to accurately predict its phonon modes using the out-of-plane normal modes of an elastic plate.
Nevertheless, the computed frequencies of the out-of-plane normal modes of the double layer model for $\theta=1.12 \degree$ ($f_{1,2,3}^{\rm out}=34$, $f_{4,5,6}^{\rm out}=104$, $f_{7,8,9}^{\rm out}=139$, and $f_{10,11,12}^{\rm out}=243$ GHz) provide a good approximation for the averaged frequencies in phonon modes (6,$\ldots$,35) with values 42, 81, 173, and 251 GHz.

The first phonon modes that have dominant in-plane deformation for the free-standing TBG are in the range (120,$\ldots$,140) at an average frequency of 1199 GHz (see \figs{fig:phonon_analy_p1q59_no_sub:2} and \figx{fig:phonon_analy_p1q59_no_sub:4}). The fact that the
 $\bar{A}_l$ terms are dominant and that $\bar{A}_1^{\rm top}+\bar{A}_2^{\rm top}+\bar{A}_3^{\rm top}$ and $\bar{A}_1^{\rm bot}+\bar{A}_2^{\rm bot}+\bar{A}_3^{\rm bot}$ have equal magnitudes of 0.5 indicates that phonon modes (120,$\ldots$,140) correspond to the in-plane normal modes of two independent elastic plates at $f_{1,2,3}^{\rm in}=1197.92$ GHz.

\begin{figure*}[t]\centering
    \begin{subfigure}[b]{0.49\textwidth}
         \centering
         \includegraphics[trim = 0mm 0mm 0mm 0mm,clip=true,width=\textwidth]{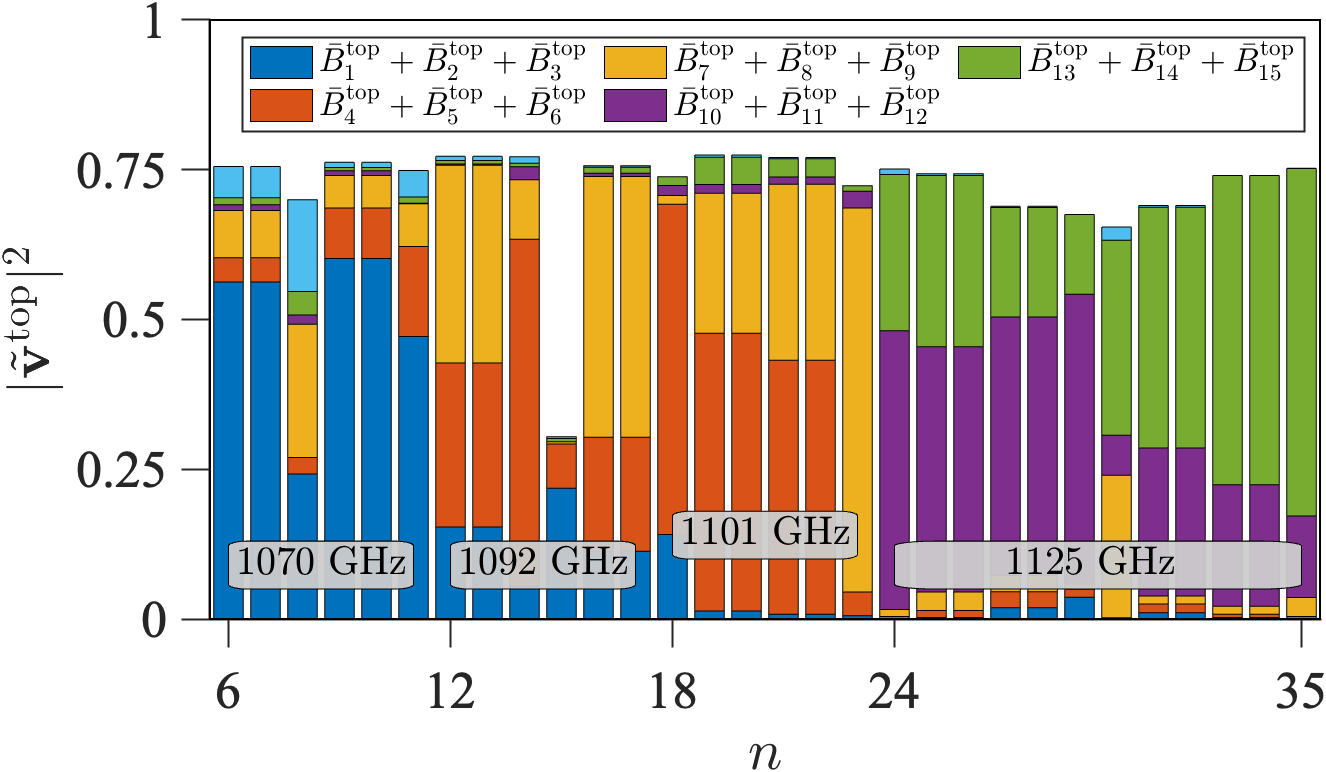}
         \caption{Modes 6 to 35 for top layer}
         \label{fig:phonon_analy_p1q59_hBN:1}
    \end{subfigure}
    \begin{subfigure}[b]{0.49\textwidth}
         \centering
         \includegraphics[trim = 0mm 0mm 00mm 0mm,clip=true,width=\textwidth]{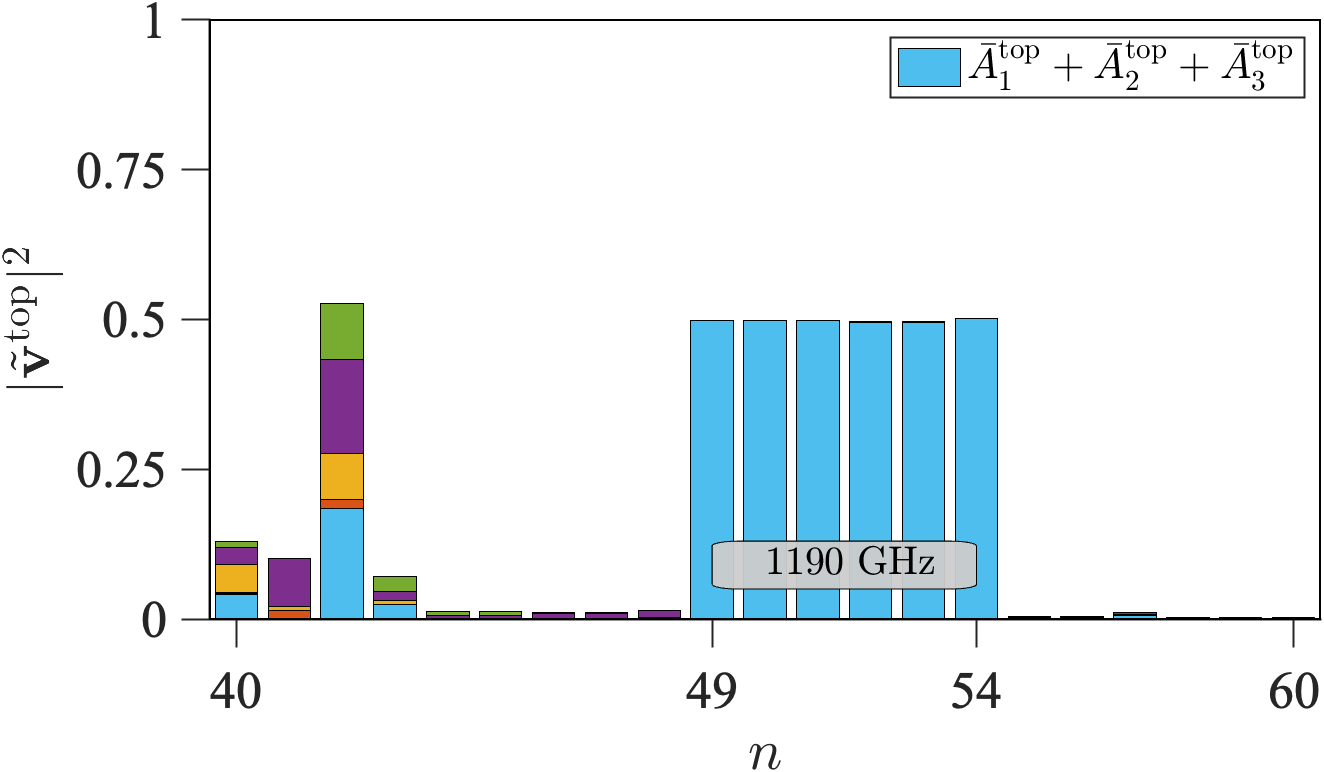}
         \caption{Modes 60 to 83 for top layer}
         \label{fig:phonon_analy_p1q59_hBN:2}
     \end{subfigure}
\\[10pt]
     \begin{subfigure}[b]{0.49\textwidth}
         \centering
         \includegraphics[trim = 0mm 0mm 00mm 0mm,clip=true,width=\textwidth]{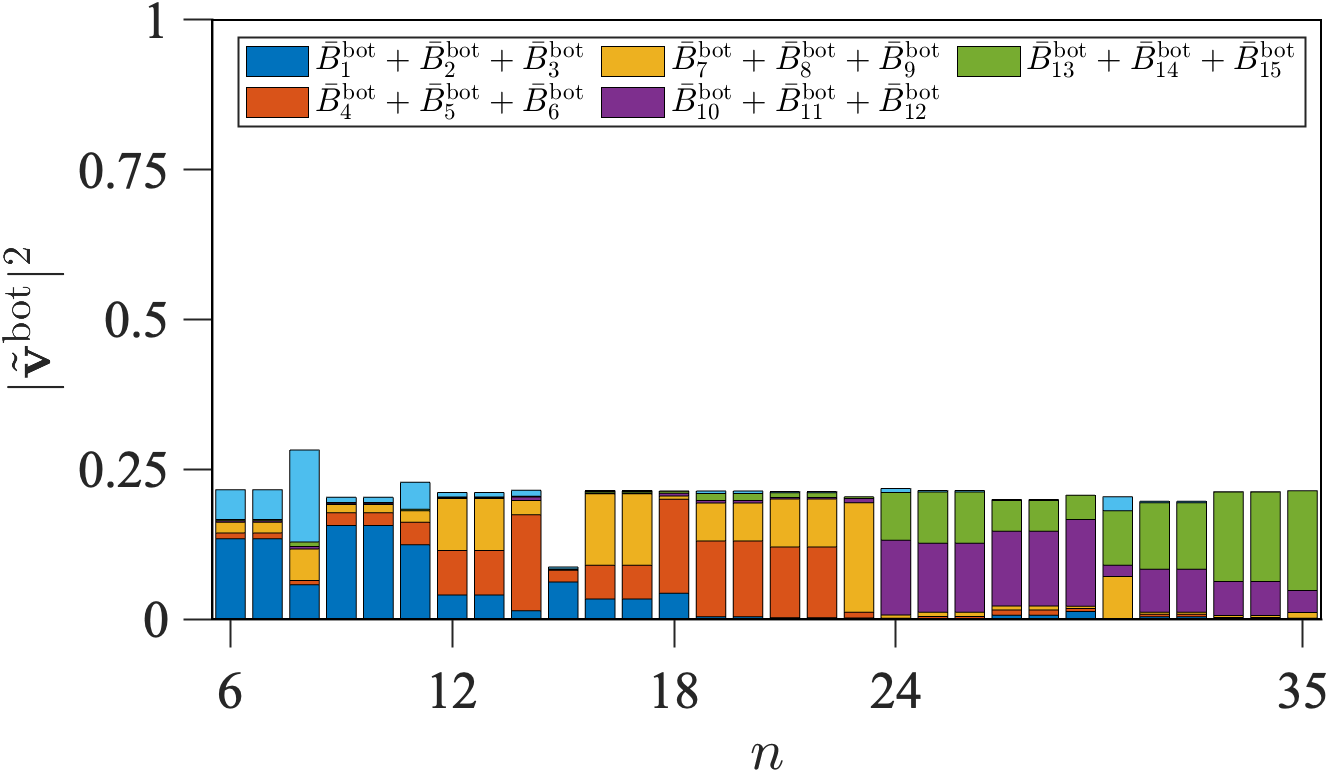}
         \caption{Modes 6 to 35 for bottom layer}
         \label{fig:phonon_analy_p1q59_hBN:3}
     \end{subfigure}
     \begin{subfigure}[b]{0.49\textwidth}
         \centering
         \includegraphics[trim = 0mm 0mm 00mm 0mm,clip=true,width=\textwidth]{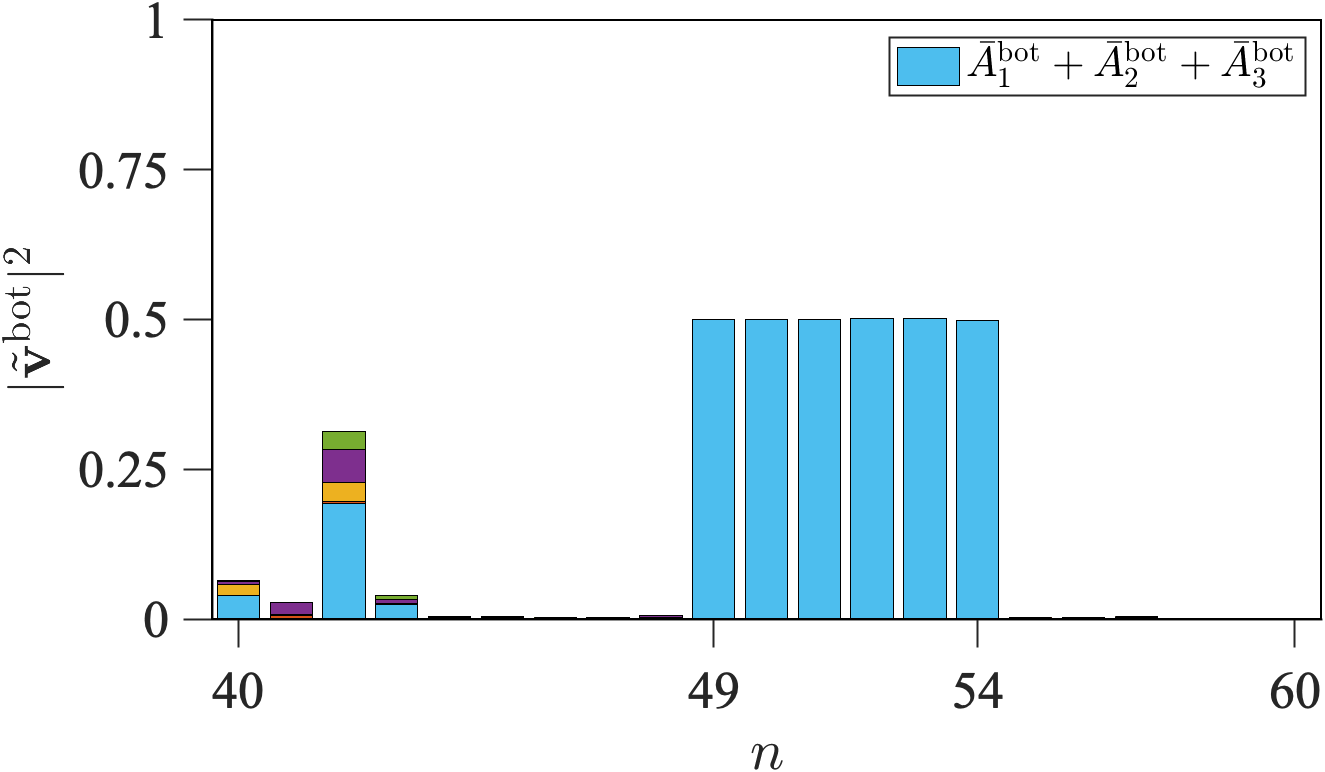}
         \caption{Modes 60 to 83 for bottom layer}
         \label{fig:phonon_analy_p1q59_hBN:4}
     \end{subfigure}
\captionsetup{justification=raggedright, singlelinecheck=false} 
\caption{Contributions to the squared magnitude of $\mathbf{v}_n(\mathbf{0})$ for different mode numbers $n$ from the top and bottom graphene layers of a relaxed TBG supported on an hBN substrate at $\theta=1.18\degree$ computed from atomistic simulation results. At each mode number, the contributions from different out-of-plane and in-plane basis terms are shown as colored bars.  Terms with the same frequencies are summed together as indicated in the legend boxes.}
\label{fig:phonon_analy_p1q59_hBN}
\end{figure*}

The $\bar{A}_k$ and $\bar{B}_l$ values for an hBN-supported TBG at $\theta=1.12\degree$ are shown in \fig{fig:phonon_analy_p1q59_hBN} for modes (6,$\ldots$,35) and (40,$\ldots$,60). Similar to phonon modes for the larger twist angle ($\theta = 2.28\degree$), the first two phonon modes ($n=1,2$) correspond to rigid-body translation along the in-plane directions, and two shearing modes (3 and 4) follow with a frequency of 32~GHz. Mode 5 is an out-of-plane translation of the two graphene layers with different squared magnitudes ($\sum_{i=1}^{\Nlay}\vert\mathbf{w}_i^{\rm top}\vert^2=0.743$ and $\sum_{j=1}^{\Nlay}\vert\mathbf{w}_j^{\rm bot}\vert^2=0.257$) with a common frequency of 1031~GHz.
For phonons in the range (6,$\ldots$,35), combinations of $\bar{B}_l$ terms associated with different frequencies contribute to the same phonon modes (as seen above for the free-standing TBG). This makes it difficult to directly represent these phonon modes using the double elastic plate model.
Nevertheless, the computed frequencies for the double elastic plate model ($f_{1,2,3}^{\rm dbl} = 1074.2$, $f_{4,5,6}^{\rm dbl}=1078$, $f_{7,8,9}^{\rm dbl} = 1082$, $f_{10,11,12}^{\rm dbl}= 1100$ GHz) for $\theta=1.12\degree$ are in good agreement with the frequencies of the phonon modes ($f_{1,2,3}^{\rm TBG}=$1061--1080, $f_{4,5,6}^{\rm TBG}=$1086--1095, $f_{7,8,9}^{\rm TBG}=$1099--1108, $f_{10,11,\ldots,15}^{\rm TBG}=$1114--1135~GHz in
\figs{fig:phonon_analy_p1q59_hBN:1} and \figx{fig:phonon_analy_p1q59_hBN:3}).
The first phonon modes where in-plane deformation is dominant appear in the range (49,$\ldots$,54), where the averaged frequency of these 6 modes (1190 GHz) is represented by $f_{1,2,3}^{\rm in}=1197.92$ since $\bar{A}_k$ for the top and bottom layers have the same magnitude of 0.5.

Overall, we find that phonon modes with dominant out-of-plane and in-plane deformation in the low frequency range are well represented by elastic plate models. Although the accuracy of this prediction decreases as the twist angle of the TBG decreases due to the increasing deformation of the graphene layers, the normal modes of the elastic plate models continue to provide a good approximation in terms of phonon mode frequency and mode shape (i.e., $\bar{B}$ and $\bar{A}$) for the dynamics of both relaxed free-standing and hBN-supported TBGs.

\section{Conclusion}\label{sec:conclusion}
A basis to describe the deformation of TBG systems during static relaxation is derived systematically from the in-plane and out-of-plane normal modes of an elastic continuum plate. It is found that a combination of only a few basis terms successfully captures the in-plane and out-of-plane relaxed TBG structure obtained using atomistic simulations across a range of twist angles computed with and without an underlying hBN and Si$3$N$_4$ substrates. The results are verified by comparing the basis representation of the molecular statics deformation with basis coefficients optimized to reproduce experimental electron diffraction pattern measurements. The Matlab code for this optimization process is provided to facilitate further research.

In addition to the static relaxation, the dynamics of TBG systems are analyzed using the phonon band structure of a TBG supercell. Phonon modes at $\mathbf{k}=\mathbf{0}$ of the relaxed TBG for small and large twist angles, with and without a supporting substrate, are analyzed using the elastic plate basis. It is found that both the frequencies and deformed shapes of the phonon modes are well predicted by in-plane and out-of-plane normal modes of single and double (stacked) elastic continuum plates.

The elastic plate basis can be used to characterize the deformation in any periodic supercell, therefore an interesting topic for future research is to extend this work to systematically quantify the deformation of arbitrary twisted multilayer vdW heterostructures obtained from either atomistic simulations or experimental electron diffraction patterns, including deformations involving symmetry breaking \cite{rakib2022}.

Another potential application is to use the elastic plate basis as a means for constructing reduced-order models (ROMs) for 2D materials in which the degrees of freedom are the in-plane and out-of-plane mode coefficients. This would enable computationally efficient atomistic simulation of the relaxation of arbitrary twisted vdW heterostructures (including TBGs). To be specific, the total energy of the atomistic system can be minimized relative to the basis coefficients with the atom displacements constrained by the elastic basis. 
Additionally, in the case of small deformation where linear elasticity holds, an analytical model can be constructed in which the total energy is decomposed into the sum of energies of the individual modes.

\section{Acknowledgement}\label{sec:acknowledgement}
The authors acknowledge partial support by the National Science Foundation (NSF) under grants OAC-1931304 and DMR-1834251, and through the University of Minnesota MRSEC under Award Number DMR-2011401. Additional partial support was provided through the Army Research Office (W911NF-14-1-0247) under the MURI program.
S.H.S.\ and R.H.\ acknowledge support from the U.S. Department of Energy, Basic
Energy Sciences, under award DE-SC0024147.

\appendix
\section{Derivation of $\phi_{\bar{\bG}_k}$ and $\phi_{\hat{\bG}_l}$ under 6-fold symmetry conditions}\label{app:phasederiv}

In order for combinations of bases to satisfy 6-fold symmetry, the displacement field must remain invariant to a 60$\degree$ rotation. Taking the first three in-plane bases as an example, the following relation must be satisfied at all positions $\br$,
\begin{equation}\label{eqn:rot_in_plane_basis}
    \sum_{k=1}^3 \bth{k}^{\rm in}(\br) 
    = \sum_{k=1}^3 \mathbf{Q}^{\rm T}\bth{k}^{\rm in}(\bQ\br),
    \quad\forall\br
\end{equation}
where $\bQ$ is a two-dimensional 60$\degree$ rotation matrix:
\begin{equation}
\bQ = \begin{bmatrix}
\cos{60\degree} & -\sin{60\degree} \\
\sin{60\degree} & \cos{60\degree}
\end{bmatrix}.
\end{equation}
\Eqn{eqn:rot_in_plane_basis} provides two equations, one for the $x$ direction, and one for the $y$ direction. To ensure that 6-fold symmetry is centered on the origin (corresponding to the center of the AA domain in the TBG), we take the phase angle to be the same for the three bases, such that $\phi_{\bar{\bG}_k}=\phi^{\rm in}\ (k=1,2,3)$. The real part of \eqn{eqn:rot_in_plane_basis} in the $x$ direction is
\begin{multline}\label{eqn:rot_in_plane_basis_x}
    -\sqrt{3}/2\cos{(\bar{\bG}_1\cdot\br+\phi^{\rm in})}+\sqrt{3}/2\cos{(\bar{\bG}_3\cdot\br+\phi^{\rm in})} \\
    = - \sqrt{3}/2 \cos{(\bar{\bG}_2\cdot \bQ\br+\phi^{\rm in})} + \sqrt{3}/2 \cos{(\bar{\bG}_3\cdot \bQ\br+\phi^{\rm in})}.
\end{multline}
The reciprocal lattice vectors $\bar{\bG}_1$, $\bar{\bG}_2$, $\bar{\bG}_3$, are related by $120\degree$ rotations, such that
\begin{equation}
\bar{\bG}_1=(\bQ^{\rm T})^2\bar{\bG}_2= \bQ^2\bar{\bG}_3.
\end{equation}
Using these relations, \eqn{eqn:rot_in_plane_basis_x} becomes
\begin{multline}\label{eqn:rot_in_plane_basis_x_2}
    \cos{(\bar{\bG}_1 \cdot \br + \phi^{\rm in})} + \cos{(-\bar{\bG}_1 \cdot \br + \phi^{\rm in})} \\ 
    - \bigl(\cos{(\bQ\bar{\bG}_1 \cdot \br + \phi^{\rm in})} + \cos{(-\bQ\bar{\bG}_1 \cdot \br + \phi^{\rm in})}\bigr) = 0,
\end{multline}
where we have used $\bQ^T\bQ=\bI$, $\bQ^3=-\bI$ (for a $60\degree$ rotation), and the identity
\begin{equation}
\bar{\bG}\cdot\bQ\br = \bQ^{\rm T}\bar{\bG}\cdot\br.
\end{equation}
Through basic trigonmetry, \eqn{eqn:rot_in_plane_basis_x_2} further simplifies to
\begin{equation}\label{eqn:rot_in_plane_basis_x_3}
    \cos{\phi^{\rm in}}\bigl(\cos{(\bar{\bG}_1 \cdot \br)}-\cos{(\bQ\bar{\bG}_1 \cdot \br)}\bigr)=0.
\end{equation}
Similary, the real part of \eqn{eqn:rot_in_plane_basis} in the $y$ direction is
\begin{multline}\label{eqn:rot_in_plane_basis_y}
    \frac{1}{2}\bigl(\cos{(\bG_1\cdot\br+\phi^{\rm in})}+\cos{(-\bG_1\cdot\br+\phi^{\rm in})} \\
    + \cos{(\bQ\bG_1\cdot\br+\phi^{\rm in})}+\cos{(-\bQ\bG_1\cdot\br+\phi^{\rm in})}\bigr) \\ 
    - \bigl(\cos{(\bQ^{\rm T}\bG_1\cdot\br+\phi^{\rm in})}+\cos{(-\bQ^{\rm T}\bG_1\cdot\br+\phi^{\rm in})}\bigr) = 0,
\end{multline}
which simplifies to 
\begin{equation}\label{eqn:rot_in_plane_basis_y_2}
    \cos{\phi^{\rm in}}\bigl(\cos{(\bar{\bG}_1 \cdot \br)}+\cos{(\bQ\bar{\bG}_1 \cdot \br)}-2\cos{(\bQ^{\rm T}\bar{\bG}_1 \cdot \br)}\bigr)=0.
\end{equation}
\Eqns{eqn:rot_in_plane_basis_x_3} and \eqnx{eqn:rot_in_plane_basis_y_2} are satisfied identically for all $\br$ provided that $\cos\phi^{\rm in}=0$, so that $\phi^{\rm in}=(n+\frac{1}{2})\pi$, where $n$ is an integer. Note that this solution for $\phi^{\rm in}$ holds for all summed basis triplets that have the same magnitude and are related through $120 \degree$ rotations (i.e., $\sum_{k=1}^3\bth{k}^{\rm in}(\br)$, $\sum_{k=4}^6\bth{k}^{\rm in}(\br)$, $\sum_{k=7}^9\bth{k}^{\rm in}(\br)$, and $\sum_{k=10}^{12}\bth{k}^{\rm in}(\br)$). 

Setting $\phi^{\rm in}=(n+\frac{1}{2})\pi$ leads to a local twist about the origin in the displacement $\sum_{k=1}^3\bth{k}^{\rm in}(\br)$. This displacement exhibits a clockwise rotation when $n$ is even (see \fig{fig:in_plane_mode}), and a counterclockwise rotation when $n$ is odd.
To achieve positive coefficients for the top layer experiencing a local counterclockwise twist during TBG relaxation, we set $\phi^{\rm in} = -\frac{\pi}{2}$ for $\sum_{k=1}^3\bth{k}^{\rm in}(\br)$. We also set $\phi^{\rm in} = -\frac{\pi}{2}$ for the remaining in-plane bases to preserve the counterclockwise rotation in the displacements of $\sum_{k=4}^6\bth{k}^{\rm in}(\br)$, $\sum_{k=7}^9\bth{k}^{\rm in}(\br)$, and $\sum_{k=10}^{12}\bth{k}^{\rm in}(\br)$. 

Similarly, for out-of-plane deformation, the requirement for 6-fold symmetry for the first three bases is 
\begin{equation}\label{eqn:rot_out_of_plane_basis}
    \sum_{l=1}^{3} \bth{l}^{\rm out}(\br) = \sum_{l=1}^{3} \bth{l}^{\rm out}(\bQ\br),
    \quad\forall\br.
\end{equation}
Again, we take the phase angle for all three bases to be the same, $\phi_{\hat{\bG}_l}=\phi^{\rm out}\ (l=1,2,3)$. The real part of \eqn{eqn:rot_out_of_plane_basis} follows as 
\begin{equation}\label{eqn:rot_out_of_plane_basis_2}
    \sum_{l=1}^3\cos{(\hat{\bG}_l\cdot\br+\phi^{\rm out})} =  \sum_{l=1}^3\cos{(\hat{\bG}_l\cdot\bQ\br+\phi^{\rm out})}.
\end{equation}
Similar to \eqns{eqn:rot_in_plane_basis_x_3} and \eqnx{eqn:rot_in_plane_basis_y_2}, \eqn{eqn:rot_out_of_plane_basis_2} simplifies to  
\begin{equation}\label{eqn:rot_out_of_plane_basis_3}
    \sin{\phi^{\rm out}}\bigl(\sin{(\hat{\bG}_1\cdot\br)}+\sin{(-\bQ^{\rm T}\hat{\bG}_1\cdot\br)}+\sin{(-\bQ\hat{\bG}_1\cdot\br}\bigr))=0.
\end{equation}
\Eqn{eqn:rot_out_of_plane_basis_3} is satisfied identically for all $\br$, for $\phi^{\rm out}=n\pi$, where $n$ is an integer. \Eqn{eqn:rot_out_of_plane_basis_3} holds for all summed basis triplets (i.e., $\sum_{l=1}^3\bth{l}^{\rm out}(\br)$, $\sum_{l=4}^6\bth{l}^{\rm out}(\br)$, $\sum_{l=7}^9\bth{l}^{\rm out}(\br)$, $\sum_{l=10}^{12}\bth{l}^{\rm out}(\br)$, and 
$\sum_{l=13}^{15}\bth{l}^{\rm out}(\br)$). 

In the displacement of all summed out-of-plane bases, a dome-shaped displacement is observed centered on the origin. When $n$ is even, this displacement is positive (see \fig{fig:out_of_plane_mode}), while it is negative when $n$ is odd. To achieve a positive coefficient for the top layer experiencing a positive out-of-plane deformation at the origin during TBG relaxation, we set $\phi^{\rm out}=0$ for all out-of-plane bases.
\vfill

\bibliography{main}

\providecommand{\noopsort}[1]{}\providecommand{\singleletter}[1]{#1}%
\begin{thebibliography}{51}%
\makeatletter
\providecommand \@ifxundefined [1]{%
 \@ifx{#1\undefined}
}%
\providecommand \@ifnum [1]{%
 \ifnum #1\expandafter \@firstoftwo
 \else \expandafter \@secondoftwo
 \fi
}%
\providecommand \@ifx [1]{%
 \ifx #1\expandafter \@firstoftwo
 \else \expandafter \@secondoftwo
 \fi
}%
\providecommand \natexlab [1]{#1}%
\providecommand \enquote  [1]{``#1''}%
\providecommand \bibnamefont  [1]{#1}%
\providecommand \bibfnamefont [1]{#1}%
\providecommand \citenamefont [1]{#1}%
\providecommand \href@noop [0]{\@secondoftwo}%
\providecommand \href [0]{\begingroup \@sanitize@url \@href}%
\providecommand \@href[1]{\@@startlink{#1}\@@href}%
\providecommand \@@href[1]{\endgroup#1\@@endlink}%
\providecommand \@sanitize@url [0]{\catcode `\\12\catcode `\$12\catcode `\&12\catcode `\#12\catcode `\^12\catcode `\_12\catcode `\%12\relax}%
\providecommand \@@startlink[1]{}%
\providecommand \@@endlink[0]{}%
\providecommand \url  [0]{\begingroup\@sanitize@url \@url }%
\providecommand \@url [1]{\endgroup\@href {#1}{\urlprefix }}%
\providecommand \urlprefix  [0]{URL }%
\providecommand \Eprint [0]{\href }%
\providecommand \doibase [0]{https://doi.org/}%
\providecommand \selectlanguage [0]{\@gobble}%
\providecommand \bibinfo  [0]{\@secondoftwo}%
\providecommand \bibfield  [0]{\@secondoftwo}%
\providecommand \translation [1]{[#1]}%
\providecommand \BibitemOpen [0]{}%
\providecommand \bibitemStop [0]{}%
\providecommand \bibitemNoStop [0]{.\EOS\space}%
\providecommand \EOS [0]{\spacefactor3000\relax}%
\providecommand \BibitemShut  [1]{\csname bibitem#1\endcsname}%
\let\auto@bib@innerbib\@empty
\bibitem [{\citenamefont {Nimbalkar}\ and\ \citenamefont {Kim}(2020)}]{nimbalkar2020}%
  \BibitemOpen
  \bibfield  {author} {\bibinfo {author} {\bibfnamefont {A.}~\bibnamefont {Nimbalkar}}\ and\ \bibinfo {author} {\bibfnamefont {H.}~\bibnamefont {Kim}},\ }\bibfield  {title} {\bibinfo {title} {Opportunities and challenges in twisted bilayer graphene: a review},\ }\href@noop {} {\bibfield  {journal} {\bibinfo  {journal} {Nano-Micro Letters}\ }\textbf {\bibinfo {volume} {12}},\ \bibinfo {pages} {1} (\bibinfo {year} {2020})}\BibitemShut {NoStop}%
\bibitem [{\citenamefont {Peng}\ \emph {et~al.}(2020)\citenamefont {Peng}, \citenamefont {Chen}, \citenamefont {Fan}, \citenamefont {Srolovitz},\ and\ \citenamefont {Lei}}]{Peng2020}%
  \BibitemOpen
  \bibfield  {author} {\bibinfo {author} {\bibfnamefont {Z.}~\bibnamefont {Peng}}, \bibinfo {author} {\bibfnamefont {X.}~\bibnamefont {Chen}}, \bibinfo {author} {\bibfnamefont {Y.}~\bibnamefont {Fan}}, \bibinfo {author} {\bibfnamefont {D.~J.}\ \bibnamefont {Srolovitz}},\ and\ \bibinfo {author} {\bibfnamefont {D.}~\bibnamefont {Lei}},\ }\bibfield  {title} {\bibinfo {title} {Strain engineering of 2{D} semiconductors and graphene: from strain fields to band-structure tuning and photonic applications},\ }\href@noop {} {\bibfield  {journal} {\bibinfo  {journal} {Light: Science \& Applications}\ }\textbf {\bibinfo {volume} {9}} (\bibinfo {year} {2020})}\BibitemShut {NoStop}%
\bibitem [{\citenamefont {Yu}\ \emph {et~al.}(2021)\citenamefont {Yu}, \citenamefont {Han}, \citenamefont {Hossain}, \citenamefont {Watanabe}, \citenamefont {Taniguchi}, \citenamefont {Ertekin}, \citenamefont {van~der Zande},\ and\ \citenamefont {Huang}}]{yu2021}%
  \BibitemOpen
  \bibfield  {author} {\bibinfo {author} {\bibfnamefont {J.}~\bibnamefont {Yu}}, \bibinfo {author} {\bibfnamefont {E.}~\bibnamefont {Han}}, \bibinfo {author} {\bibfnamefont {M.~A.}\ \bibnamefont {Hossain}}, \bibinfo {author} {\bibfnamefont {K.}~\bibnamefont {Watanabe}}, \bibinfo {author} {\bibfnamefont {T.}~\bibnamefont {Taniguchi}}, \bibinfo {author} {\bibfnamefont {E.}~\bibnamefont {Ertekin}}, \bibinfo {author} {\bibfnamefont {A.~M.}\ \bibnamefont {van~der Zande}},\ and\ \bibinfo {author} {\bibfnamefont {P.~Y.}\ \bibnamefont {Huang}},\ }\bibfield  {title} {\bibinfo {title} {Designing the bending stiffness of 2{D} material heterostructures},\ }\href@noop {} {\bibfield  {journal} {\bibinfo  {journal} {Advanced Materials}\ }\textbf {\bibinfo {volume} {33}},\ \bibinfo {pages} {2007269} (\bibinfo {year} {2021})}\BibitemShut {NoStop}%
\bibitem [{\citenamefont {Lemme}\ \emph {et~al.}(2022)\citenamefont {Lemme}, \citenamefont {Akinwande}, \citenamefont {Huyghebaert},\ and\ \citenamefont {Stampfer}}]{lemme2022}%
  \BibitemOpen
  \bibfield  {author} {\bibinfo {author} {\bibfnamefont {M.~C.}\ \bibnamefont {Lemme}}, \bibinfo {author} {\bibfnamefont {D.}~\bibnamefont {Akinwande}}, \bibinfo {author} {\bibfnamefont {C.}~\bibnamefont {Huyghebaert}},\ and\ \bibinfo {author} {\bibfnamefont {C.}~\bibnamefont {Stampfer}},\ }\bibfield  {title} {\bibinfo {title} {2{D} materials for future heterogeneous electronics},\ }\href@noop {} {\bibfield  {journal} {\bibinfo  {journal} {Nature Communications}\ }\textbf {\bibinfo {volume} {13}},\ \bibinfo {pages} {1} (\bibinfo {year} {2022})}\BibitemShut {NoStop}%
\bibitem [{\citenamefont {Mounet}\ \emph {et~al.}(2018)\citenamefont {Mounet}, \citenamefont {Gibertini}, \citenamefont {Schwaller}, \citenamefont {Campi}, \citenamefont {Merkys}, \citenamefont {Marrazzo}, \citenamefont {Sohier}, \citenamefont {Castelli}, \citenamefont {Cepellotti}, \citenamefont {Pizzi},\ and\ \citenamefont {Marzari}}]{mounet:gibertini:2018}%
  \BibitemOpen
  \bibfield  {author} {\bibinfo {author} {\bibfnamefont {N.}~\bibnamefont {Mounet}}, \bibinfo {author} {\bibfnamefont {M.}~\bibnamefont {Gibertini}}, \bibinfo {author} {\bibfnamefont {P.}~\bibnamefont {Schwaller}}, \bibinfo {author} {\bibfnamefont {D.}~\bibnamefont {Campi}}, \bibinfo {author} {\bibfnamefont {A.}~\bibnamefont {Merkys}}, \bibinfo {author} {\bibfnamefont {A.}~\bibnamefont {Marrazzo}}, \bibinfo {author} {\bibfnamefont {T.}~\bibnamefont {Sohier}}, \bibinfo {author} {\bibfnamefont {I.~E.}\ \bibnamefont {Castelli}}, \bibinfo {author} {\bibfnamefont {A.}~\bibnamefont {Cepellotti}}, \bibinfo {author} {\bibfnamefont {G.}~\bibnamefont {Pizzi}},\ and\ \bibinfo {author} {\bibfnamefont {N.}~\bibnamefont {Marzari}},\ }\bibfield  {title} {\bibinfo {title} {Two-dimensional materials from high-throughput computational exfoliation of experimentally known compounds},\ }\href {https://doi.org/10.1038/s41565-017-0035-5} {\bibfield  {journal} {\bibinfo  {journal} {Nature Nanotech.}\ }\textbf {\bibinfo {volume}
  {13}},\ \bibinfo {pages} {246} (\bibinfo {year} {2018})}\BibitemShut {NoStop}%
\bibitem [{\citenamefont {Cui}\ \emph {et~al.}(2015)\citenamefont {Cui}, \citenamefont {Lee}, \citenamefont {Kim}, \citenamefont {Arefe}, \citenamefont {Huang}, \citenamefont {Lee}, \citenamefont {Chenet}, \citenamefont {Zhang}, \citenamefont {Wang}, \citenamefont {Ye} \emph {et~al.}}]{cui2015}%
  \BibitemOpen
  \bibfield  {author} {\bibinfo {author} {\bibfnamefont {X.}~\bibnamefont {Cui}}, \bibinfo {author} {\bibfnamefont {G.-H.}\ \bibnamefont {Lee}}, \bibinfo {author} {\bibfnamefont {Y.~D.}\ \bibnamefont {Kim}}, \bibinfo {author} {\bibfnamefont {G.}~\bibnamefont {Arefe}}, \bibinfo {author} {\bibfnamefont {P.~Y.}\ \bibnamefont {Huang}}, \bibinfo {author} {\bibfnamefont {C.-H.}\ \bibnamefont {Lee}}, \bibinfo {author} {\bibfnamefont {D.~A.}\ \bibnamefont {Chenet}}, \bibinfo {author} {\bibfnamefont {X.}~\bibnamefont {Zhang}}, \bibinfo {author} {\bibfnamefont {L.}~\bibnamefont {Wang}}, \bibinfo {author} {\bibfnamefont {F.}~\bibnamefont {Ye}}, \emph {et~al.},\ }\bibfield  {title} {\bibinfo {title} {Multi-terminal transport measurements of {MoS$_2$} using a {van der Waals} heterostructure device platform},\ }\href@noop {} {\bibfield  {journal} {\bibinfo  {journal} {Nature nanotechnology}\ }\textbf {\bibinfo {volume} {10}},\ \bibinfo {pages} {534} (\bibinfo {year} {2015})}\BibitemShut {NoStop}%
\bibitem [{\citenamefont {Glavin}\ \emph {et~al.}(2020)\citenamefont {Glavin}, \citenamefont {Rao}, \citenamefont {Varshney}, \citenamefont {Bianco}, \citenamefont {Apte}, \citenamefont {Roy}, \citenamefont {Ringe},\ and\ \citenamefont {Ajayan}}]{glavin2020}%
  \BibitemOpen
  \bibfield  {author} {\bibinfo {author} {\bibfnamefont {N.~R.}\ \bibnamefont {Glavin}}, \bibinfo {author} {\bibfnamefont {R.}~\bibnamefont {Rao}}, \bibinfo {author} {\bibfnamefont {V.}~\bibnamefont {Varshney}}, \bibinfo {author} {\bibfnamefont {E.}~\bibnamefont {Bianco}}, \bibinfo {author} {\bibfnamefont {A.}~\bibnamefont {Apte}}, \bibinfo {author} {\bibfnamefont {A.}~\bibnamefont {Roy}}, \bibinfo {author} {\bibfnamefont {E.}~\bibnamefont {Ringe}},\ and\ \bibinfo {author} {\bibfnamefont {P.~M.}\ \bibnamefont {Ajayan}},\ }\bibfield  {title} {\bibinfo {title} {Emerging applications of elemental 2{D} materials},\ }\href@noop {} {\bibfield  {journal} {\bibinfo  {journal} {Advanced Materials}\ }\textbf {\bibinfo {volume} {32}},\ \bibinfo {pages} {1904302} (\bibinfo {year} {2020})}\BibitemShut {NoStop}%
\bibitem [{\citenamefont {Zhang}\ and\ \citenamefont {Tadmor}(2018)}]{zhang2018}%
  \BibitemOpen
  \bibfield  {author} {\bibinfo {author} {\bibfnamefont {K.}~\bibnamefont {Zhang}}\ and\ \bibinfo {author} {\bibfnamefont {E.~B.}\ \bibnamefont {Tadmor}},\ }\bibfield  {title} {\bibinfo {title} {Structural and electron diffraction scaling of twisted graphene bilayers},\ }\href@noop {} {\bibfield  {journal} {\bibinfo  {journal} {Journal of the Mechanics and Physics of Solids}\ }\textbf {\bibinfo {volume} {112}},\ \bibinfo {pages} {225} (\bibinfo {year} {2018})}\BibitemShut {NoStop}%
\bibitem [{\citenamefont {Yoo}\ \emph {et~al.}(2019)\citenamefont {Yoo}, \citenamefont {Engelke}, \citenamefont {Carr}, \citenamefont {Fang}, \citenamefont {Zhang}, \citenamefont {Cazeaux}, \citenamefont {Sung}, \citenamefont {Hovden}, \citenamefont {Tsen}, \citenamefont {Taniguchi} \emph {et~al.}}]{yoo2019}%
  \BibitemOpen
  \bibfield  {author} {\bibinfo {author} {\bibfnamefont {H.}~\bibnamefont {Yoo}}, \bibinfo {author} {\bibfnamefont {R.}~\bibnamefont {Engelke}}, \bibinfo {author} {\bibfnamefont {S.}~\bibnamefont {Carr}}, \bibinfo {author} {\bibfnamefont {S.}~\bibnamefont {Fang}}, \bibinfo {author} {\bibfnamefont {K.}~\bibnamefont {Zhang}}, \bibinfo {author} {\bibfnamefont {P.}~\bibnamefont {Cazeaux}}, \bibinfo {author} {\bibfnamefont {S.~H.}\ \bibnamefont {Sung}}, \bibinfo {author} {\bibfnamefont {R.}~\bibnamefont {Hovden}}, \bibinfo {author} {\bibfnamefont {A.~W.}\ \bibnamefont {Tsen}}, \bibinfo {author} {\bibfnamefont {T.}~\bibnamefont {Taniguchi}}, \emph {et~al.},\ }\bibfield  {title} {\bibinfo {title} {Atomic and electronic reconstruction at the {van der Waals} interface in twisted bilayer graphene},\ }\href@noop {} {\bibfield  {journal} {\bibinfo  {journal} {Nature materials}\ }\textbf {\bibinfo {volume} {18}},\ \bibinfo {pages} {448} (\bibinfo {year} {2019})}\BibitemShut {NoStop}%
\bibitem [{\citenamefont {Alden}\ \emph {et~al.}(2013)\citenamefont {Alden}, \citenamefont {Tsen}, \citenamefont {Huang}, \citenamefont {Hovden}, \citenamefont {Brown}, \citenamefont {Park}, \citenamefont {Muller},\ and\ \citenamefont {McEuen}}]{alden2013}%
  \BibitemOpen
  \bibfield  {author} {\bibinfo {author} {\bibfnamefont {J.~S.}\ \bibnamefont {Alden}}, \bibinfo {author} {\bibfnamefont {A.~W.}\ \bibnamefont {Tsen}}, \bibinfo {author} {\bibfnamefont {P.~Y.}\ \bibnamefont {Huang}}, \bibinfo {author} {\bibfnamefont {R.}~\bibnamefont {Hovden}}, \bibinfo {author} {\bibfnamefont {L.}~\bibnamefont {Brown}}, \bibinfo {author} {\bibfnamefont {J.}~\bibnamefont {Park}}, \bibinfo {author} {\bibfnamefont {D.~A.}\ \bibnamefont {Muller}},\ and\ \bibinfo {author} {\bibfnamefont {P.~L.}\ \bibnamefont {McEuen}},\ }\bibfield  {title} {\bibinfo {title} {Strain solitons and topological defects in bilayer graphene},\ }\href@noop {} {\bibfield  {journal} {\bibinfo  {journal} {Proceedings of the National Academy of Sciences}\ }\textbf {\bibinfo {volume} {110}},\ \bibinfo {pages} {11256} (\bibinfo {year} {2013})}\BibitemShut {NoStop}%
\bibitem [{\citenamefont {van Wijk}\ \emph {et~al.}(2015)\citenamefont {van Wijk}, \citenamefont {Schuring}, \citenamefont {Katsnelson},\ and\ \citenamefont {Fasolino}}]{wijk:schuring:2015}%
  \BibitemOpen
  \bibfield  {author} {\bibinfo {author} {\bibfnamefont {M.~M.}\ \bibnamefont {van Wijk}}, \bibinfo {author} {\bibfnamefont {A.}~\bibnamefont {Schuring}}, \bibinfo {author} {\bibfnamefont {M.~I.}\ \bibnamefont {Katsnelson}},\ and\ \bibinfo {author} {\bibfnamefont {A.}~\bibnamefont {Fasolino}},\ }\bibfield  {title} {\bibinfo {title} {Relaxation of {M}oir{\'{e}} patterns for slightly misaligned identical lattices: graphene on graphite},\ }\href {https://doi.org/10.1088/2053-1583/2/3/034010} {\bibfield  {journal} {\bibinfo  {journal} {2D Materials}\ }\textbf {\bibinfo {volume} {2}},\ \bibinfo {pages} {034010} (\bibinfo {year} {2015})}\BibitemShut {NoStop}%
\bibitem [{\citenamefont {Guinea}\ and\ \citenamefont {Walet}(2019)}]{guinea2019}%
  \BibitemOpen
  \bibfield  {author} {\bibinfo {author} {\bibfnamefont {F.}~\bibnamefont {Guinea}}\ and\ \bibinfo {author} {\bibfnamefont {N.~R.}\ \bibnamefont {Walet}},\ }\bibfield  {title} {\bibinfo {title} {Continuum models for twisted bilayer graphene: effect of lattice deformation and hopping parameters},\ }\href@noop {} {\bibfield  {journal} {\bibinfo  {journal} {Physical Review B}\ }\textbf {\bibinfo {volume} {99}},\ \bibinfo {pages} {205134} (\bibinfo {year} {2019})}\BibitemShut {NoStop}%
\bibitem [{\citenamefont {Sung}\ \emph {et~al.}(2022)\citenamefont {Sung}, \citenamefont {Goh}, \citenamefont {Yoo}, \citenamefont {Engelke}, \citenamefont {Xie}, \citenamefont {Zhang}, \citenamefont {Li}, \citenamefont {Ye}, \citenamefont {Deotare}, \citenamefont {Tadmor} \emph {et~al.}}]{sung2022}%
  \BibitemOpen
  \bibfield  {author} {\bibinfo {author} {\bibfnamefont {S.~H.}\ \bibnamefont {Sung}}, \bibinfo {author} {\bibfnamefont {Y.~M.}\ \bibnamefont {Goh}}, \bibinfo {author} {\bibfnamefont {H.}~\bibnamefont {Yoo}}, \bibinfo {author} {\bibfnamefont {R.}~\bibnamefont {Engelke}}, \bibinfo {author} {\bibfnamefont {H.}~\bibnamefont {Xie}}, \bibinfo {author} {\bibfnamefont {K.}~\bibnamefont {Zhang}}, \bibinfo {author} {\bibfnamefont {Z.}~\bibnamefont {Li}}, \bibinfo {author} {\bibfnamefont {A.}~\bibnamefont {Ye}}, \bibinfo {author} {\bibfnamefont {P.~B.}\ \bibnamefont {Deotare}}, \bibinfo {author} {\bibfnamefont {E.~B.}\ \bibnamefont {Tadmor}}, \emph {et~al.},\ }\bibfield  {title} {\bibinfo {title} {Torsional periodic lattice distortions and diffraction of twisted 2{D} materials},\ }\href@noop {} {\bibfield  {journal} {\bibinfo  {journal} {Nature communications}\ }\textbf {\bibinfo {volume} {13}},\ \bibinfo {pages} {7826} (\bibinfo {year} {2022})}\BibitemShut {NoStop}%
\bibitem [{\citenamefont {Nam}\ and\ \citenamefont {Koshino}(2017)}]{nam2017}%
  \BibitemOpen
  \bibfield  {author} {\bibinfo {author} {\bibfnamefont {N.~N.~T.}\ \bibnamefont {Nam}}\ and\ \bibinfo {author} {\bibfnamefont {M.}~\bibnamefont {Koshino}},\ }\bibfield  {title} {\bibinfo {title} {Lattice relaxation and energy band modulation in twisted bilayer graphene},\ }\href@noop {} {\bibfield  {journal} {\bibinfo  {journal} {Physical Review B}\ }\textbf {\bibinfo {volume} {96}},\ \bibinfo {pages} {075311} (\bibinfo {year} {2017})}\BibitemShut {NoStop}%
\bibitem [{\citenamefont {Pathak}\ \emph {et~al.}(2022)\citenamefont {Pathak}, \citenamefont {Rakib}, \citenamefont {Hou}, \citenamefont {Nevidomskyy}, \citenamefont {Ertekin}, \citenamefont {Johnson},\ and\ \citenamefont {Wagner}}]{pathak2022}%
  \BibitemOpen
  \bibfield  {author} {\bibinfo {author} {\bibfnamefont {S.}~\bibnamefont {Pathak}}, \bibinfo {author} {\bibfnamefont {T.}~\bibnamefont {Rakib}}, \bibinfo {author} {\bibfnamefont {R.}~\bibnamefont {Hou}}, \bibinfo {author} {\bibfnamefont {A.}~\bibnamefont {Nevidomskyy}}, \bibinfo {author} {\bibfnamefont {E.}~\bibnamefont {Ertekin}}, \bibinfo {author} {\bibfnamefont {H.~T.}\ \bibnamefont {Johnson}},\ and\ \bibinfo {author} {\bibfnamefont {L.~K.}\ \bibnamefont {Wagner}},\ }\bibfield  {title} {\bibinfo {title} {Accurate tight-binding model for twisted bilayer graphene describes topological flat bands without geometric relaxation},\ }\href@noop {} {\bibfield  {journal} {\bibinfo  {journal} {Physical Review B}\ }\textbf {\bibinfo {volume} {105}},\ \bibinfo {pages} {115141} (\bibinfo {year} {2022})}\BibitemShut {NoStop}%
\bibitem [{\citenamefont {Cao}\ \emph {et~al.}(2018)\citenamefont {Cao}, \citenamefont {Fatemi}, \citenamefont {Fang}, \citenamefont {Watanabe}, \citenamefont {Taniguchi}, \citenamefont {Kaxiras},\ and\ \citenamefont {Jarillo-Herrero}}]{cao2018}%
  \BibitemOpen
  \bibfield  {author} {\bibinfo {author} {\bibfnamefont {Y.}~\bibnamefont {Cao}}, \bibinfo {author} {\bibfnamefont {V.}~\bibnamefont {Fatemi}}, \bibinfo {author} {\bibfnamefont {S.}~\bibnamefont {Fang}}, \bibinfo {author} {\bibfnamefont {K.}~\bibnamefont {Watanabe}}, \bibinfo {author} {\bibfnamefont {T.}~\bibnamefont {Taniguchi}}, \bibinfo {author} {\bibfnamefont {E.}~\bibnamefont {Kaxiras}},\ and\ \bibinfo {author} {\bibfnamefont {P.}~\bibnamefont {Jarillo-Herrero}},\ }\bibfield  {title} {\bibinfo {title} {Unconventional superconductivity in magic-angle graphene superlattices},\ }\href@noop {} {\bibfield  {journal} {\bibinfo  {journal} {Nature}\ }\textbf {\bibinfo {volume} {556}},\ \bibinfo {pages} {43} (\bibinfo {year} {2018})}\BibitemShut {NoStop}%
\bibitem [{\citenamefont {Lu}\ \emph {et~al.}(2019)\citenamefont {Lu}, \citenamefont {Stepanov}, \citenamefont {Yang}, \citenamefont {Xie}, \citenamefont {Aamir}, \citenamefont {Das}, \citenamefont {Urgell}, \citenamefont {Watanabe}, \citenamefont {Taniguchi}, \citenamefont {Zhang} \emph {et~al.}}]{lu2019}%
  \BibitemOpen
  \bibfield  {author} {\bibinfo {author} {\bibfnamefont {X.}~\bibnamefont {Lu}}, \bibinfo {author} {\bibfnamefont {P.}~\bibnamefont {Stepanov}}, \bibinfo {author} {\bibfnamefont {W.}~\bibnamefont {Yang}}, \bibinfo {author} {\bibfnamefont {M.}~\bibnamefont {Xie}}, \bibinfo {author} {\bibfnamefont {M.~A.}\ \bibnamefont {Aamir}}, \bibinfo {author} {\bibfnamefont {I.}~\bibnamefont {Das}}, \bibinfo {author} {\bibfnamefont {C.}~\bibnamefont {Urgell}}, \bibinfo {author} {\bibfnamefont {K.}~\bibnamefont {Watanabe}}, \bibinfo {author} {\bibfnamefont {T.}~\bibnamefont {Taniguchi}}, \bibinfo {author} {\bibfnamefont {G.}~\bibnamefont {Zhang}}, \emph {et~al.},\ }\bibfield  {title} {\bibinfo {title} {Superconductors, orbital magnets and correlated states in magic-angle bilayer graphene},\ }\href@noop {} {\bibfield  {journal} {\bibinfo  {journal} {Nature}\ }\textbf {\bibinfo {volume} {574}},\ \bibinfo {pages} {653} (\bibinfo {year} {2019})}\BibitemShut {NoStop}%
\bibitem [{\citenamefont {Yankowitz}\ \emph {et~al.}(2019)\citenamefont {Yankowitz}, \citenamefont {Chen}, \citenamefont {Polshyn}, \citenamefont {Zhang}, \citenamefont {Watanabe}, \citenamefont {Taniguchi}, \citenamefont {Graf}, \citenamefont {Young},\ and\ \citenamefont {Dean}}]{yankowitz2019}%
  \BibitemOpen
  \bibfield  {author} {\bibinfo {author} {\bibfnamefont {M.}~\bibnamefont {Yankowitz}}, \bibinfo {author} {\bibfnamefont {S.}~\bibnamefont {Chen}}, \bibinfo {author} {\bibfnamefont {H.}~\bibnamefont {Polshyn}}, \bibinfo {author} {\bibfnamefont {Y.}~\bibnamefont {Zhang}}, \bibinfo {author} {\bibfnamefont {K.}~\bibnamefont {Watanabe}}, \bibinfo {author} {\bibfnamefont {T.}~\bibnamefont {Taniguchi}}, \bibinfo {author} {\bibfnamefont {D.}~\bibnamefont {Graf}}, \bibinfo {author} {\bibfnamefont {A.~F.}\ \bibnamefont {Young}},\ and\ \bibinfo {author} {\bibfnamefont {C.~R.}\ \bibnamefont {Dean}},\ }\bibfield  {title} {\bibinfo {title} {Tuning superconductivity in twisted bilayer graphene},\ }\href@noop {} {\bibfield  {journal} {\bibinfo  {journal} {Science}\ }\textbf {\bibinfo {volume} {363}},\ \bibinfo {pages} {1059} (\bibinfo {year} {2019})}\BibitemShut {NoStop}%
\bibitem [{\citenamefont {de~Jong}\ \emph {et~al.}(2022)\citenamefont {de~Jong}, \citenamefont {Benschop}, \citenamefont {Chen}, \citenamefont {Krasovskii}, \citenamefont {de~Dood}, \citenamefont {Tromp}, \citenamefont {Allan},\ and\ \citenamefont {Van~der Molen}}]{de2022}%
  \BibitemOpen
  \bibfield  {author} {\bibinfo {author} {\bibfnamefont {T.~A.}\ \bibnamefont {de~Jong}}, \bibinfo {author} {\bibfnamefont {T.}~\bibnamefont {Benschop}}, \bibinfo {author} {\bibfnamefont {X.}~\bibnamefont {Chen}}, \bibinfo {author} {\bibfnamefont {E.~E.}\ \bibnamefont {Krasovskii}}, \bibinfo {author} {\bibfnamefont {M.~J.}\ \bibnamefont {de~Dood}}, \bibinfo {author} {\bibfnamefont {R.~M.}\ \bibnamefont {Tromp}}, \bibinfo {author} {\bibfnamefont {M.~P.}\ \bibnamefont {Allan}},\ and\ \bibinfo {author} {\bibfnamefont {S.~J.}\ \bibnamefont {Van~der Molen}},\ }\bibfield  {title} {\bibinfo {title} {Imaging moir{\'e} deformation and dynamics in twisted bilayer graphene},\ }\href@noop {} {\bibfield  {journal} {\bibinfo  {journal} {Nature Communications}\ }\textbf {\bibinfo {volume} {13}},\ \bibinfo {pages} {1} (\bibinfo {year} {2022})}\BibitemShut {NoStop}%
\bibitem [{\citenamefont {Gadelha}\ \emph {et~al.}(2021)\citenamefont {Gadelha}, \citenamefont {Ohlberg}, \citenamefont {Rabelo}, \citenamefont {Neto}, \citenamefont {Vasconcelos}, \citenamefont {Campos}, \citenamefont {Lemos}, \citenamefont {Ornelas}, \citenamefont {Miranda}, \citenamefont {Nadas} \emph {et~al.}}]{gadelha2021}%
  \BibitemOpen
  \bibfield  {author} {\bibinfo {author} {\bibfnamefont {A.~C.}\ \bibnamefont {Gadelha}}, \bibinfo {author} {\bibfnamefont {D.~A.}\ \bibnamefont {Ohlberg}}, \bibinfo {author} {\bibfnamefont {C.}~\bibnamefont {Rabelo}}, \bibinfo {author} {\bibfnamefont {E.~G.}\ \bibnamefont {Neto}}, \bibinfo {author} {\bibfnamefont {T.~L.}\ \bibnamefont {Vasconcelos}}, \bibinfo {author} {\bibfnamefont {J.~L.}\ \bibnamefont {Campos}}, \bibinfo {author} {\bibfnamefont {J.~S.}\ \bibnamefont {Lemos}}, \bibinfo {author} {\bibfnamefont {V.}~\bibnamefont {Ornelas}}, \bibinfo {author} {\bibfnamefont {D.}~\bibnamefont {Miranda}}, \bibinfo {author} {\bibfnamefont {R.}~\bibnamefont {Nadas}}, \emph {et~al.},\ }\bibfield  {title} {\bibinfo {title} {Localization of lattice dynamics in low-angle twisted bilayer graphene},\ }\href@noop {} {\bibfield  {journal} {\bibinfo  {journal} {Nature}\ }\textbf {\bibinfo {volume} {590}},\ \bibinfo {pages} {405} (\bibinfo {year} {2021})}\BibitemShut {NoStop}%
\bibitem [{\citenamefont {Cantele}\ \emph {et~al.}(2020)\citenamefont {Cantele}, \citenamefont {Alfe}, \citenamefont {Conte}, \citenamefont {Cataudella}, \citenamefont {Ninno},\ and\ \citenamefont {Lucignano}}]{cantele2020}%
  \BibitemOpen
  \bibfield  {author} {\bibinfo {author} {\bibfnamefont {G.}~\bibnamefont {Cantele}}, \bibinfo {author} {\bibfnamefont {D.}~\bibnamefont {Alfe}}, \bibinfo {author} {\bibfnamefont {F.}~\bibnamefont {Conte}}, \bibinfo {author} {\bibfnamefont {V.}~\bibnamefont {Cataudella}}, \bibinfo {author} {\bibfnamefont {D.}~\bibnamefont {Ninno}},\ and\ \bibinfo {author} {\bibfnamefont {P.}~\bibnamefont {Lucignano}},\ }\bibfield  {title} {\bibinfo {title} {Structural relaxation and low-energy properties of twisted bilayer graphene},\ }\href@noop {} {\bibfield  {journal} {\bibinfo  {journal} {Physical Review Research}\ }\textbf {\bibinfo {volume} {2}},\ \bibinfo {pages} {043127} (\bibinfo {year} {2020})}\BibitemShut {NoStop}%
\bibitem [{\citenamefont {Leconte}\ \emph {et~al.}(2022)\citenamefont {Leconte}, \citenamefont {Javvaji}, \citenamefont {An}, \citenamefont {Samudrala},\ and\ \citenamefont {Jung}}]{leconte2022}%
  \BibitemOpen
  \bibfield  {author} {\bibinfo {author} {\bibfnamefont {N.}~\bibnamefont {Leconte}}, \bibinfo {author} {\bibfnamefont {S.}~\bibnamefont {Javvaji}}, \bibinfo {author} {\bibfnamefont {J.}~\bibnamefont {An}}, \bibinfo {author} {\bibfnamefont {A.}~\bibnamefont {Samudrala}},\ and\ \bibinfo {author} {\bibfnamefont {J.}~\bibnamefont {Jung}},\ }\bibfield  {title} {\bibinfo {title} {Relaxation effects in twisted bilayer graphene: a multiscale approach},\ }\href@noop {} {\bibfield  {journal} {\bibinfo  {journal} {Physical Review B}\ }\textbf {\bibinfo {volume} {106}},\ \bibinfo {pages} {115410} (\bibinfo {year} {2022})}\BibitemShut {NoStop}%
\bibitem [{\citenamefont {Lu}\ \emph {et~al.}(2022)\citenamefont {Lu}, \citenamefont {Zhu}, \citenamefont {Angeli}, \citenamefont {Larson},\ and\ \citenamefont {Kaxiras}}]{lu2022}%
  \BibitemOpen
  \bibfield  {author} {\bibinfo {author} {\bibfnamefont {J.~Z.}\ \bibnamefont {Lu}}, \bibinfo {author} {\bibfnamefont {Z.}~\bibnamefont {Zhu}}, \bibinfo {author} {\bibfnamefont {M.}~\bibnamefont {Angeli}}, \bibinfo {author} {\bibfnamefont {D.~T.}\ \bibnamefont {Larson}},\ and\ \bibinfo {author} {\bibfnamefont {E.}~\bibnamefont {Kaxiras}},\ }\bibfield  {title} {\bibinfo {title} {Low-energy moir{\'e} phonons in twisted bilayer van der waals heterostructures},\ }\href@noop {} {\bibfield  {journal} {\bibinfo  {journal} {Physical Review B}\ }\textbf {\bibinfo {volume} {106}},\ \bibinfo {pages} {144305} (\bibinfo {year} {2022})}\BibitemShut {NoStop}%
\bibitem [{\citenamefont {Annevelink}\ \emph {et~al.}(2020)\citenamefont {Annevelink}, \citenamefont {Johnson},\ and\ \citenamefont {Ertekin}}]{annevelink2020}%
  \BibitemOpen
  \bibfield  {author} {\bibinfo {author} {\bibfnamefont {E.}~\bibnamefont {Annevelink}}, \bibinfo {author} {\bibfnamefont {H.~T.}\ \bibnamefont {Johnson}},\ and\ \bibinfo {author} {\bibfnamefont {E.}~\bibnamefont {Ertekin}},\ }\bibfield  {title} {\bibinfo {title} {Topologically derived dislocation theory for twist and stretch moir{\'e} superlattices in bilayer graphene},\ }\href@noop {} {\bibfield  {journal} {\bibinfo  {journal} {Physical Review B}\ }\textbf {\bibinfo {volume} {102}},\ \bibinfo {pages} {184107} (\bibinfo {year} {2020})}\BibitemShut {NoStop}%
\bibitem [{\citenamefont {Stuart}\ \emph {et~al.}(2000)\citenamefont {Stuart}, \citenamefont {Tutein},\ and\ \citenamefont {Harrison}}]{stuart2000}%
  \BibitemOpen
  \bibfield  {author} {\bibinfo {author} {\bibfnamefont {S.~J.}\ \bibnamefont {Stuart}}, \bibinfo {author} {\bibfnamefont {A.~B.}\ \bibnamefont {Tutein}},\ and\ \bibinfo {author} {\bibfnamefont {J.~A.}\ \bibnamefont {Harrison}},\ }\bibfield  {title} {\bibinfo {title} {A reactive potential for hydrocarbons with intermolecular interactions},\ }\href@noop {} {\bibfield  {journal} {\bibinfo  {journal} {The Journal of chemical physics}\ }\textbf {\bibinfo {volume} {112}},\ \bibinfo {pages} {6472} (\bibinfo {year} {2000})}\BibitemShut {NoStop}%
\bibitem [{\citenamefont {Wen}\ \emph {et~al.}(2018)\citenamefont {Wen}, \citenamefont {Carr}, \citenamefont {Fang}, \citenamefont {Kaxiras},\ and\ \citenamefont {Tadmor}}]{wen2018}%
  \BibitemOpen
  \bibfield  {author} {\bibinfo {author} {\bibfnamefont {M.}~\bibnamefont {Wen}}, \bibinfo {author} {\bibfnamefont {S.}~\bibnamefont {Carr}}, \bibinfo {author} {\bibfnamefont {S.}~\bibnamefont {Fang}}, \bibinfo {author} {\bibfnamefont {E.}~\bibnamefont {Kaxiras}},\ and\ \bibinfo {author} {\bibfnamefont {E.~B.}\ \bibnamefont {Tadmor}},\ }\bibfield  {title} {\bibinfo {title} {Dihedral-angle-corrected registry-dependent interlayer potential for multilayer graphene structures},\ }\href@noop {} {\bibfield  {journal} {\bibinfo  {journal} {Physical Review B}\ }\textbf {\bibinfo {volume} {98}},\ \bibinfo {pages} {235404} (\bibinfo {year} {2018})}\BibitemShut {NoStop}%
\bibitem [{\citenamefont {Wen}\ and\ \citenamefont {Tadmor}(2019)}]{wen2019}%
  \BibitemOpen
  \bibfield  {author} {\bibinfo {author} {\bibfnamefont {M.}~\bibnamefont {Wen}}\ and\ \bibinfo {author} {\bibfnamefont {E.~B.}\ \bibnamefont {Tadmor}},\ }\bibfield  {title} {\bibinfo {title} {Hybrid neural network potential for multilayer graphene},\ }\href@noop {} {\bibfield  {journal} {\bibinfo  {journal} {Physical Review B}\ }\textbf {\bibinfo {volume} {100}},\ \bibinfo {pages} {195419} (\bibinfo {year} {2019})}\BibitemShut {NoStop}%
\bibitem [{\citenamefont {Larsen}\ \emph {et~al.}(2017)\citenamefont {Larsen}, \citenamefont {Mortensen}, \citenamefont {Blomqvist}, \citenamefont {Castelli}, \citenamefont {Christensen}, \citenamefont {Du{\l}ak}, \citenamefont {Friis}, \citenamefont {Groves}, \citenamefont {Hammer}, \citenamefont {Hargus} \emph {et~al.}}]{larsen2017}%
  \BibitemOpen
  \bibfield  {author} {\bibinfo {author} {\bibfnamefont {A.~H.}\ \bibnamefont {Larsen}}, \bibinfo {author} {\bibfnamefont {J.~J.}\ \bibnamefont {Mortensen}}, \bibinfo {author} {\bibfnamefont {J.}~\bibnamefont {Blomqvist}}, \bibinfo {author} {\bibfnamefont {I.~E.}\ \bibnamefont {Castelli}}, \bibinfo {author} {\bibfnamefont {R.}~\bibnamefont {Christensen}}, \bibinfo {author} {\bibfnamefont {M.}~\bibnamefont {Du{\l}ak}}, \bibinfo {author} {\bibfnamefont {J.}~\bibnamefont {Friis}}, \bibinfo {author} {\bibfnamefont {M.~N.}\ \bibnamefont {Groves}}, \bibinfo {author} {\bibfnamefont {B.}~\bibnamefont {Hammer}}, \bibinfo {author} {\bibfnamefont {C.}~\bibnamefont {Hargus}}, \emph {et~al.},\ }\bibfield  {title} {\bibinfo {title} {The atomic simulation environment—a python library for working with atoms},\ }\href@noop {} {\bibfield  {journal} {\bibinfo  {journal} {Journal of Physics: Condensed Matter}\ }\textbf {\bibinfo {volume} {29}},\ \bibinfo {pages} {273002} (\bibinfo {year} {2017})}\BibitemShut {NoStop}%
\bibitem [{\citenamefont {Shallcross}\ \emph {et~al.}(2010)\citenamefont {Shallcross}, \citenamefont {Sharma}, \citenamefont {Kandelaki},\ and\ \citenamefont {Pankratov}}]{shallcross2010}%
  \BibitemOpen
  \bibfield  {author} {\bibinfo {author} {\bibfnamefont {S.}~\bibnamefont {Shallcross}}, \bibinfo {author} {\bibfnamefont {S.}~\bibnamefont {Sharma}}, \bibinfo {author} {\bibfnamefont {E.}~\bibnamefont {Kandelaki}},\ and\ \bibinfo {author} {\bibfnamefont {O.~A.}\ \bibnamefont {Pankratov}},\ }\bibfield  {title} {\bibinfo {title} {Electronic structure of turbostratic graphene},\ }\href@noop {} {\bibfield  {journal} {\bibinfo  {journal} {Physical Review B}\ }\textbf {\bibinfo {volume} {81}},\ \bibinfo {pages} {165105} (\bibinfo {year} {2010})}\BibitemShut {NoStop}%
\bibitem [{\citenamefont {Cooper}\ \emph {et~al.}(2012)\citenamefont {Cooper}, \citenamefont {D'Anjou}, \citenamefont {Ghattamaneni}, \citenamefont {Harack}, \citenamefont {Hilke}, \citenamefont {Horth}, \citenamefont {Majlis}, \citenamefont {Massicotte}, \citenamefont {Vandsburger}, \citenamefont {Whiteway},\ and\ \citenamefont {Yu}}]{cooper2012}%
  \BibitemOpen
  \bibfield  {author} {\bibinfo {author} {\bibfnamefont {D.~R.}\ \bibnamefont {Cooper}}, \bibinfo {author} {\bibfnamefont {B.}~\bibnamefont {D'Anjou}}, \bibinfo {author} {\bibfnamefont {N.}~\bibnamefont {Ghattamaneni}}, \bibinfo {author} {\bibfnamefont {B.}~\bibnamefont {Harack}}, \bibinfo {author} {\bibfnamefont {M.}~\bibnamefont {Hilke}}, \bibinfo {author} {\bibfnamefont {A.}~\bibnamefont {Horth}}, \bibinfo {author} {\bibfnamefont {N.}~\bibnamefont {Majlis}}, \bibinfo {author} {\bibfnamefont {M.}~\bibnamefont {Massicotte}}, \bibinfo {author} {\bibfnamefont {L.}~\bibnamefont {Vandsburger}}, \bibinfo {author} {\bibfnamefont {E.}~\bibnamefont {Whiteway}},\ and\ \bibinfo {author} {\bibfnamefont {V.}~\bibnamefont {Yu}},\ }\bibfield  {title} {\bibinfo {title} {Experimental review of graphene},\ }\href@noop {} {\bibfield  {journal} {\bibinfo  {journal} {{ISRN} Condensed Matter Physics}\ }\textbf {\bibinfo {volume} {2012}},\ \bibinfo {pages} {1} (\bibinfo {year} {2012})}\BibitemShut {NoStop}%
\bibitem [{\citenamefont {Plimpton}(1995)}]{plimpton1995}%
  \BibitemOpen
  \bibfield  {author} {\bibinfo {author} {\bibfnamefont {S.}~\bibnamefont {Plimpton}},\ }\bibfield  {title} {\bibinfo {title} {Fast parallel algorithms for short-range molecular dynamics},\ }\href@noop {} {\bibfield  {journal} {\bibinfo  {journal} {Journal of computational physics}\ }\textbf {\bibinfo {volume} {117}},\ \bibinfo {pages} {1} (\bibinfo {year} {1995})}\BibitemShut {NoStop}%
\bibitem [{\citenamefont {Gu{\'e}nol{\'e}}\ \emph {et~al.}(2020)\citenamefont {Gu{\'e}nol{\'e}}, \citenamefont {N{\"o}hring}, \citenamefont {Vaid}, \citenamefont {Houll{\'e}}, \citenamefont {Xie}, \citenamefont {Prakash},\ and\ \citenamefont {Bitzek}}]{guenole2020}%
  \BibitemOpen
  \bibfield  {author} {\bibinfo {author} {\bibfnamefont {J.}~\bibnamefont {Gu{\'e}nol{\'e}}}, \bibinfo {author} {\bibfnamefont {W.~G.}\ \bibnamefont {N{\"o}hring}}, \bibinfo {author} {\bibfnamefont {A.}~\bibnamefont {Vaid}}, \bibinfo {author} {\bibfnamefont {F.}~\bibnamefont {Houll{\'e}}}, \bibinfo {author} {\bibfnamefont {Z.}~\bibnamefont {Xie}}, \bibinfo {author} {\bibfnamefont {A.}~\bibnamefont {Prakash}},\ and\ \bibinfo {author} {\bibfnamefont {E.}~\bibnamefont {Bitzek}},\ }\bibfield  {title} {\bibinfo {title} {Assessment and optimization of the fast inertial relaxation engine (fire) for energy minimization in atomistic simulations and its implementation in lammps},\ }\href@noop {} {\bibfield  {journal} {\bibinfo  {journal} {Computational Materials Science}\ }\textbf {\bibinfo {volume} {175}},\ \bibinfo {pages} {109584} (\bibinfo {year} {2020})}\BibitemShut {NoStop}%
\bibitem [{\citenamefont {Cai}\ and\ \citenamefont {Yu}(2021)}]{cai2021}%
  \BibitemOpen
  \bibfield  {author} {\bibinfo {author} {\bibfnamefont {L.}~\bibnamefont {Cai}}\ and\ \bibinfo {author} {\bibfnamefont {G.}~\bibnamefont {Yu}},\ }\bibfield  {title} {\bibinfo {title} {Fabrication strategies of twisted bilayer graphenes and their unique properties},\ }\href@noop {} {\bibfield  {journal} {\bibinfo  {journal} {Advanced Materials}\ }\textbf {\bibinfo {volume} {33}},\ \bibinfo {pages} {2004974} (\bibinfo {year} {2021})}\BibitemShut {NoStop}%
\bibitem [{\citenamefont {Cao}\ \emph {et~al.}(2021)\citenamefont {Cao}, \citenamefont {Wu},\ and\ \citenamefont {Sun}}]{cao2021}%
  \BibitemOpen
  \bibfield  {author} {\bibinfo {author} {\bibfnamefont {C.}~\bibnamefont {Cao}}, \bibinfo {author} {\bibfnamefont {T.}~\bibnamefont {Wu}},\ and\ \bibinfo {author} {\bibfnamefont {Y.}~\bibnamefont {Sun}},\ }\bibfield  {title} {\bibinfo {title} {A review of assembly techniques for fabricating twisted bilayer graphene},\ }\href@noop {} {\bibfield  {journal} {\bibinfo  {journal} {Journal of Micromechanics and Microengineering}\ }\textbf {\bibinfo {volume} {31}},\ \bibinfo {pages} {114004} (\bibinfo {year} {2021})}\BibitemShut {NoStop}%
\bibitem [{\citenamefont {Leven}\ \emph {et~al.}(2016)\citenamefont {Leven}, \citenamefont {Maaravi}, \citenamefont {Azuri}, \citenamefont {Kronik},\ and\ \citenamefont {Hod}}]{leven:maaravi:2016}%
  \BibitemOpen
  \bibfield  {author} {\bibinfo {author} {\bibfnamefont {I.}~\bibnamefont {Leven}}, \bibinfo {author} {\bibfnamefont {T.}~\bibnamefont {Maaravi}}, \bibinfo {author} {\bibfnamefont {I.}~\bibnamefont {Azuri}}, \bibinfo {author} {\bibfnamefont {L.}~\bibnamefont {Kronik}},\ and\ \bibinfo {author} {\bibfnamefont {O.}~\bibnamefont {Hod}},\ }\bibfield  {title} {\bibinfo {title} {Interlayer potential for {G}raphene/h-{BN} heterostructures},\ }\href@noop {} {\bibfield  {journal} {\bibinfo  {journal} {Journal of Chemical Theory and Computation}\ }\textbf {\bibinfo {volume} {12}},\ \bibinfo {pages} {2896} (\bibinfo {year} {2016})}\BibitemShut {NoStop}%
\bibitem [{\citenamefont {Baowan}\ and\ \citenamefont {Hill}(2007)}]{baowan2007}%
  \BibitemOpen
  \bibfield  {author} {\bibinfo {author} {\bibfnamefont {D.}~\bibnamefont {Baowan}}\ and\ \bibinfo {author} {\bibfnamefont {J.~M.}\ \bibnamefont {Hill}},\ }\bibfield  {title} {\bibinfo {title} {Nested boron nitride and carbon-boron nitride nanocones},\ }\href@noop {} {\bibfield  {journal} {\bibinfo  {journal} {Micro \& Nano Letters}\ }\textbf {\bibinfo {volume} {2}},\ \bibinfo {pages} {46} (\bibinfo {year} {2007})}\BibitemShut {NoStop}%
\bibitem [{\citenamefont {Mayo}\ \emph {et~al.}(1990)\citenamefont {Mayo}, \citenamefont {Olafson},\ and\ \citenamefont {Goddard}}]{mayo1990}%
  \BibitemOpen
  \bibfield  {author} {\bibinfo {author} {\bibfnamefont {S.~L.}\ \bibnamefont {Mayo}}, \bibinfo {author} {\bibfnamefont {B.~D.}\ \bibnamefont {Olafson}},\ and\ \bibinfo {author} {\bibfnamefont {W.~A.}\ \bibnamefont {Goddard}},\ }\bibfield  {title} {\bibinfo {title} {Dreiding: a generic force field for molecular simulations},\ }\href@noop {} {\bibfield  {journal} {\bibinfo  {journal} {Journal of Physical chemistry}\ }\textbf {\bibinfo {volume} {94}},\ \bibinfo {pages} {8897} (\bibinfo {year} {1990})}\BibitemShut {NoStop}%
\bibitem [{\citenamefont {Heinz}\ \emph {et~al.}(2013)\citenamefont {Heinz}, \citenamefont {Lin}, \citenamefont {Kishore~Mishra},\ and\ \citenamefont {Emami}}]{heinz2013}%
  \BibitemOpen
  \bibfield  {author} {\bibinfo {author} {\bibfnamefont {H.}~\bibnamefont {Heinz}}, \bibinfo {author} {\bibfnamefont {T.-J.}\ \bibnamefont {Lin}}, \bibinfo {author} {\bibfnamefont {R.}~\bibnamefont {Kishore~Mishra}},\ and\ \bibinfo {author} {\bibfnamefont {F.~S.}\ \bibnamefont {Emami}},\ }\bibfield  {title} {\bibinfo {title} {Thermodynamically consistent force fields for the assembly of inorganic, organic, and biological nanostructures: the interface force field},\ }\href@noop {} {\bibfield  {journal} {\bibinfo  {journal} {Langmuir}\ }\textbf {\bibinfo {volume} {29}},\ \bibinfo {pages} {1754} (\bibinfo {year} {2013})}\BibitemShut {NoStop}%
\bibitem [{\citenamefont {Arreola-Lucas}\ \emph {et~al.}(2015)\citenamefont {Arreola-Lucas}, \citenamefont {Franco-Villafa{\~n}e}, \citenamefont {B{\'a}ez},\ and\ \citenamefont {M{\'e}ndez-S{\'a}nchez}}]{arreola2015}%
  \BibitemOpen
  \bibfield  {author} {\bibinfo {author} {\bibfnamefont {A.}~\bibnamefont {Arreola-Lucas}}, \bibinfo {author} {\bibfnamefont {J.}~\bibnamefont {Franco-Villafa{\~n}e}}, \bibinfo {author} {\bibfnamefont {G.}~\bibnamefont {B{\'a}ez}},\ and\ \bibinfo {author} {\bibfnamefont {R.}~\bibnamefont {M{\'e}ndez-S{\'a}nchez}},\ }\bibfield  {title} {\bibinfo {title} {In-plane vibrations of a rectangular plate: plane wave expansion modeling and experiment},\ }\href@noop {} {\bibfield  {journal} {\bibinfo  {journal} {Journal of Sound and Vibration}\ }\textbf {\bibinfo {volume} {342}},\ \bibinfo {pages} {168} (\bibinfo {year} {2015})}\BibitemShut {NoStop}%
\bibitem [{\citenamefont {Griffin}\ \emph {et~al.}(1974)\citenamefont {Griffin}, \citenamefont {Nagel},\ and\ \citenamefont {Koshel}}]{griffin1974}%
  \BibitemOpen
  \bibfield  {author} {\bibinfo {author} {\bibfnamefont {P.}~\bibnamefont {Griffin}}, \bibinfo {author} {\bibfnamefont {P.}~\bibnamefont {Nagel}},\ and\ \bibinfo {author} {\bibfnamefont {R.}~\bibnamefont {Koshel}},\ }\bibfield  {title} {\bibinfo {title} {The plane-wave expansion method},\ }\href@noop {} {\bibfield  {journal} {\bibinfo  {journal} {Journal of Mathematical Physics}\ }\textbf {\bibinfo {volume} {15}},\ \bibinfo {pages} {1913} (\bibinfo {year} {1974})}\BibitemShut {NoStop}%
\bibitem [{\citenamefont {Manzanares-Mart{\'\i}nez}\ \emph {et~al.}(2010)\citenamefont {Manzanares-Mart{\'\i}nez}, \citenamefont {Flores}, \citenamefont {Guti{\'e}rrez}, \citenamefont {M{\'e}ndez-S{\'a}nchez}, \citenamefont {Monsivais}, \citenamefont {Morales},\ and\ \citenamefont {Ramos-Mendieta}}]{manzanares2010}%
  \BibitemOpen
  \bibfield  {author} {\bibinfo {author} {\bibfnamefont {B.}~\bibnamefont {Manzanares-Mart{\'\i}nez}}, \bibinfo {author} {\bibfnamefont {J.}~\bibnamefont {Flores}}, \bibinfo {author} {\bibfnamefont {L.}~\bibnamefont {Guti{\'e}rrez}}, \bibinfo {author} {\bibfnamefont {R.}~\bibnamefont {M{\'e}ndez-S{\'a}nchez}}, \bibinfo {author} {\bibfnamefont {G.}~\bibnamefont {Monsivais}}, \bibinfo {author} {\bibfnamefont {A.}~\bibnamefont {Morales}},\ and\ \bibinfo {author} {\bibfnamefont {F.}~\bibnamefont {Ramos-Mendieta}},\ }\bibfield  {title} {\bibinfo {title} {Flexural vibrations of a rectangular plate for the lower normal modes},\ }\href@noop {} {\bibfield  {journal} {\bibinfo  {journal} {Journal of sound and vibration}\ }\textbf {\bibinfo {volume} {329}},\ \bibinfo {pages} {5105} (\bibinfo {year} {2010})}\BibitemShut {NoStop}%
\bibitem [{\citenamefont {Choi}\ \emph {et~al.}(2023)\citenamefont {Choi}, \citenamefont {Pasetto}, \citenamefont {Shen}, \citenamefont {Tadmor},\ and\ \citenamefont {Kamensky}}]{choi2023}%
  \BibitemOpen
  \bibfield  {author} {\bibinfo {author} {\bibfnamefont {M.-k.}\ \bibnamefont {Choi}}, \bibinfo {author} {\bibfnamefont {M.}~\bibnamefont {Pasetto}}, \bibinfo {author} {\bibfnamefont {Z.}~\bibnamefont {Shen}}, \bibinfo {author} {\bibfnamefont {E.~B.}\ \bibnamefont {Tadmor}},\ and\ \bibinfo {author} {\bibfnamefont {D.}~\bibnamefont {Kamensky}},\ }\bibfield  {title} {\bibinfo {title} {Atomistically-informed continuum modeling and isogeometric analysis of 2{D} materials over holey substrates},\ }\href@noop {} {\bibfield  {journal} {\bibinfo  {journal} {Journal of the Mechanics and Physics of Solids}\ }\textbf {\bibinfo {volume} {170}},\ \bibinfo {pages} {105100} (\bibinfo {year} {2023})}\BibitemShut {NoStop}%
\bibitem [{\citenamefont {Lee}\ \emph {et~al.}(2008)\citenamefont {Lee}, \citenamefont {Wei}, \citenamefont {Kysar},\ and\ \citenamefont {Hone}}]{lee2008}%
  \BibitemOpen
  \bibfield  {author} {\bibinfo {author} {\bibfnamefont {C.}~\bibnamefont {Lee}}, \bibinfo {author} {\bibfnamefont {X.}~\bibnamefont {Wei}}, \bibinfo {author} {\bibfnamefont {J.~W.}\ \bibnamefont {Kysar}},\ and\ \bibinfo {author} {\bibfnamefont {J.}~\bibnamefont {Hone}},\ }\bibfield  {title} {\bibinfo {title} {Measurement of the elastic properties and intrinsic strength of monolayer graphene},\ }\href@noop {} {\bibfield  {journal} {\bibinfo  {journal} {science}\ }\textbf {\bibinfo {volume} {321}},\ \bibinfo {pages} {385} (\bibinfo {year} {2008})}\BibitemShut {NoStop}%
\bibitem [{\citenamefont {Zhang}\ and\ \citenamefont {Pan}(2012)}]{zhang2012}%
  \BibitemOpen
  \bibfield  {author} {\bibinfo {author} {\bibfnamefont {Y.}~\bibnamefont {Zhang}}\ and\ \bibinfo {author} {\bibfnamefont {C.}~\bibnamefont {Pan}},\ }\bibfield  {title} {\bibinfo {title} {Measurements of mechanical properties and number of layers of graphene from nano-indentation},\ }\href@noop {} {\bibfield  {journal} {\bibinfo  {journal} {Diamond and related materials}\ }\textbf {\bibinfo {volume} {24}},\ \bibinfo {pages} {1} (\bibinfo {year} {2012})}\BibitemShut {NoStop}%
\bibitem [{\citenamefont {Han}\ \emph {et~al.}(2020)\citenamefont {Han}, \citenamefont {Yu}, \citenamefont {Annevelink}, \citenamefont {Son}, \citenamefont {Kang}, \citenamefont {Watanabe}, \citenamefont {Taniguchi}, \citenamefont {Ertekin}, \citenamefont {Huang},\ and\ \citenamefont {van~der Zande}}]{han2020}%
  \BibitemOpen
  \bibfield  {author} {\bibinfo {author} {\bibfnamefont {E.}~\bibnamefont {Han}}, \bibinfo {author} {\bibfnamefont {J.}~\bibnamefont {Yu}}, \bibinfo {author} {\bibfnamefont {E.}~\bibnamefont {Annevelink}}, \bibinfo {author} {\bibfnamefont {J.}~\bibnamefont {Son}}, \bibinfo {author} {\bibfnamefont {D.~A.}\ \bibnamefont {Kang}}, \bibinfo {author} {\bibfnamefont {K.}~\bibnamefont {Watanabe}}, \bibinfo {author} {\bibfnamefont {T.}~\bibnamefont {Taniguchi}}, \bibinfo {author} {\bibfnamefont {E.}~\bibnamefont {Ertekin}}, \bibinfo {author} {\bibfnamefont {P.~Y.}\ \bibnamefont {Huang}},\ and\ \bibinfo {author} {\bibfnamefont {A.~M.}\ \bibnamefont {van~der Zande}},\ }\bibfield  {title} {\bibinfo {title} {Ultrasoft slip-mediated bending in few-layer graphene},\ }\href@noop {} {\bibfield  {journal} {\bibinfo  {journal} {Nature materials}\ }\textbf {\bibinfo {volume} {19}},\ \bibinfo {pages} {305} (\bibinfo {year} {2020})}\BibitemShut {NoStop}%
\bibitem [{\citenamefont {Blees}\ \emph {et~al.}(2015)\citenamefont {Blees}, \citenamefont {Barnard}, \citenamefont {Rose}, \citenamefont {Roberts}, \citenamefont {McGill}, \citenamefont {Huang}, \citenamefont {Ruyack}, \citenamefont {Kevek}, \citenamefont {Kobrin}, \citenamefont {Muller} \emph {et~al.}}]{blees2015}%
  \BibitemOpen
  \bibfield  {author} {\bibinfo {author} {\bibfnamefont {M.~K.}\ \bibnamefont {Blees}}, \bibinfo {author} {\bibfnamefont {A.~W.}\ \bibnamefont {Barnard}}, \bibinfo {author} {\bibfnamefont {P.~A.}\ \bibnamefont {Rose}}, \bibinfo {author} {\bibfnamefont {S.~P.}\ \bibnamefont {Roberts}}, \bibinfo {author} {\bibfnamefont {K.~L.}\ \bibnamefont {McGill}}, \bibinfo {author} {\bibfnamefont {P.~Y.}\ \bibnamefont {Huang}}, \bibinfo {author} {\bibfnamefont {A.~R.}\ \bibnamefont {Ruyack}}, \bibinfo {author} {\bibfnamefont {J.~W.}\ \bibnamefont {Kevek}}, \bibinfo {author} {\bibfnamefont {B.}~\bibnamefont {Kobrin}}, \bibinfo {author} {\bibfnamefont {D.~A.}\ \bibnamefont {Muller}}, \emph {et~al.},\ }\bibfield  {title} {\bibinfo {title} {Graphene kirigami},\ }\href@noop {} {\bibfield  {journal} {\bibinfo  {journal} {Nature}\ }\textbf {\bibinfo {volume} {524}},\ \bibinfo {pages} {204} (\bibinfo {year} {2015})}\BibitemShut {NoStop}%
\bibitem [{SM()}]{SM}%
  \BibitemOpen
  \href@noop {} {}\bibinfo {note} {See Supplemental Material}\BibitemShut {NoStop}%
\bibitem [{\citenamefont {Lee}\ \emph {et~al.}(2012)\citenamefont {Lee}, \citenamefont {Yoon},\ and\ \citenamefont {Cheong}}]{lee2012}%
  \BibitemOpen
  \bibfield  {author} {\bibinfo {author} {\bibfnamefont {J.-U.}\ \bibnamefont {Lee}}, \bibinfo {author} {\bibfnamefont {D.}~\bibnamefont {Yoon}},\ and\ \bibinfo {author} {\bibfnamefont {H.}~\bibnamefont {Cheong}},\ }\bibfield  {title} {\bibinfo {title} {Estimation of young’s modulus of graphene by raman spectroscopy},\ }\href@noop {} {\bibfield  {journal} {\bibinfo  {journal} {Nano letters}\ }\textbf {\bibinfo {volume} {12}},\ \bibinfo {pages} {4444} (\bibinfo {year} {2012})}\BibitemShut {NoStop}%
\bibitem [{\citenamefont {Hovden}\ \emph {et~al.}(2016)\citenamefont {Hovden}, \citenamefont {Tsen}, \citenamefont {Liu}, \citenamefont {Savitzky}, \citenamefont {El~Baggari}, \citenamefont {Liu}, \citenamefont {Lu}, \citenamefont {Sun}, \citenamefont {Kim}, \citenamefont {Pasupathy} \emph {et~al.}}]{hovden2016}%
  \BibitemOpen
  \bibfield  {author} {\bibinfo {author} {\bibfnamefont {R.}~\bibnamefont {Hovden}}, \bibinfo {author} {\bibfnamefont {A.~W.}\ \bibnamefont {Tsen}}, \bibinfo {author} {\bibfnamefont {P.}~\bibnamefont {Liu}}, \bibinfo {author} {\bibfnamefont {B.~H.}\ \bibnamefont {Savitzky}}, \bibinfo {author} {\bibfnamefont {I.}~\bibnamefont {El~Baggari}}, \bibinfo {author} {\bibfnamefont {Y.}~\bibnamefont {Liu}}, \bibinfo {author} {\bibfnamefont {W.}~\bibnamefont {Lu}}, \bibinfo {author} {\bibfnamefont {Y.}~\bibnamefont {Sun}}, \bibinfo {author} {\bibfnamefont {P.}~\bibnamefont {Kim}}, \bibinfo {author} {\bibfnamefont {A.~N.}\ \bibnamefont {Pasupathy}}, \emph {et~al.},\ }\bibfield  {title} {\bibinfo {title} {Atomic lattice disorder in charge-density-wave phases of exfoliated dichalcogenides {(1T-TaS$_2$)}},\ }\href@noop {} {\bibfield  {journal} {\bibinfo  {journal} {Proceedings of the National Academy of Sciences}\ }\textbf {\bibinfo {volume} {113}},\ \bibinfo {pages} {11420} (\bibinfo {year} {2016})}\BibitemShut {NoStop}%
\bibitem [{\citenamefont {Kong}(2011)}]{kong2011}%
  \BibitemOpen
  \bibfield  {author} {\bibinfo {author} {\bibfnamefont {L.~T.}\ \bibnamefont {Kong}},\ }\bibfield  {title} {\bibinfo {title} {Phonon dispersion measured directly from molecular dynamics simulations},\ }\href@noop {} {\bibfield  {journal} {\bibinfo  {journal} {Computer Physics Communications}\ }\textbf {\bibinfo {volume} {182}},\ \bibinfo {pages} {2201} (\bibinfo {year} {2011})}\BibitemShut {NoStop}%
\bibitem [{\citenamefont {Rakib}\ \emph {et~al.}(2022)\citenamefont {Rakib}, \citenamefont {Pochet}, \citenamefont {Ertekin},\ and\ \citenamefont {Johnson}}]{rakib2022}%
  \BibitemOpen
  \bibfield  {author} {\bibinfo {author} {\bibfnamefont {T.}~\bibnamefont {Rakib}}, \bibinfo {author} {\bibfnamefont {P.}~\bibnamefont {Pochet}}, \bibinfo {author} {\bibfnamefont {E.}~\bibnamefont {Ertekin}},\ and\ \bibinfo {author} {\bibfnamefont {H.~T.}\ \bibnamefont {Johnson}},\ }\bibfield  {title} {\bibinfo {title} {Corrugation-driven symmetry breaking in magic-angle twisted bilayer graphene},\ }\href@noop {} {\bibfield  {journal} {\bibinfo  {journal} {Communications Physics}\ }\textbf {\bibinfo {volume} {5}},\ \bibinfo {pages} {242} (\bibinfo {year} {2022})}\BibitemShut {NoStop}%
\end{thebibliography}%
\end{document}



\title{Supplementary Material for ``Elastic plate basis for the deformation and electron diffraction of twisted bilayer graphene on a substrate''}

\author{Moon-ki Choi}
\affiliation{Department of Aerospace Engineering and Mechanics, University of Minnesota, Minneapolis, Minnesota 55455, USA}

\author{Suk Hyun Sung}
\affiliation{Department of Materials Science and Engineering, University of Michigan, Ann Arbor, Michigan 48109, USA}

\author{Robert Hovden}
\affiliation{Department of Materials Science and Engineering, University of Michigan, Ann Arbor, Michigan 48109, USA}

\author{Ellad B. Tadmor}
\email[Corresponding author: ]{tadmor@umn.edu}
\affiliation{Department of Aerospace Engineering and Mechanics, University of Minnesota, Minneapolis, Minnesota 55455, USA}

\maketitle

\onecolumngrid
\section{Orthogonality of in-plane and out-of-plane modes}\label{sec:app:ortho_basis}

The inner product of the real parts of in-plane modes $k_1$ and $k_2$ (\eqn{eqn:in_plane_mode} in main text) is defined as an integral over the supercell domain,
\begin{equation}\label{eqn:int_ortho_in_1}
    I_{\rm in} = \bd_{k_1}\cdot\bd_{k_2}\int_S  \cos{(\bar{\bG}_{k_1}\cdot\mathbf{r}+\phi_{\bar{\bG}_{k_1}})}
    \cos{(\bar{\bG}_{k_2}\cdot\mathbf{r}+\phi_{\bar{\bG}_{k_2}})}\,dS,
\end{equation}
where $S$ is the TBG supercell domain. To perform the integration, two variables $u=x$ and $v=y+\frac{1}{\sqrt{3}}x$ are introduced. Using $u$ and $v$, $I_{\rm in}$ is reformulated as
\begin{multline}\label{eqn:int_ortho_in_2}
    I_{\rm in} =  \bd_{k_1}\cdot\bd_{k_2}\int_{0}^{L}\int_{0}^{\frac{\sqrt{3}}{2}L}
    \cos{(\bar{G}_{k_1}^x u + \bar{G}_{k_1}^y (v-\frac{1}{\sqrt{3}}u) +\phi_{\bar{\bG}_{k_1}})}
    \\
    \cos{(\bar{G}_{k_2}^x u + \bar{G}_{k_2}^y (v-\frac{1}{\sqrt{3}}u)+\phi_{\bar{\bG}_{k_2}})}J\,du\,dv,
\end{multline}
where $\bar{G}_{k_1}^x$, $\bar{G}_{k_1}^y$, $\bar{G}_{k_2}^x$, and $\bar{G}_{k_2}^y$ are the $x$ and $y$ components of $\bar{\bG}_{k_1}$ and $\bar{\bG}_{k_2}$. The Jacobian $J$ is
\begin{equation}
    J = \frac{\partial(x,y)}{\partial(u,v)} = \frac{1}{\begin{vmatrix}
\frac{\partial u}{\partial x} & \frac{\partial u}{\partial y} \\
\frac{\partial v}{\partial x} & \frac{\partial v}{\partial y}
\end{vmatrix}} = 1.
\end{equation}
Substituting \eqn{eqn:Gmn} in main text for modes $k_1$ and $k_2$ into $I_{\rm in}$ gives
\begin{multline}\label{eqn:int_ortho_in_3}
    I_{\rm in} = \bd_{k_1}\cdot\bd_{k_2}\int_{0}^{L}\int_{0}^{\frac{\sqrt{3}}{2}L}
    \cos{\biggl(\frac{2\pi}{\sqrt{3}L}(2m_1u+\sqrt{3}n_1v)
    +\phi_{\bar{\bG}_{k_1}}\biggr)} \\
    \cos{\biggl(\frac{2\pi}{\sqrt{3}L}(2m_2u+\sqrt{3}n_2v)
    +\phi_{\bar{\bG}_{k_2}}\biggr)}\,du\,dv.
\end{multline}
When $k_1\neq k_2$ (i.e., $m_1 \neq m_2$ and/or $n_1 \neq n_2$), the closed-form solution of $I_{\rm in}$ is
\begin{multline}\label{eqn:int_ortho_in_sol1}
I_{\rm in} = \bd_{k_1}\cdot\bd_{k_2} \frac{\sqrt{3} L^2}{{4 \pi ^2}}\biggl(\frac{\sin (\pi  (m_1-m_2)) \sin (\pi  (n_1-n_2)) \cos (\pi  (m_1-m_2+n_1-n_2)+\phi_{\bar{\bG}_{k_1}}-\phi_{\bar{\bG}_{k_2}})}{(m_1-m_2) (n_1-n_2)}\\
+\frac{\sin (\pi  (m_1+m_2)) \sin (\pi  (n_1+n_2)) \cos (\pi  (m_1+m_2+n_1+n_2)+\phi_{\bar{\bG}_{k_1}}+\phi_{\bar{\bG}_{k_2}})}{(m_1+m_2) (n_1+n_2)}\biggr).
\end{multline}
Since $\sin (\pi  (m_1-m_2)) \sin (\pi  (n_1-n_2))$ and $\sin (\pi  (m_1+m_2)) \sin (\pi  (n_1+n_2))$ in \eqn{eqn:int_ortho_in_sol1} are zero for any set of $m_1$, $n_1$, $m_2$, and $n_2$, $I_{\rm in}$ becomes zero when $k_1\neq k_2$.
For modes $k_1$ and $k_2$ that have the same $\bG$ and $\bd$ (i.e., $m_1=m_2$, $n_1=n_2$, and $\bd_{k_1}=\bd_{k_2}$), the closed-form solution of $I_{\rm in}$ (denoted as $\bar{I}_{\rm in}$) is
\begin{equation}\label{eqn:int_ortho_in_sol2}
    \bar{I}_{\rm in}(\phi_{\bar{\bG}_{k_1}},\phi_{\bar{\bG}_{k_2}}) = \frac{\sqrt{3}L^2}{4}\cos{(\phi_{\bar{\bG}_{k_1}}-\phi_{\bar{\bG}_{k_2}})}.
\end{equation}
Note that $\bar{I}_{\rm in}(\phi_{\bar{\bG}_{k_1}},\phi_{\bar{\bG}_{k_2}})$ is determined by the difference between $\phi_{\bar{\bG}_{k_1}}$ and $\phi_{\bar{\bG}_{k_2}}$.

Similar to $I_{\rm in}$, the inner product of real parts of out-of-plane modes $l_1$ and $l_2$ (\eqn{eqn:out_of_plane_mode} in main text) is given by
\begin{equation}\label{eqn:int_ortho_out_1}
    I_{\rm out} = \int_{S} \cos{(\hat{\bG}_{l_1}\cdot\mathbf{r}+\phi_{\bar{\bG}_{l_1}})}
    \cos{(\hat{\bG}_{l_2}\cdot\mathbf{r}+\phi_{\hat{\bG}_{l_2}})}\,dS.
\end{equation}
The closed-form solution of $I_{\rm  out}$ when $l_1\neq l_2$ is
\begin{multline}\label{eqn:int_ortho_out_sol1}
I_{\rm out} = \frac{\sqrt{3} L^2}{{4 \pi ^2}}\biggl(\frac{\sin (\pi  (m_1-m_2)) \sin (\pi  (n_1-n_2)) \cos (\pi  (m_1-m_2+n_1-n_2)+\phi_{\bar{\bG}_{k_1}}-\phi_{\bar{\bG}_{k_2}})}{(m_1-m_2) (n_1-n_2)}\\
+\frac{\sin (\pi  (m_1+m_2)) \sin (\pi  (n_1+n_2)) \cos (\pi  (m_1+m_2+n_1+n_2)+\phi_{\bar{\bG}_{k_1}}+\phi_{\bar{\bG}_{k_2}})}{(m_1+m_2) (n_1+n_2)}\biggr).
\end{multline}
Similar to \eqn{eqn:int_ortho_in_sol1}, \eqn{eqn:int_ortho_out_sol1} is zero for any set of $m_1$, $n_1$, $m_2$, and $n_2$.
For modes $l_1$ and $l_2$ have the same $\bG$ (i.e., $m_1=m_2$ and $n_1=n_2$), the closed-form solution of $\bar{I}_{\rm out}$ (denoted as $\bar{I}_{\rm out}$) is
\begin{equation}\label{eqn:int_ortho_out_sol2}
    \bar{I}_{\rm out}(\phi_{\hat{\bG}_{l_1}},\phi_{\hat{\bG}_{l_2}}) = \frac{\sqrt{3}L^2}{4}\cos{(\phi_{\hat{\bG}_{l_1}}-\phi_{\hat{\bG}_{l_2}})}.
\end{equation}

\section{Gradients of $\phi$ with respect to $A_{1,2,3}^{\rm top}$, $A_{4,5,6}^{\rm top}$, $A_{7,8,9}^{\rm top}$, and $A_{10,11,12}^{\rm top}$}\label{sec:grad_I}
A derivative of $\phi$ (\eqn{eqn:phi} in main text) with respect to $A_{1,2,3}^{\rm top}$ is expressed as
\begin{multline}\label{eqn:dphidA}
    \frac{\partial\phi}{\partial A_{1,2,3}^{\rm top}} = \sum_{i=1}^{3} \biggl   [2(\hat{J}_{\rm first}^i/\hat{J}_{\rm Bragg}^i-J_{\rm first}^{i, \rm exp}/J_{\rm Bragg}^{i, \rm exp})
    (1/\hat{J}_{\rm Bragg}^i\frac{\partial \hat{J}_{\rm first}^i}{\partial A_{1,2,3}^{\rm top}}-\hat{J}_{\rm first}^i/(\hat{J}_{\rm Bragg}^i)^2\frac{\partial \hat{J}_{\rm Bragg}^i}{\partial A_{1,2,3}^{\rm top}}
    )\\
    +  2(\hat{J}_{\rm second}^i/\hat{J}_{\rm Bragg}^i-J_{\rm second}^{i, \rm exp}/J_{\rm Bragg}^{i, \rm exp})
    (1/\hat{J}_{\rm Bragg}^i\frac{\partial \hat{J}_{\rm second}^i}{\partial A_{1,2,3}^{\rm top}}-\hat{J}_{\rm second}^i/(\hat{J}_{\rm Bragg}^i)^2\frac{\partial \hat{J}_{\rm Bragg}^i}{\partial A_{1,2,3}^{\rm top}}\biggr].
\end{multline}
To compute $\frac{\partial \hat{J}_{\rm first}}{\partial A_{1,2,3}^{\rm top}}$, $\frac{\partial \hat{J}_{\rm second}}{\partial A_{1,2,3}^{\rm top}}$, and $\frac{\partial \hat{J}_{\rm Bragg}}{\partial A_{1,2,3}^{\rm top}}$ in \eqn{eqn:dphidA}, the derivative $\hat{J}(\mathbf{k})$ (\eqn{eqn:intensity2} in main text) with respect to $A_{1,2,3}^{\rm top}$ is expressed as
\begin{equation}\label{eqn:dIdA}
    \frac{\partial\hat{J}}{\partial A_{1,2,3}^{\rm top}}(\mathbf{k}) = C \left(\hat{V}^{\rm top}+\hat{V}^{\rm bot}\right)\left(\frac{\partial\hat{V}^{\rm top}}{\partial A_{1,2,3}^{\rm top}}+\frac{\partial\hat{V}^{\rm bot}}{\partial A_{1,2,3}^{\rm top}}\right),
\end{equation}
where $C$ is a constant for the proportionality in \eqn{eqn:intensity2} in main text, and derivatives of $\hat{V}^{\rm top}$ and $\hat{V}^{\rm bot}$ (\eqns{eqn:v_hat} and \eqnx{eqn:v_hat_bot} in main text) with respect to $A_{1,2,3}^{\rm in}$ in \eqn{eqn:dIdA} are expressed as
\begin{align}\label{eqn:dVdA}
    \frac{\partial \hat{V}^{\rm top}}{\partial A_{1,2,3}^{\rm top}}(\mathbf{k}) &= \sum_{j=1}^{\Nlay}i\mathbf{k}\cdot(\bth{1}^{\rm in}(\br_0^j)+\bth{2}^{\rm in}(\br_0^j)+\bth{3}^{\rm in}(\br_0^j))\exp{(i\mathbf{k}\cdot(\br_0+\sum_k A_k^{\rm top}
    \bth{k}^{\rm in}(\br_0^j))}. \\
    \frac{\partial \hat{V}^{\rm bot}}{\partial A_{1,2,3}^{\rm top}}(\mathbf{k}) &= \sum_{j=1}^{\Nlay}i\mathbf{k}\cdot(\bth{1}^{\rm in}(\br_0^j)+\bth{2}^{\rm in}(\br_0^j)+\bth{3}^{\rm in}(\br_0^j))\exp{(i\mathbf{k}\cdot(\br_0+\sum_k A_k^{\rm bot}\bth{k}^{\rm in}(\br_0^j))}.
\end{align}
Therefore, $\frac{\partial \hat{J}_{\rm second}}{\partial A_{1,2,3}^{\rm top}}$, $\frac{\partial \hat{J}_{\rm second}}{\partial A_{1,2,3}^{\rm top}}$, and $\frac{\partial \hat{J}_{\rm Bragg}}{\partial A_{1,2,3}^{\rm top}}$ are computed as
\begin{align}
    \frac{\partial \hat{J}_{\rm first}^i}{\partial A_{1,2,3}^{\rm top}} &= \frac{\partial\hat{J}}{\partial A_{1,2,3}^{\rm top}}(\mathbf{k}_{\rm first}^i), \label{eqn:dIdA_first}\\
    \frac{\partial \hat{J}_{\rm second}^i}{\partial A_{1,2,3}^{\rm top}} &= \frac{\partial\hat{J}^i}{\partial A_{1,2,3}^{\rm top}}(\mathbf{k}_{\rm second}^i), \label{eqn:dIdA_second}\\
    \frac{\partial \hat{J}_{\rm Bragg}^i}{\partial A_{1,2,3}^{\rm top}} &= \frac{\partial\hat{J}}{\partial A_{1,2,3}^{\rm top}}(\mathbf{k}_{\rm Bragg}^i), \label{eqn:dIdA_Bragg}
\end{align}
where $\mathbf{k}_{\rm first}^i$, $\mathbf{k}_{\rm second}^i$, and $\mathbf{k}_{\rm Bragg}^i$  are reciprocal positions of the first and second superlattice and Bragg peaks in the $i$th Bragg spot. Substituting \eqns{eqn:dIdA_first}, \eqnx{eqn:dIdA_second}, and \eqnx{eqn:dIdA_Bragg} into \eqn{eqn:dphidA} provides $\frac{\partial\phi}{\partial A_{1,2,3}^{\rm top}}$. Note that we use the real part of \eqn{eqn:dphidA} for the gradients. Similar to $\frac{\partial\phi}{\partial A_{1,2,3}^{\rm top}}$, terms $\frac{\partial\phi}{\partial A_{4,5,6}^{\rm top}}$, $\frac{\partial\phi}{\partial A_{7,8,9}^{\rm top}}$, and $\frac{\partial\phi}{\partial A_{10,11,12}^{\rm top}}$ are computed and used to minimize  $\phi$.
\section{Double plate model for a TBG}\label{sec:double_plate_model}
The out-of-plane wave equations for a stack of two parallel plates (double plate model) are
\begin{align}\label{eqn:wave_double:1}
    D\nabla^4 w_1 + k_{\rm{int}} \left(w_1-w_2\right)+k_1w_1 = \rho h \frac{\partial^2w_1}{\partial t^2}, \\
    D\nabla^4 w_2 + k_{\rm{int}} \left(w_2-w_1\right)+k_2w_2 = \rho h \frac{\partial^2w_2}{\partial t^2}, \label{eqn:wave_double:2}
\end{align}
where $w_1$ and $w_2$ are the out-of-plane displacements of the top and bottom elastic plates, and $k_{\rm int}$, $k_{1}$, and $k_{2}$ are spring constants per unit area for the interlayer and substrate--layer interactions of the top and bottom graphene layers.
The spring constant $k_{\rm{int}}$ is computed from a central difference approximation of the second derivative of the total energy of a relaxed TBG by translating the top graphene layer along the out-of-plane direction obtained from molecular statics calculations.
The constants $k_1$ and $k_2$ are computed from the second derivative of \eqn{eqn:ext_energy} in main text,
\begin{equation}\label{eqn:kz}
k(z) = \rho_{\rm num} \frac{\partial^2U_{\rm ext}(z_0)}{\partial z_0^2}\bigg|_{z_0 = z},
\end{equation}
where $\rho_{\rm num}= 4/(3\sqrt{3}a^2)$ is the number density of monolayer graphene computed as 0.38 num/$\angstrom^2$ for $a = 1.42$~\angstrom.
Using \eqn{eqn:kz}, $k_1=k(\bar{z}^{\rm top})$ and $k_2=k(\bar{z}^{\rm bot})$, where $\bar{z}^{\rm top}$ and $\bar{z}^{\rm bot}$ are the average heights of the atoms in the top and bottom layers of the relaxed TBG relative to the substrate.
For the relaxed free-standing TBG, $k_1=k_2=0$.
The spring constants per unit area ($k_{\rm{int}}$, $k_1$, and $k_2$) for
relaxed free-standing and hBN-supported TBGs for a twist angle of $2.28\degree$
are presented in \tab{tab:double_plate_parameters}.

\begin{figure}[!htbp]\centering
         \includegraphics[trim = 0mm 0mm 0mm 0mm,clip,width=0.5\textwidth]{deg_vs_k.png}
\caption{Computed $k_{\rm int}$ of a free-standing TBG (free) and an hBN-supported TBG (hBN) as a function of twist angle.}
\label{fig:deg_k}
\end{figure}

\fig{fig:deg_k} shows the computed $k_{\rm int}$ values for a free-standing TBG and a TBG supported on an hBN substrate for a range of twist angles. For all twist angles, $k_{\rm int}$ for the hBN-supported TBG is higher than that of the free-standing TBG due to a decrease in interlayer spacing (see \fig{fig:inter_dis}) resulting from the presence of the substrate. For the hBN-supported TBG, $k_{\rm int}$ is almost constant across the range of twist angles considered with a value of approximately 53.8 meV$\angstrom^4$. In contrast, $k_{\rm int}$ of the free-standing TBG shows a somewhat stronger angular dependence, decreasing with increasing twist angle, with an overall reduction of about 3\% over the range of tested twist angles (from $0.48\degree$ to $3.48\degree$).

\begin{table}\centering
\begin{tabular}{c c c c}
\hline
 & $k_{\rm int}$ (meV/$\angstrom^4$) & $k_1$ (meV/$\angstrom^4$) & $k_2$ (meV/$\angstrom^4$)\\
\hline
free & 48.73 & 0.0 & 0.0\\
\hline
hBN & 53.94 & $-0.77$ & 65.96\\
\hline
\end{tabular}
\caption{Spring constants ($k_{\rm int}$, $k_{\rm 1}$, and $k_{\rm 2}$) for the double plate model (stack of two plates). The TBG is relaxed with IP$_2$ (hNN) for $\theta = 2.28\degree$. Results are presented for a free-standing TBG (free), and a TBG supported on an hBN substrate (hBN).}
\label{tab:double_plate_parameters}
\end{table}

To obtain the normal modes of the double plate model, we follow a procedure
similar to that in \sect{sec:in_plane_basis}, representing $w_1$ and $w_2$ as
\begin{align}\label{eqn:out_of_plane_displacement_TBG:1}
    w_1(\br,t) = \exp{(-i\omega^{\rm dbl}t)}\sum_{\bG}\eta_{\bG}^{1}\exp{(i(\bG\cdot\br+\phi_{\bG}))},\\
    w_2(\br,t) = \exp{(-i\omega^{\rm dbl}t)}\sum_{\bG}\eta_{\bG}^{2}\exp{(i(\bG\cdot\br+\phi_{\bG}))}, \label{eqn:out_of_plane_displacement_TBG:2}
\end{align}
where $\omega^{\rm dbl}$ is the double plate angular frequency (related to the ordinary frequency $f^{\rm dbl}$ through $\omega^{\rm dbl}=2\pi f^{\rm dbl}$), $\eta_{\bG}^{1}$ and $\eta_{\bG}^{2}$ are coefficients for $w_1$ and $w_2$ associated with $\bG$.
Substituting \eqns{eqn:out_of_plane_displacement_TBG:1} and \eqnx{eqn:out_of_plane_displacement_TBG:2} into \eqns{eqn:wave_double:1} and \eqnx{eqn:wave_double:2} results in
\begin{equation}\label{eqn:mat_G_double}
\begin{bmatrix}
\vert\bG\vert^4+(k_{\rm int}+k_1) && -k_{\rm int} \\
-k_{\rm int} && \vert\bG\vert^4+(k_{\rm int}+k_2)
\end{bmatrix}
\begin{bmatrix}
\eta_{\bG}^{1} \\
\eta_{\bG}^{2}
\end{bmatrix}
= \rho h (\omega^{\rm dbl})^2
\begin{bmatrix}
\eta_{\bG}^{1} \\
\eta_{\bG}^{2}
\end{bmatrix}.
\end{equation}
The eigenproblem in \eqn{eqn:mat_G_double} gives two eigenvalues and eigenvectors for each vector $\bG(m,n)$. The eigenvalues are:
\begin{multline}\label{eqn:eig_val_double}
\mu(m,n,o) =  \rho h(\omega^{\rm dbl}(m,n,o))^2 =  \\
\begin{cases} \frac{1}{2} (2 (G_x^2+G_y^2)^2-\sqrt{(k_1-k_2)^2+4 k_{\rm int}^2}+k_1+k_2+2 k_{\rm int}) & \text{for}\ o = 1 \\[6pt]
\frac{1}{2} (2 (G_x^2+G_y^2)^2+\sqrt{(k_1-k_2)^2+4 k_{\rm int}^2}+k_1+k_2+2 k_{\rm int}) & \text{for}\ o = 2,
\end{cases}
\end{multline}
and the eigenvectors are
\begin{multline}\label{eqn:bg}
\mathbf{g}(m,n,o) = [\eta_{\bG(m,n)}^{1}(o),\eta_{\bG(m,n)}^{2}(o)]^{\rm T}
= \\
\begin{cases}
\biggl(\left\vert \frac{-k_1+k_2+\sqrt{(k_1-k_2)^2+4 (k_{\rm int})^2}}{k_{\rm int}}\right\vert^2+4\biggr)^{-1/2}\biggl[
\frac{\sqrt{(k_1-k_2)^2+4(k_{\rm int})^2}-k_1+k_2}{k_{\rm int}}, 2\biggr]^{\rm T}
 & \text{for}\ o=1, \\[12pt]
\biggl(\left\vert \frac{-k_1+k_2+\sqrt{(k_1-k_2)^2+4 (k_{\rm int})^2}}{k_{\rm int}}\right\vert^2+4\biggr)^{-1/2}\biggl[
\frac{-\sqrt{(k_1-k_2)^2+4(k_{\rm int})^2}-k_1+k_2}{k_{\rm int}},2\biggr]^{\rm T}
 & \text{for}\ o=2.
\end{cases}
\end{multline}
After computing the eigenvalues and eigenvectors using the 60 $\bG$ values given in \sect{sec:in_plane_basis},
the results are ordered in increasing eigenvalue order using the simplified notation $\mu_k^{\rm dbl}$ (and $\omega^{\rm dbl}_k$) with associated eigenvectors $\mathbf{g}_k^{\rm dbl}$, where $k\in[1,120]$.
A specific $\bG(m,n)$ associated with $k$ is denoted as $\tilde{\bG}_k$.
We note that the eigenvalue ($\mu_0^{\rm dbl}$) associated with $\bG(0,0)$ has a non-zero value for $o=2$ in \eqn{eqn:bg} unlike the eigenvalue of the single plate model in \sect{sec:out_of_plane_basis} where the eigenvalue is zero for $\bG(0,0)$ (i.e., rigid translation). Therefore, the eigenvalue for $\bG(0,0)$ (i.e., $\mu(0,0,2)$) is denoted $\mu_0^{\rm dbl}$, and $\bG(0,0)$ is denoted as $\tilde{\bG}_0$.
The normal mode displacements for each $\bG$ for the top and bottom layers at position $\br_0$ follow as
\begin{align}\label{eqn:mode_w1}
    \bar{w}_1(\br_0) &= \eta_\bG^{1} \exp{(i(\bG\cdot\br_0+\phi_\bG))},\\
    \bar{w}_2(\br_0) &= \eta_\bG^{2} \exp{(i(\bG\cdot\br_0+\phi_\bG))}.
    \label{eqn:mode_w2}
\end{align}